\documentclass[aps,PRB,reprint,superscriptaddress]{revtex4-2}
\usepackage{soul}
\usepackage{graphicx}
\graphicspath{ {images2} }
\usepackage{amsmath}
\usepackage{amssymb}
\usepackage{float}
\usepackage[hidelinks]{hyperref}
\hypersetup{
  colorlinks,
  citecolor=magenta,
  linkcolor=blue,
  urlcolor=blue}
\usepackage[capitalise]{cleveref}
\usepackage{nicefrac}
\usepackage{xcolor}
\usepackage{epstopdf}

\newcommand{\dE}{{\partial^2{E_\text{HF}}}}
\newcommand{\bK}{{\mathbf K}}
\newcommand{\bq}{{\mathbf q}}
\newcommand{\br}{{\mathbf r}}
\newcommand{\bn}{{\mathbf n}}
\newcommand{\bs}{{\mathbf s}}
\newcommand{\bk}{{\mathbf k}}
\newcommand{\bp}{{\mathbf p}}
\newcommand{\ba}{{\mathbf a}}

\newcommand{\bA}{{\mathbf A}}
\newcommand{\bR}{{\mathbf R}}
\newcommand{\he}{{\hat e}}

\begin{document}

\title{Global phase diagram of charge neutral graphene in the quantum Hall regime for generic interactions}
	
\author{Suman Jyoti De}
\email{sumanjyotide@gmail.com}
\affiliation{Harish-Chandra Research Institute, A CI of Homi Bhabha National Institute, Chhatnag  Road, Jhunsi, Prayagraj 211019, India}
\author{Ankur Das}
\email{ankur.das@weizmann.ac.il}
\affiliation{Department of Condensed Matter Physics, Weizmann Institute of Science, Rehovot, 76100 Israel}
\author{Sumathi Rao}
\email{sumathi.rao@icts.res.in}
\affiliation{International Centre for Theoretical Sciences (ICTS-TIFR),
Shivakote, Hesaraghatta Hobli, Bangalore 560089, India}
\author{Ribhu K. Kaul}
\email{ribhu.kaul@psu.edu}
\affiliation{Department of Physics, The Pennsylvania State University, University Park, PA 16802, USA}
\author{Ganpathy Murthy}
\email{murthy@g.uky.edu}
\affiliation{Department of Physics \& Astronomy, University of Kentucky, Lexington, KY 40506, USA}	
	
\begin{abstract}
Monolayer graphene at charge neutrality in a quantizing magnetic field is a quantum Hall
ferromagnet. Due to the spin and valley (near) degeneracies, there is a plethora of possible
ground states. Previous theoretical work, based on a stringent ultra short-range assumption on the
symmetry-allowed interactions, predicts a phase diagram with distinct regions of spin-polarized,
canted antiferromagnetic, inter-valley coherent, and charge density wave order. While early
experiments suggested that the system was in the canted antiferromagnetic phase at a perpendicular
field, recent scanning tunneling studies universally find Kekul\'e bond order, and sometimes also
charge density wave order. Recently, it was found that if one relaxes the stringent assumption mentioned
above, a phase with coexisting canted antiferromagnetic and Kekul\'e
order exists in the region of the phase diagram believed to correspond to real samples. In this work, starting from the continuum limit appropriate for experiments,  we
present the complete phase diagram of $\nu=0$ graphene in the Hartree-Fock approximation, using
generic symmetry-allowed interactions, assuming translation invariant ground states up to an
intervalley coherence.  Allowing for a sublattice potential (valley Zeeman coupling), we find
numerous phases with different types of coexisting order. We conclude with a discussion of
the physical signatures of the various states.  
\end{abstract}
\maketitle

\section{Introduction}
\vspace*{0.5pt}
The quantum Hall effects (QHE) \cite{PrangeGirvin1990,dassarma1996:qhe}, discovered four decades ago \cite{IQHE_Discovery_1980} in semiconductor heterostructures, embody many phenomena observed there for the first time, but later found in many systems. The QHE represents the first and simplest topological insulator \cite{Kane_Mele_2DTI_PhysRevLett.95.146802,TIs_RevModPhys.82.3045}, as a consequence of which the electric and thermal Hall conductances are quantized. The bulk is insulating; charge and heat are carried by edge modes \cite{Halperin_Edge_1982} which are robust against disorder. Due to the quantization of kinetic energy into discrete values, Landau levels are also the first example of truly flat bulk bands. As a result, the bulk physics is controlled entirely by electron-electron interactions in a partially filled Landau level. Notably, this leads to the fractional QHE (FQHE) states \cite{FQHE_Discovery_1982}, which host excitations with fractional charge and statistics \cite{Laughlin_1983}. Due to spin or other internal degeneracies (such as valley or layer), ground states at some integer fillings are also controlled by interactions. Typically, interactions lead to ferromagnetism, as exemplified by the single layer $\nu=1$ spin ferromagnet \cite{Shivaji_Skyrmion}  or the bilayer $\nu=1$ state in $GaAs$ quantum wells \cite{QHFM_Fertig_1989,QHFM_Yang_etal_1994,QHFM_Moon_etal_1995}. Such quantum Hall ferromagnets also have interesting topological charged excitations such as skyrmions \cite{Shivaji_Skyrmion} or merons \cite{QHFM_Yang_etal_1994,QHFM_Moon_etal_1995}. 

Graphene \cite{Berger_etal_2004,Novoselov_etal_2004,Zhang_etal_2005, neto2009:rmp} is a single layer of Carbon atoms arranged in a honeycomb lattice with two sites ($A$ and $B$) in each unit cell. Near charge neutrality, low-energy electrons in graphene occur in two valleys at the two inequivalent zone corners $\bK$ and $\bK'$, and obey a Dirac equation in each valley. In a quantizing perpendicular magnetic field $B$, the Dirac spectrum produces particle-hole symmetric Landau levels $n=0,\pm1,\pm2\cdots$ with energy $E_{\pm n}\propto \pm\sqrt{B|n|}$ in each valley (ignoring the Zeeman splitting). Each Landau level is (almost) four-fold degenerate, with the four states being labelled by spin and valley. The $n=0$ Landau level is special; states in each valley are localized on one sublattice. Furthermore, since the states are comprised of a superposition of particle-like and hole-like states of the $B=0$ problem, there is valley mixing at the edge, leading to one edge mode with a particle-like dispersion and one with a hole-like dispersion \cite{Graphene_Edge_Herb_2006}. 

It was recognized very early \cite{alicea2006:gqhe,KYang_SU4_Skyrmion_2006,Herbut1,Herbut2}that partial filling of the $n=0$ manifold of Landau levels (called the zero-LLs or ZLLs) would lead to a rich set  of quantum Hall ferromagnets at integer fillings, the most interesting of which is the $\nu=0$ case, when two of the four ZLLs are filled. The particular superpositions of the spin and valley LLs that are occupied determine the nature of the ground state. 

In the non-interacting limit, the orbital part of the kinetic energy has an $SU(4)$ spin/valley symmetry in the continuum limit. The Zeeman coupling $E_Z$ splits the spin $\uparrow$ and spin $\downarrow$ Landau levels. The ground state is fully spin-polarized \cite{abanin2006:nu0}, with the $\bK\uparrow$, and $\bK'\uparrow$ Landau levels occupied. The chemical potential lies between the $\uparrow$-spin and $\downarrow$-spin sets of ZLLs. Due to the nature of the edge states of the ZLLs mentioned above \cite{Graphene_Edge_Herb_2006}, at each edge, the chemical potential intersects a particle-like level and a hole-like level of opposite spin. Since the system is symmetric under $U(1)$ spin rotations around the total $B$ field, these two modes cannot back-scatter, even if potential disorder is present. Therefore the system should be in a quantum spin Hall (QSH) phase \cite{QSH_Zhang_2004,Kane_Mele_2DTI_PhysRevLett.95.146802,abanin2006:nu0}. 

Now let us add electron-electron interactions. The dominant long-range part of the Coulomb interaction does not discriminate between spin and valley indices, being $SU(4)$ symmetric. The Coulomb interaction changes the single-particle spectrum at the edge, but leaves the two gapless, opposite-spin, charge-carrying edge modes intact, preserving the QSH phase \cite{brey2006:nu0}. In transport, such a state should show a two-terminal conductance of $2e^2/h$.

Initial experiments \cite{zhang2006:nu0,jiang2007:nu0,young2012:nu0,Maher_Kim_etal_2013} saw a trivial insulating state at $\nu=0$ without any protected edges. A seminal experiment measured the two-terminal conductance in tilted field \cite{young2014:nu0}, which allowed independent tuning of the Zeeman coupling. It was found that at large $E_Z$, the two-terminal conductance does indeed tend asymptotically to $2e^2/h$. Below a critical Zeeman energy $E_Z^*$, the system remains a trivial insulator. The gap at the edge vanishes continuously as one approaches  $E_Z\to E_Z^*$, indicating a second-order phase transition. 

The fact that the ground state at purely perpendicular field is not a QSH state means that  interactions beyond the $SU(4)$-symmetric Coulomb interactions must play an important role \cite{alicea2006:gqhe,KYang_SU4_Skyrmion_2006,Herbut1,Herbut2}. The ground state must be chosen by $SU(4)$-anisotropic residual interactions, arising from lattice-scale couplings. Low-energy effective symmetries inherited from the $B=0$ problem can be used to deduce a $U(1)$ valley symmetry at the four-Fermi level (reduced to a $Z_3$ symmetry when higher-Fermi interactions are included) \cite{alicea2006:gqhe}. One can classify the four-Fermi anisotropic residual interactions into two types; an Ising-like coupling in the valley space $v_z(\bq)$ and an $xy$-like coupling in the valley space $v_{xy}(\bq)$. See \cref{subsec:Hamiltonian} for the full definition of $v_z(\bq)$, $v_{xy}(\bq)$ and the interaction Hamiltonian. 

Since the residual interactions arise from lattice-scale couplings, and the magnetic length $\ell=\sqrt{\frac{\hbar}{eB}}$ is two orders of magnitude larger than the lattice spacing ($\ell\gg a$), the bare interactions are ultra-short-range (USR). The first step in  obtaining the effective interactions in the ZLL manifold is to project the bare interactions to this manifold. Operationally, this assumption implies that $v_{\mu}$ are independent of $\bq$ in the ZLL. It should be noted that if one projects USR bare interactions to a $N\neq0$ manifold in graphene, the effective interactions will not be USR \cite{Stefanides_Sodemann2022}.

Using the USR assumption for the interactions and building upon previous work \cite{alicea2006:gqhe,KYang_SU4_Skyrmion_2006,Herbut1,Herbut2}, Kharitonov \cite{kharitonov2012:nu0} found the phase diagram in the Hartree-Fock approximation for $\nu=0$ graphene. There are four phases: A fully polarized phase $F$, an antiferromagnetic phase $AF$ (which becomes a canted AF, or CAF in the presence of the Zeeman coupling), a charge density wave (CDW) phase, and a phase with inter-valley coherence (IVC), sometimes also called a Kekul\'e distorted or bond-ordered (KD/BO) phase. All the phase boundaries are first-order, except for the CAF to F transition, which is second-order. Upon the addition of a valley Zeeman coupling, the CDW and bond order coexist, leading to a partially sublattice polarized (PSP) phase \cite{Young_2018}, but the transition between the PSP and the CAF phases remains first-order.  

This picture indeed reproduces the phenomenology of $\nu=0$ graphene in tilted field \cite{young2014:nu0}, assuming that the anisotropic couplings are such that the system (in perpendicular field) is in the CAF phase \cite{kharitonov2012:nu0}. Subsequent magnon transmission experiments \cite{Magnon_transport_Yacoby_2018,Young_Skyrmion_Solid_Graphene_2019,Magnon_Transport_Assouline_2021,Magnon_transport_Zhou_2022} through a  $\nu=0$ region surrounded by ferromagnetic $\nu=1$ regions also confirm that coherently propagating magnetic excitations are present in the $\nu=0$ state. In the CAF phase the natural candidate is the gapless Goldstone mode associated with the spontaneous symmetry breaking of the $U(1)$ spin symmetry. 

More recently, three scanning tunneling studies \cite{li2019:stm,STM_Yazdani2021visualizing,STM_Coissard_2022} on $\nu=0$ graphene  perpendicular $B$ field have introduced more complexity into this picture. While they cannot directly confirm or rule out CAF order, all three see Kekul\'e bond order, and two of the three \cite{STM_Yazdani2021visualizing,STM_Coissard_2022} see CDW order as well. Partial alignment of the graphene layer with the hexagonal Boron Nitride (HBN) substrate is known \cite{KV_expt_2013,KV_DGG_2013,KV_HBN_Jung2015,KV_HBN_Jung2017}  to produce a sublattice potential/valley Zeeman coupling $E_V$, which can lead to CDW order. This origin of the CDW order was confirmed in one of the experiments \cite{STM_Yazdani2021visualizing}, while it remains unclear in the other experiment \cite{STM_Coissard_2022}. 

It is possible that the samples used in the scanning tunneling experiments are in a different phase from those used in transport experiments. However, the most parsimonious explanation is that CAF and bond order coexist in all samples, the CAF being undetected in STM experiments and the bond order being undetected in transport. 

Since there is no coexistence between CAF and BO in Kharitonov's phase diagram \cite{kharitonov2012:nu0}, one is led to re-examine the assumption of ultra-short-range interactions. From the point of view of an effective model in the ZLLs, there is no reason to assume any particular form for the symmetry-allowed interactions, because integrating out high energy Landau levels will lead to renormalizations in the form of the interactions \cite{RG_Murthy_Shankar_2002,RG_Bishara_Nayak_2009,RG_Sodemann_MacD2013,RG_Peterson_Nayak_2013,RG_Peterson_Nayak_2014}. In a recent work, three of us \cite{Das_Kaul_Murthy_2022} used this reasoning, generalizing the symmetry-allowed interactions $v_z(\bq),\ v_{xy}(\bq)$ to be arbitrary functions of $\bq$. This might seem to introduce an infinite number of new couplings.  Remarkably, in the continuum Hartree-Fock (HF) approximation, assuming translation symmetry is preserved up to an inter-valley coherence, only two independent numbers per type of coupling suffice \cite{Das_Kaul_Murthy_2022}, namely the Hartree and the Fock couplings, defined as 
\begin{equation}
    g_{\mu,H}=\frac{v_\mu(\bq=0)}{2\pi\ell^2};\ \  g_{\mu,F}=\int\frac{d^2q}{(2\pi)^2} v_{\mu}(\bq)e^{-q^2\ell^2/2}
    \label{eq:gmuhf}
\end{equation}
where $\mu=z,xy$. For ultra-short-range (USR) interactions $g_{\mu,H}=g_{\mu,F}$. Indeed, it was found \cite{Das_Kaul_Murthy_2022} that given certain inequalities between the Hartree and Fock couplings, there was a region of couplings which showed generic coexistence between the CAF and BO orders. We will refer to this as the B/CAF phase. 

It should be noted that non-USR interactions have implicitly been introduced earlier by Goerbig and collaborators in the context of effective nonlinear sigma models for $\nu=\pm1$ in graphene \cite{Lian_Rosch_Goerbig_2016,Lian_Goerbig_2017}. In $\nu=-1$ for example, only a single ZLL (some linear combination of the four possibilities) is occupied. As in any ferromagnet with single occupancy, the many-body wave function is completely antisymmetric in space. Thus, USR interactions cannot contribute to the energy of the state and the entire physics is controlled by the non-USR couplings \cite{Atteia_Goerbig_2021}. The case of $\nu=1$ is related to that of $\nu=-1$ by particle-hole symmetry. More recently, the connection of the parameters of the effective theory to the non-USR nature of the microscopic anisotropic interactions was made explicit by Atteia and Goerbig \cite{Atteia_Goerbig_2021}, once again in the context of $\nu=\pm1$. 

There is a different line of reasoning coming from Bernal-stacked bilayer graphene (BLG) in the quantum Hall regime \cite{Murthy_BLG_2017} which also leads to the non-USR condition $g_{\mu,H}\neq g_{\mu,F}$. Assuming solely nearest neighbor hoppings, the $B=0$ dispersion has quadratic band touchings at $\bK$ and $\bK'$. However, upon including the symmetry-allowed trigonal warping (a hopping between non-Bernal stacked sites in different layers) the quadratic band touching reconstructs into four Dirac cones \cite{McCann_Falko_trigonal_2006}. The inclusion of trigonal warping has a profound effect \cite{McCann_Falko_trigonal_2006} on the eight-fold (nearly) degenerate manifold of states near charge neutrality in a quantizing $B$. The upshot is that the the symmetry-allowed interactions, when projected into the low-energy manifold, now acquire structure on the scale of $\ell$ and no longer satisfy the USR condition $g_{\mu,H}=g_{\mu,F}$. This fact is crucial in obtaining phases which show coexistence between different kinds of order in BLG at $\nu=0$ \cite{Murthy_BLG_2017}. There is a deep analogy between MLG and BLG, as we will see; the states that we will uncover in the full phase diagram of MLG are identical to a subset of states found earlier in BLG \cite{Murthy_BLG_2017}.  

The purpose of this paper is to find the complete HF phase diagram of monolayer graphene in the continuum approximation, assuming translation invariance up to an inter-valley coherence. Our previous work \cite{Das_Kaul_Murthy_2022} was motivated by the STM experiments \cite{li2019:stm,STM_Yazdani2021visualizing,STM_Coissard_2022}, and confined to values of couplings thought to apply to real samples. Furthermore, the valley Zeeman coupling was ignored. We will explore the full phase diagram, in the presence of nonzero $E_Z,\ E_V$ for all possible $g_{z,H},g_{z,F},g_{xy,H},g_{xy,F}$. 

The majority of the results in the main text are for the case when the Hartree and Fock parts of a given coupling have the same sign: 
\begin{equation}
    \frac{g_{z,F}}{g_{z,H}}>0;\ \ \ \ \frac{g_{xy,F}}{g_{xy,H}}>0
\end{equation}
This seems natural for weak LL-mixing, when the renormalizations from integrating out the higher energy states are expected to be small compared to the bare values of the couplings. However, for strong LL-mixing, one may well have situations when the Hartree and Fock parts of a given coupling have opposite signs. We will present some interesting results in this case as well.  

To give a brief preview of our results. We find three coexistence phases: (i) The coexistence phase occurring near the BO/CAF boundary in the USR model, which was found earlier \cite{Das_Kaul_Murthy_2022}. This phase, which we call the B/CAF phase, also has a spin-valley entangled order even at $E_V=0$, which we label as SVEY (we will explain the notation shortly in \cref{subsec:ansatz_Hessian}). (ii) A phase occuring near the CDW/FM boundary in the USR model. This phase displays the coexistence between CDW and FM order, mediated by a spin-valley entangled order we label SVE+ (explained in \cref{subsec:ansatz_Hessian}).   (iii) When $\frac{g_{z,F}}{g_{z,H}}<0$ we find a phase where FM and SVEX/SVEY order (explained in \cref{subsec:ansatz_Hessian}) coexist, without any other order being present at $E_V=0$. We call this the FSVE phase. When $E_V>0$ many of these phases acquire a CDW order parameter, but remain largely unchanged otherwise. In sum, the full phase diagram of monolayer graphene for generic interactions is much richer than was previously believed.    

The plan of the paper is as follows: In \cref{sec:model} we will briefly review the previous work on the effective model for MLG in the continuum approximation. We will generalize the interactions to be non-USR, discuss the HF approximation and find the ground state energy. Also in \cref{sec:model}, we present a parameterization of translation-invariant $\nu=0$ states  \cite{Doucot_Goerbig_Skyrmion_2008,Lian_Goerbig_2017,Goerbig_nu0_Skyrmion_Zoo_2021} whose energy depends on four angles.  It turns out that the states that have been found in the USR limit \cite{kharitonov2012:nu0} can be characterized in terms of a single angle. Instabilities of these states, which can be computed analytically,  will signal the occurrence of more complex states with coexisting order parameters. In general, the actual ground state in any region of coexistence has to be found numerically. In \cref{sec:results} we present our results; since there are six independent tuning parameters, we will present many two-dimensional sections through the space of coupling constants. Each two-dimensional section will satisfy different inequalities between the Hartree and Fock couplings.  We end with our summary, conclusions, and open questions in \cref{sec:conclusions}. The appendices contain the details of our calculations, analytical expressions for various instabilities, and sample results for strong Landau-level mixing.

\section{Model Hamiltonian, parameters, and methods}
\label{sec:model}

We choose the primitive translation vectors for graphene as $\ba_1=a\he_x$, $\ba_2=a(\frac{\he_x}{2}+\frac{\sqrt{3}\he_y}{2})$, with the general Bravais lattice site $\bR=n_1\ba_1+n_2\ba_2$. The noninteracting Hamiltonian of graphene at zero magnetic field, suppressing the spin index for the moment, is 
\begin{align}
    H_0&=-t\sum\limits_{\bR,j} c^{\dagger}_{A{\bR}}c_{B{\bR}+{\bf d}_j}
    +h.c.
\label{eq:H-realspace}
\end{align}
where $t$ is the nearest-neighbor hopping matrix element, $c_{A\bR},c_{B\bR}$ destroy electrons at the $A$ and $B$ sublattice sites of the Bravais site $\bR$, the sum on $j=1,2,3$ with ${\bf d_1}=0,{\bf d_2}=\ba_1-\ba_2,{\bf d_3}=-\ba_2$ and $h.c.$ stands for hermitian conjugate. Note that there is no spin-orbit coupling in the Hamiltonian of \cref{eq:H-realspace}. First-principles \cite{SOC_Graphene_MacD2006,SOC_Graphene_Yao2007} and tight-binding calculations \cite{SOC_Graphene_Paco2006,SOC_Graphene_Sandler2007} show that the spin-orbit coupling in graphene is of the order of tens  of $\mu eV$, smaller than any other energy scale in the problem. We will set the spin-orbit coupling to zero here and henceforth. Fourier transforming \cref{eq:H-realspace}, we obtain the Bloch Hamiltonian at wave-vector $\bk$ as a matrix in the sublattice space 
\begin{equation}
    H(\bk)=-t\left(\begin{array}{cc}
    0&f(\bk)\\
    f^*(\bk)&0
    \end{array}\right)
\end{equation}
where $f(\bk)=1+e^{i\bk\cdot(\ba_1-\ba_2)}+e^{-i\bk\cdot\ba_2}$. It is easily checked that $f(\bk)$ vanishes at the two inequivalent zone corners (valleys) $\bK=\he_x \frac{4\pi}{3a}=-\bK'$ indicating Dirac crossings. The low-energy effective Hamiltonian in the $\bK$ valley ($\bk=\bK+\bp$, $|\bp|\ll 2\pi/a$) can be obtained by expansion. 
\begin{equation}
    H_{\bK}(\bp)=\frac{ta\sqrt{3}}{2}\left(\begin{array}{cc}
    0&p_x-ip_y\\
    p_x+ip_y&0
    \end{array}\right)
\end{equation}
The low-energy effective Hamiltonian at the $\bK'$ valley can be obtained by the identity $H_{\bK'}(\bp)=(H_{\bK}(-\bp))^*$. 

In the continuum limit, we turn on a weak perpendicular magnetic field $B_{\perp}$ by allowing the Hamiltonian to act on slowly varying envelope functions. Operationally, this involves promoting $p_i\to -i\partial_i\to -i\partial_i+eA_i(\br)$, where the electron's charge is $-e$, and $\bA(\br)$ is the vector potential satisfying $\nabla\times\bA=\he_zB_{\perp}$. In order for the continuum limit to be justified, the magnetic length has to be much larger than the lattice spacing; $\ell=\sqrt{\frac{\hbar}{eB_{\perp}}}\gg a$. This is extremely well-satisfied for realistic fields. 

Next, one chooses Landau gauge $\bA=B_{\perp} x\he_y$, and imposes periodic boundary conditions in the $y$-direction with a length $L_y$. Let us define the Landau level wavefunctions as
\begin{equation}
    \langle x,y|n,k\rangle=\frac{e^{iky}}{\sqrt{L_y}}\Phi_n\left(\frac{x-k\ell^2}{\ell}\right)
\end{equation}
where $\Phi_n$ are the normalized harmonic oscillator wavefunctions. Note that, here and henceforth, $k$ is a one-dimensional guiding center label, and not a two-dimensional momentum. 
Now it is straightforward to see that the ZLL states are $\left(0,|n=0,k\rangle\right)^T$ in the $\bK$ valley, and $\left(|n=0,k\rangle,0\right)^T$ in the $\bK'$ valley. Thus, in the ZLLs, valley and sublattice are locked together. 
Now we are ready to present our model Hamiltonian. 

\subsection{Hamiltonian and the Hartree-Fock Approximation}
\label{subsec:Hamiltonian}

In what follows, we will index the fermion operators with a valley index $\alpha,\beta$, which can be $\bK\equiv0$ or $\bK'\equiv1$, and a spin index $s=\uparrow\equiv0$ or $s=\downarrow\equiv1$. In this notation, the non-interacting Hamiltonian of the ZLLs is 
\begin{equation}
    H_{1b}=-\sum\limits_{\alpha,s,k} \left(E_Z (-1)^s+E_V(-1)^\alpha\right) c^{\dagger}_{\alpha,s,k}c_{\alpha,s,k}
\end{equation}
which introduces the Zeeman energy $E_Z$ and the valley Zeeman/sublattice potential $E_V$.

Turning to interactions, Alicea and Fisher \cite{alicea2006:gqhe} noted that, in addition to the $SU(4)$-symmetric Coulomb interaction, two other types of low-energy effective interactions were allowed by $SU(2)$ spin-rotation symmetry and momentum conservation in the $B=0$ problem. We recall that the spin-orbit coupling is negligible \cite{SOC_Graphene_Paco2006,SOC_Graphene_MacD2006,SOC_Graphene_Yao2007,SOC_Graphene_Sandler2007} and has been neglected. A $U(1)$ symmetry in the valley space (separate conservation of the number of electrons in each valley) emerges when restricting oneself to four-Fermi interactions. Upon including six-Fermi terms this is reduced to a $Z_3$ symmetry because $3(\bK-\bK')$ is a reciprocal lattice vector. The full interaction Hamiltonian for the ZLLs in monolayer graphene, in the Landau gauge basis discussed earlier, is 
\begin{widetext}
\begin{align}
H=&H_{1b}+H_{int}\\
H_{int}=&H_{Coul}+H_Z+H_{xy}\\
H_Z=&\frac{1}{2L_xL_y}\sum\limits_{k,k',\bq}v_z(\bq)e^{-iq_x(k-k'-q_y)\ell^2}e^{-(q\ell)^2/2} :c^{\dagger}_{k-q_y}\tau_Zc_k c^\dagger_{k'+q_y}\tau_Zc_{k'}:\\
H_{xy}=&\frac{1}{2L_xL_y}\sum\limits_{k,k',\bq}v_{xy}(\bq)e^{-iq_x(k-k'-q_y)\ell^2}e^{-(q\ell)^2/2} \left(:c^{\dagger}_{k-q_y}\tau_xc_k c^\dagger_{k'+q_y}\tau_xc_{k'}:+:c^{\dagger}_{k-q_y}\tau_yc_k c^\dagger_{k'+q_y}\tau_yc_{k'}:\right)
\label{eq:Ham}
\end{align}
\end{widetext}
We have used a shorthand notation where the spin/valley indices are suppressed and summed, and $\tau_i$ is a Pauli matrix in the valley space (it acts as the identity in the spin space). Furthermore, since the Coulomb interaction has no role to play in selecting the ground state, we drop it henceforth. In the generic case, when $E_Z,\ E_V>0$, the Hamiltonian has a $U(1)_s$ spin-rotation symmetry generated by total $\sigma_z$, a $U(1)_v$ valley-rotation symmetry generated by total $\tau_z$ (which is also the difference between the number of electrons in the $\bK$ and $\bK'$ valleys), and an entangled spin-valley $U(1)_{sv}$ symmetry generated by total $\tau_z\sigma_z$. In the fine-tuned case $E_Z=0$ (not realizable in experimental samples) the spin-rotation symmetry is enhanced to $SU(2)_s$. In the fine-tuned case $E_V=0$, the valley symmetry is enhanced to a $U(1)_v\otimes Z_{2v}$, where the $Z_{2v}$ represents the symmetry exchanging the two valleys. 
 
In the HF approximation, one looks for the single Slater determinant that has the right electron filling and minimizes the energy. Such a state, symbolically written as $|HF\rangle$, can be completely characterized by the set of 1-body expectation values.
\begin{equation}
    \Delta_{\alpha\beta}^{ss'}(k,k')=\langle HF|c^{\dagger}_{\alpha,s,k}c_{\beta,s',k'}|HF\rangle
\end{equation}
We will restrict the space of HF states to those obeying translation invariance, up to an intervalley coherence. This means that the $\Delta$ becomes diagonal in $k$ and independent of it. 
\begin{equation}
    \Delta_{\alpha\beta}^{ss'}(k,k')=\delta_{kk'} \Delta_{\alpha\beta}^{ss'}
    \label{eq:IVC}
\end{equation}
Let us make the idea of translation invariance up to an intervalley coherence more explicit. The $\Delta$ we have assumed allows nonzero averages of the form $\langle HF|c^{\dagger}_{\bK,s}c_{\bK's'}|HF\rangle$. Clearly these break lattice translations since $\bK$ and $\bK'$ are not identical up to a reciprocal lattice vector. Allowing such averages introduces a new set of reciprocal lattice vectors which are $\bK-\bK'$ and all their rotated versions, and leads to translation invariance with an enlarged unit cell of size $\sqrt{3}\times\sqrt{3}$ as compared to the original. Indeed, this is exactly what is seen STM experiments \cite{li2019:stm,STM_Yazdani2021visualizing,STM_Coissard_2022}.  The ansatz of Eq. \ref{eq:IVC} makes sure that there is no translation symmetry breaking beyond the minimal one implied by intervalley coherence. 

 In general, the matrix $\Delta$ is the projector on to the linear space of the occupied states. Given that two orthogonal linear combinations of the four ZLLs (call them $|f_1\rangle$ and $|f_2\rangle$) are occupied at $\nu=0$ we can write 
\begin{equation}
    \Delta=|f_1\rangle\langle f_1|+|f_2\rangle\langle f_2|
\end{equation}
We can now express the HF energy of the Hamiltonian of \cref{eq:Ham} per guiding center in terms of $\Delta$, with $N_\phi=\frac{L_xL_y}{2\pi\ell^2}$, as
\begin{widetext}
\begin{align}
\frac{E_{HF}}{N_\phi}=&-E_Z Tr\left[\sigma_Z\Delta\right]-E_VTr\left[\tau_Z\Delta\right]+\frac{g_{z,H}}{2}\left(Tr[\tau_Z\Delta]\right)^2-\frac{g_{z,F}}{2} Tr\left[\tau_Z\Delta\tau_Z\Delta\right]\nonumber\\
&+\frac{g_{xy,H}}{2}\left\{\left(Tr[\tau_x\Delta]\right)^2+\left(Tr[\tau_y\Delta]\right)^2\right\}-
\frac{g_{xy,F}}{2}\left(Tr[\tau_x\Delta\tau_x\Delta]+Tr[\tau_y\Delta\tau_y\Delta]\right)
\end{align}
Note that $g_{\mu,h}$ and $g_{\mu,F}$ are defined as in \cref{eq:gmuhf}.
\end{widetext}
\subsection{Ansatz for States, Instabilities, and Order Parameters}
\label{subsec:ansatz_Hessian}

We will start with an efficient parameterization  \cite{Doucot_Goerbig_Skyrmion_2008,Lian_Goerbig_2017,Goerbig_nu0_Skyrmion_Zoo_2021} for the two orthogonal occupied states $|f_1\rangle$ and $|f_2\rangle$. This parameterization has been used not only for uniform states but also for describing skyrmions \cite{Goerbig_nu0_Skyrmion_Zoo_2021}.
\begin{align}
    |f_1\rangle=&\cos{\frac{\alpha_1}{2}}|\bn\rangle\otimes|\bs\rangle+e^{i\beta_1}\sin{\frac{\alpha_1}{2}}|-\bn\rangle\otimes|-\bs\rangle\\
    |f_2\rangle=&\cos{\frac{\alpha_2}{2}}|\bn\rangle\otimes|-\bs\rangle+e^{i\beta_2}\sin{\frac{\alpha_2}{2}}|-\bn\rangle\otimes|\bs\rangle
\end{align}
where $\bn=\sin{\theta_p}\cos{\phi_p}\he_x+\sin{\theta_p}\sin{\phi_p}\he_y+\cos{\theta_p}\he_z$, and $\bs=\sin{\theta_s}\cos{\phi_s}\he_x+\sin{\theta_s}\sin{\phi_s}\he_y+\cos{\theta_s}\he_z$ are unit vectors indicating the directions of the state on the valley and spin Bloch spheres respectively. The spinors $|\bn\rangle$ and $|\bs\rangle$ are defined in the standard way
\begin{equation}
    |\bn\rangle=\left(\begin{array}{c}
    \cos{\frac{\theta_p}{2}}\\
    e^{i\phi_p}\sin{\frac{\theta_p}{2}}\end{array}\right);\ \ |\bs\rangle=\left(\begin{array}{c}
    \cos{\frac{\theta_s}{2}}\\
    e^{i\phi_s}\sin{\frac{\theta_s}{2}}\end{array}\right)
\end{equation}
In going from $|\bn\rangle\to|-\bn\rangle$ one substitutes $\theta_p\to\pi-\theta_p$ and $\phi_p\to\phi_p+\pi$, and likewise for $\bs$. 

Given this ansatz, which depends on eight angles, we compute the HF energy. 
\begin{widetext}
\begin{align}
    \text{E}_{HF}=&-E_Z \cos \theta_s \left[\cos \alpha_1-\cos \alpha_2\right]-E_V \cos \theta_p [\cos \alpha_1+\cos \alpha_2] + \frac{g_{z,H}}{2} \cos ^2 \theta_p \left[\cos \alpha_1+\cos \alpha_2\right]^2\nonumber\\
&-\frac{g_{z,F}}{16} \left[4 \cos ^2\theta_p \left(\cos (2 \alpha_1)+\cos (2 \alpha_2)\right)-8 \sin ^2\theta_p \left(\cos \alpha_1 \cos \alpha_2-\sin \alpha_1
\sin \alpha_2 \cos (\beta_1+\beta_2)\right)+8\right]\nonumber\\
&+\frac{g_{xy,H}}{2} \sin ^2\theta_p \left[\cos \alpha_1+\cos \alpha_2\right]^2
-\frac{g_{xy,F}}{16} \bigg[8 \sin ^2\theta_p \left(\cos \alpha_1 \cos \alpha_2-\sin \alpha_1 \sin \alpha_2 \cos
(\beta_1+\beta_2)\right)\nonumber\\
&+4 \sin ^2\theta_p \left(\cos (2 \alpha_1)+ \cos (2 \alpha_2)\right)-16\left(\cos \alpha_1 \cos \alpha_2-1\right)\bigg]
\end{align}
\end{widetext}

There are three noteworthy features of this energy. The first is that it is independent of $\phi_p$ and $\phi_s$. This results from the $U(1)_v$ and  $U(1)_s$ symmetries of the Hamiltonian. Therefore we can set $\phi_s=\phi_p=0$ without loss of generality. The second feature is that the  dependence of the energy on $\beta_1,\beta_2$ occurs only in the interacting part, and only in the combination $\beta_1+\beta_2$. The reason the one-body HF energy does not involve $\beta_1,\beta_2$ is that the averages of $\tau_z,\ \sigma_z$ do not involve $\beta_i$.
\begin{align}
\langle f_i|\tau_z|f_i\rangle=&\cos{\alpha_i}\cos{\theta_p}\\
\langle f_i|\sigma_z|f_i\rangle=&-(-1)^i\cos{\alpha_i}\cos{\theta_s}
\end{align}
The dependence of the interaction energy solely on $\beta_1+\beta_2$ arises from the $SU(2)_{spin}$ symmetry of the interactions, which implies that the $U(1)$ rotation $|\bs\rangle\rightarrow e^{i\chi/2}|\bs\rangle;\ |-\bs\rangle\rightarrow e^{-i\chi/2}|-\bs\rangle$ cannot change the interaction energy. 
Ignoring overall phase factors, this rotation has the net effect $\beta_1\rightarrow\beta_1-\chi;\ \beta_2\rightarrow\beta_2+\chi$. This demonstrates that the energy can only depend on $\beta_1+\beta_2$. Thirdly, the dependence on $\beta_1+\beta_2$ occurs via the term $\cos(\beta_1+\beta_2)$, which appears linearly. Depending on the sign of its coefficient, the minimum energy will occur at $\cos(\beta_1+\beta_2)=\pm1$.

The bottom line is that the minimum of the HF energy for uniform states can be found in a subspace in which $|f_1\rangle,|f_2\rangle$ can both be chosen real. 

Below, we will call states which have Kekul\'e/BO, and/or CDW order B/CO states (because they have both bond order and/or charge order). The states originally found by Kharitonov \cite{kharitonov2012:nu0} can be represented in terms of the above angles as follows (details in Appendix A),  
\begin{widetext}
\begin{align}
& |FM\rangle=|\alpha_1=0, \alpha_2=\pi, \theta_p=\pi/2, \theta_s=0, \beta_1=\beta_2=\pi\rangle\\
& |CAF\rangle=|\alpha_1=\theta_{CAF}, \alpha_2=\pi-\alpha_1, \theta_p=\pi/2, \theta_s=0,\beta_1=\beta_2=\pi\rangle; \ g_{xy,F}<0\\
& |B/CO\rangle=|\alpha_1=\alpha_2=0,\theta_p=\theta_{B/CO},\theta_s=0,\beta_1=\beta_2=0\rangle
\end{align}
\end{widetext}
where 
\begin{align}
    &\theta_{CAF}=\cos^{-1}\left(\frac{E_Z}{2|g_{xy,F}|}\right); \ \ g_{xy,F}<0\\ 
    &\theta_{B/CO}=\cos^{-1}\left(\frac{E_V}{g_V}\right); \ \ g_V>E_V\\
    &\theta_{B/CO}=0; \ \ g_V<E_V\\
    &g_V=2g_{z,H}-g_{z,F}-2g_{xy,H}+g_{xy,F}
\end{align}\
If $\theta_{B/CO}=0$ the state is a pure CDW, if $0<\theta_{B/CO}<\pi/2$ it has coexisting BO and CDW order, and if $\theta_{B/CO}=\pi/2$ the system is in a pure Kekul\'e state. The reason we call these states ``simple" is that they can all be described by at most a single nontrivial angle, which can be analytically computed as a function of the couplings. Generic states may depend on more than one nontrivial angle, in which case it is not possible to solve for the angles or the ground state energy analytically. 

The ground state energies of the simple states are 
\begin{align}
E_\text{FM}=&-2 (E_Z+g_{xy,F})-g_{z,F}\nonumber\\
E_\text{CAF}=&\frac{E_Z^2}{2 g_{xy,F}}-g_{z,F};\ 0<E_Z<-2g_{xy,F}\nonumber\\
E_\text{B/CO}=&-\frac{E_V^2}{g_V}-g_{xy,F}+2 g_{xy,H};\ 0<E_V<g_V\nonumber\\
E_\text{CDW}=&-2 E_V-g_{z,F}+2 g_{z,H}
\label{eq:E_simple}
\end{align}

Our strategy is to examine the stability of these ``simple" states by finding the eigenvalues of the Hessian matrix of second derivatives of the energy functional with respect to the four angles $\alpha_1,\alpha_2,\theta_p,\theta_s$. 
\begin{equation}
    {\cal E}_{ij}=\frac{\partial^2 E_{HF}}{\partial\chi_i\partial\chi_j}
\end{equation}
where $\chi_i$ represents all four angles. For the ``simple" states one can compute the entire Hessian matrix analytically, and also obtain the eigenvalues analytically. 

All eigenvalues being positive means the state is stable to arbitrary small deformations. As the coupling constants are varied, a formerly positive eigenvalue may vanish, indicating an instability of the state in question. This allows us to map out the regions of stability of the ``simple" states in our coupling constant space. 

It can happen that when some of the angles take particular values, the projector on to the occupied subspace becomes independent of certain other angles. This occurs in the FM and B/CO phases. Consequently, certain rows and columns ${\cal E}_{ij}$ vanish, which means that one eigenvalue always vanishes in that state independent of the coupling constants. In such cases, the instability is marked by the vanishing of an eigenvalue that does depend on coupling constants. Once the region of possible coexistence has been found, we use numerical self-consistent Hartree-Fock to obtain the ground state and confirm the coexistence predicted by the Hessian. 

In preparation for showing the results, let us list all the order parameters which we will encounter and the symmetries they break. We have chosen a parameterization in which the projector matrix of the occupied states is real. Thus, out of all possible hermitian matrices that represent order parameters, only real matrices will have nonzero expectation values. 
\begin{align}
FM=&\langle \sigma_z\rangle/2\nonumber\\
CAF=&\langle \tau_z\sigma_x\rangle/2\nonumber\\
BO=&\langle\tau_x\rangle/2\nonumber\\
CDW=&\langle\tau_z\rangle/2\nonumber\\
SVEX=&\langle\tau_x\sigma_x\rangle/2\nonumber\\
SVEY=&\langle\tau_y\sigma_y\rangle/2\nonumber\\
SVE\pm=&\langle\tau_x\sigma_x\pm\tau_y\sigma_y\rangle/2
\label{eq:orderparameters}
\end{align}
The SVE (spin-valley entangled) type of order parameters are so called because they break the spin and valley symmetries simultaneously in an entangled way. 

Let us examine the symmetries broken by the various order parameters. In the fine-tuned case $E_Z=0$ (not realizable in experiment) the FM order parameter spontaneously breaks the $SU(2)_s$ symmetry. In the fine-tuned case $E_V=0$ (which is potentially realizable in experiment) the CDW order spontaneously breaks the $Z_{2v}$ symmetry.  In the generic case $E_Z,\ E_V\neq0$ the FM and CDW order parameters do not break any symmetries of the Hamiltonian. The CAF order parameter breaks $U(1)_s$ and $U(1)_{sv}$, but preserves $U(1)_v$. Bond order breaks $U(1)_v$ and $U(1)_{sv}$ but preserves $U(1)_s$. The SVE order parameters break $U(1)_s$ and $U(1)_v$ but preserve $U(1)_{sv}$. All three $U(1)$s are spontaneously broken in the B/CAF phase.

We emphasize that while the $U(1)_s$ symmetry is protected by the assumed vanishing of the spin-orbit coupling in graphene (\cref{eq:H-realspace}), there is no such protection for $U(1)_v$ or $U(1)_{sv}$. As mentioned in the introduction, since $3(\bK-\bK')$ is a reciprocal lattice vector, any $U(1)$ related to the valley space will be broken down to a $Z_3$ upon including six-Fermi and higher interactions. Thus, there are no Goldstone modes associated with the spontaneous breaking of the $U(1)_v$ symmetry. However, a spontaneous breaking of the $U(1)_s$  and/or the $U(1)_{sv}$ symmetries will lead to a phase with a Goldstone mode. 

\section{Results}
\label{sec:results}

There are six coupling constants in our Hamiltonian, $E_Z,E_V,g_{z,H},g_{z,F},g_{xy,H},g_{xy,F}$. Since the full six-dimensional phase diagram is impossible to visualize, we will be forced to take two-dimensional cuts. 

We will often draw a correspondence between MLG and Bernal-stacked BLG \cite{Kharitonov_BLG1_2012,Kharitonov_BLG2_2012}. In BLG, in addition to spin and valley, an orbital index $n=0,1$ also appears \cite{McCann_Falko_trigonal_2006}. For BLG states that have no orbital mixing, and are orbitally symmetric, there is a one-to-one correspondence with states in MLG. Specifially, the role of $E_V$ in MLG is played by the perpendicular electric field $D$ in BLG \cite{Murthy_BLG_2017}. 

While RG arguments tell us that generic effective interactions must have nonzero range, one does not know precisely how the LL-mixing and the intergration of high-energy states affects the Hartree and Fock parts of the couplings in the ZLL manifold. For weak LL-mixing, a natural assumption is that the sign of $g_{\mu,H},g_{\mu,F}$ are the same, but the magnitudes could be different. The majority of the results we present in the main text assume this to be true. For strong LL-mixing, it is conceivable that the effective $g_{\mu,H}$ and $g_{\mu,F}$ have opposite signs. For completeness we have analyzed this case as well, though most of the details are relegated to the appendices. 

In \cref{subsec:evzero}, we will present results $E_V=0$, restricting the Hartree and Fock parts of the couplings to have the same sign. This will allow us to examine how the original Kharitonov phase diagram \cite{kharitonov2012:nu0} changes when we relax the USR assumption. We confirm the existence of the B/CAF  phase found earlier \cite{Das_Kaul_Murthy_2022} for $g_{xy,F}/g_{xy,H}>1$. This phase also has the order parameter SVEY (\cref{eq:orderparameters}). Thus, this phase breaks $U(1)_s$, $U(1)_v$, and $U(1)_{sv}$ spontaneously.  Another coexistence phase occurs near the boundary between the FM and CDW phases when $g_{z,F}/g_{z,H}>1$. We call this the SVE+ phase because the corresponding order parameter (\cref{eq:orderparameters}) is nonzero in this phase. This phase breaks $U(1)_s$ and $U(1)_v$ but preserves $U(1)_{sv}$. The bilayer analog of this phase was found earlier in Bernal-stacked BLG \cite{Murthy_BLG_2017}, although at nonzero $E_V$. 

Next, in \cref{subsec:evnon0_ratioplus} we will keep the restriction that the Hartree and Fock parts of the couplings have the same sign, but turn on $E_V$. Already, for USR interactions, one finds a phase with coexisting CDW and Kekul\'e order \cite{Goerbig_nu0_Skyrmion_Zoo_2021}, which we call the B/CO phase. Guided by the intuition that new phases are most likely to appear near the transition lines of the original USR model, we will examine these carefully. When $0>g_{xy,H}>g_{xy,F}$, the B/CAF phase shrinks as $E_V$ increases (all other couplings remaining constant), and vanishes for large enough $E_V$. For any $E_V>0$, both SVEX and SVEY order parameters are nonzero in the B/CAF phase. However, the symmetries that are  spontaneously broken do not change in any of the phases at small $E_V$. 

Going to larger $E_V$ is even more interesting: in certain cases, even if coexistence does not occur for $E_V=0$, it can occur for intermediate $E_V$, and disappear for large $E_V$. For example, for $g_{z,F}=0.75g_{z,H},\ \ g_{xy,F}=0.75g_{xy,H}$, both the B/CAF and SVE+  phases are absent at $E_V=0$. However, both phases are present for a range of intermediate $E_V$. This is consistent with earlier results for BLG \cite{Murthy_BLG_2017}, which also found coexistence phases for intermediate values of the perpendicular electric field $D$, which plays the same role in BLG as $E_V$ plays in MLG. 

From the point of view of experiment, the most interesting phases is the B/CAF phase, which occurs  for $0>g_{xy,H}>g_{xy,F}$. At any nonzero $E_V$, the B/CAF phase evolves into one with coexisting CAF, Kekul\'e, and CDW order. An analogous phase breaking all three $U(1)$ symmetries was found earlier in BLG \cite{Murthy_BLG_2017}. (Of course, the $U(1)_v$ symmetry will be reduced to a $Z_3$ upon the inclusion of higher-Fermi interactions, and thus does not lead to a Goldstone mode). Such a phase would be consistent with observations of magnon transmission \cite{Magnon_transport_Yacoby_2018,Magnon_Transport_Assouline_2021,Magnon_transport_Zhou_2022} through the $\nu=0$ state, which implies some kind of magnetic order, and the STM experiments \cite{li2019:stm,STM_Yazdani2021visualizing,STM_Coissard_2022} which show Kekul\'e and CDW order. 

Finally, in  \cref{subsec:evzero_ratiominus} we will present some results for the signs of $g_{\mu,H}$ and $g_{\mu,F}$ being different for either or both of $g_z$ and $g_{xy}$. As may be expected, the topology of the phase diagram can change considerably in such cases. However, a new type of coexistence phase appears, which is not present when the signs of the H and F parts of both couplings are identical. This phase, which we call the FSVE phase, has coexisting FM and SVEY order, without any other order being present at $E_V=0$. Such a state breaks $U(1)_s$ and $U(1)_v$ while preserving $U(1)_{sv}$. As usual, many phases acquire nonzero CDW order when $E_V>0$. A more detailed set of results for the H and F parts of $g_{xy/z}$ being of opposite signs are presented in the appendices. 

\subsection{Vanishing Valley Zeeman Coupling with $g_{\mu.F}/g_{\mu,H}>0$}
\label{subsec:evzero}

\begin{figure}[H]
    \centering
    \includegraphics[width=0.875\columnwidth]{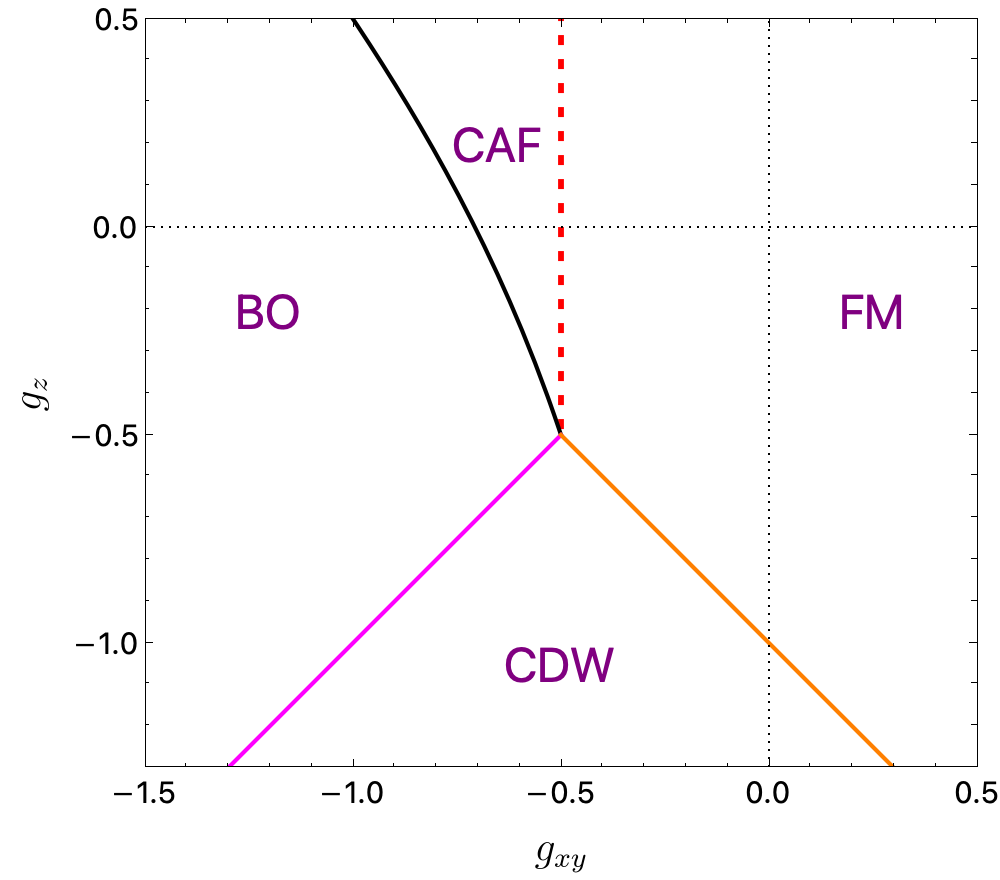}
    \caption{Parameters are $E_Z=1.0,g_{z,F}=g_{z,H}=g_z,g_{xy,F}=g_{xy,H}=g_{xy},E_V=0.0$,this is the Kharitonov's Ultra short range limit.Here as one can see their is no coexistence phase.}
    \label{fig:PDK}
\end{figure}

\begin{figure}[H]
    \centering
    \includegraphics[width=0.875\columnwidth]{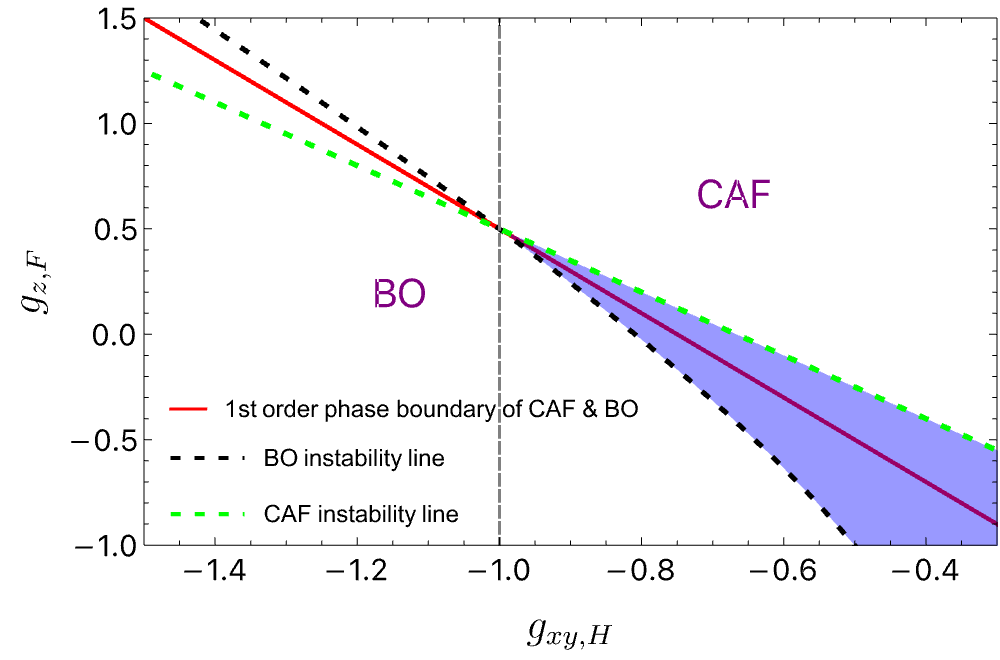}
    \caption{Instability lines for the BO and CAF phases as a function of $g_{xy,H}$ and $g_{z,F}$ for $E_Z=1.0,g_{z,H}=2.0,g_{xy,F}=-1.0,E_V=0.0$. The BO phase is unstable above the black line, while the CAF phase is unstable below the green line. The red line is where the energies of the BO and CAF phases cross. The vertical dashed line represents $g_{xy,H}=g_{xy,F}$. One can see that coexistence between BO and CAF state only occurs when $0>g_{xy,H} > g_{xy,F}$. The coexistence phase is shaded blue. }
    \label{fig:gxy_inequality}
\end{figure}

\begin{figure}[h]
    \centering
    \includegraphics[width=0.875\columnwidth]{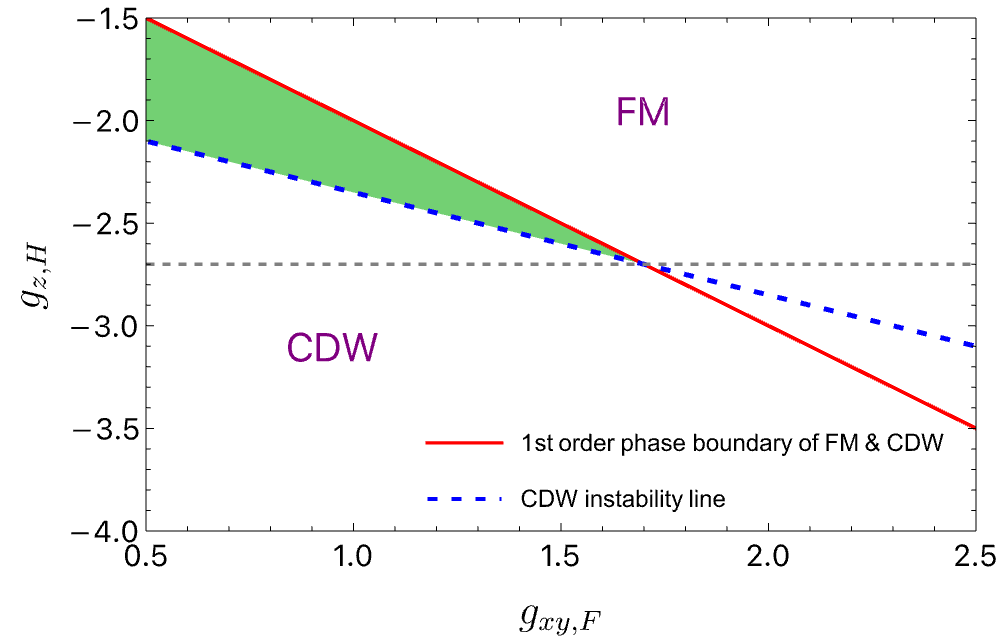}
    \caption{Instability lines for the CDW and FM phases as a function of $g_{xy,F}$ and $g_{z,H}$ for  $E_Z=1.0,E_V=0.0,g_{z,F}=-2.7$. The FM phase is unstable below the red line, while the CDW phase is unstable above the blue dotted line. The dashed horizontal line represents $g_{z,H}=g_{z,F}$. The coexistence between CDW and FM (the SVE+ phase, shaded green) occurs only for $0>g_{z,H}>g_{z,F}$.}
    \label{fig:gz_inequality}
\end{figure}

Throughout this section we will assume that we are ``close" to the USR model in the sense that the Hartree and Fock couplings of a given type have the same sign ($g_{\mu,F}/g_{\mu,H}>0$), and that there is perfect sublattice symmetry $E_V=0$. 
In order to enable comparisons with the USR model, we show the Kharitonov phase diagram \cite{kharitonov2012:nu0} in \cref{fig:PDK}. Here we choose $E_Z=1.0$ (in arbitrary units) as a fixed parameter.  All the lines between phases are first-order transitions, except for transition between the CAF and FM phases, which is second-order. Our convention here and henceforth is that solid lines represent first-order phase transitions while dashed lines represent continuous phase transitions. 
When one relaxes the USR assumption, it turns out the inequalities $g_{xy,F}/g_{xy,H}\lessgtr 1$ and $g_{z,F}/g_{z,H}\lessgtr 1$ play a crucial role in determining whether coexistence occurs at $E_V=0$ (we will see later that nonzero $E_V$ overcomes this limitation). Briefly, coexistence between Kekul\'e and CAF order occurs only when $g_{xy,F}/g_{xy,H}>1$, while coexistence between CDW and FM order occurs only when $g_{z,F}/g_{z,H}>1$. This is illustrated in \cref{fig:gxy_inequality,fig:gz_inequality} for specific values of other couplings, and turns out to be generic as long as $E_V=0$. 

In what follows, we will fix the ratio of $g_{\mu,F}$ to $g_{\mu,H}$, $\mu=xy,\ z$, and plot the phase diagrams with the axes being $g_{xy,H}$ and $g_{z,H}$. This should merely be thought of as taking a certain two-dimensional section of the full space of coupling constants, and does not represent any physical assumption about the proportionality between $g_{\mu,F}$ and $g_{\mu,H}$, say, as $B$ varies.

\begin{figure}[h]
    \centering
    \includegraphics[width=0.875\columnwidth]{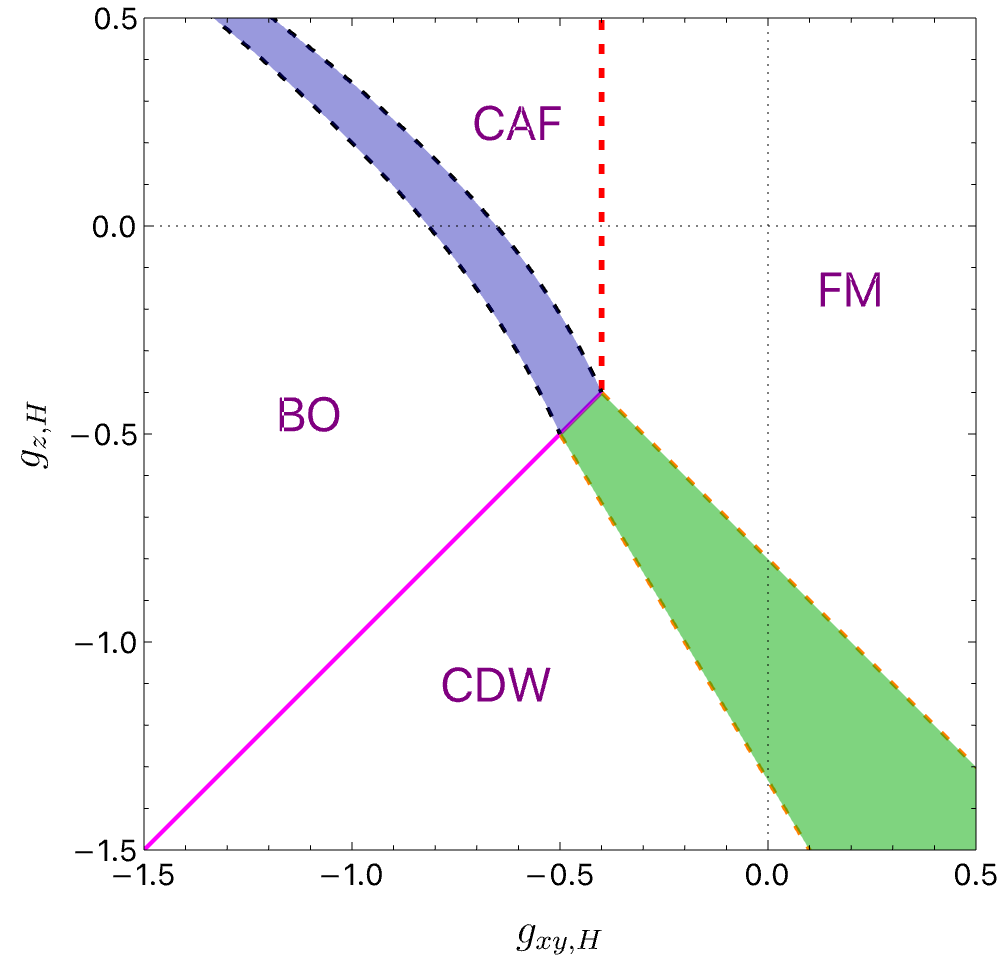}
    \caption{Phase diagram for  $E_Z=1.0,g_{z,F}=1.25g_{z,H},g_{xy,F}=1.25g_{xy,H},E_V=0.0$. The ratios $g_{\mu,F}/g_{\mu,H}>1$ for both types of couplings. There are two different coexistence phases:  The B/CAF coexistence phase (shaded blue) was already found previously \cite{Das_Kaul_Murthy_2022}, and also has SVEY order. The first-order transition between the FM and CDW phases for USR interactions has broadened into the SVE+ phase (shaded green) with coexistence between CDW and FM order parameters, so called because it has a spontaneous spin-valley entangled order parameter. }
    \label{fig:PD1}
\end{figure}

\cref{fig:PD1} shows the phase diagram for the case $g_{xy,F}=1.25g_{xy,H}$, $g_{z,F}=1.25g_{z,H}$. As found earlier \cite{Das_Kaul_Murthy_2022}, there is a region of coexistence between Kekul\'e and CAF order near the first-order line of the USR model. This phase, which we call B/CAF, also has SVEY order. 
In addition, there is a region of coexistence between  CDW and FM order (the SVE+ phase) in the neighborhood of the phase transition between the CDW and FM phases in the USR case. The transitions bordering the coexistence regions of the phase diagram are second-order, represented by dashed lines. As in \cref{fig:PDK}, $E_Z=1$. 

\begin{figure}[H]
    \centering
    \includegraphics[width=0.875\columnwidth]{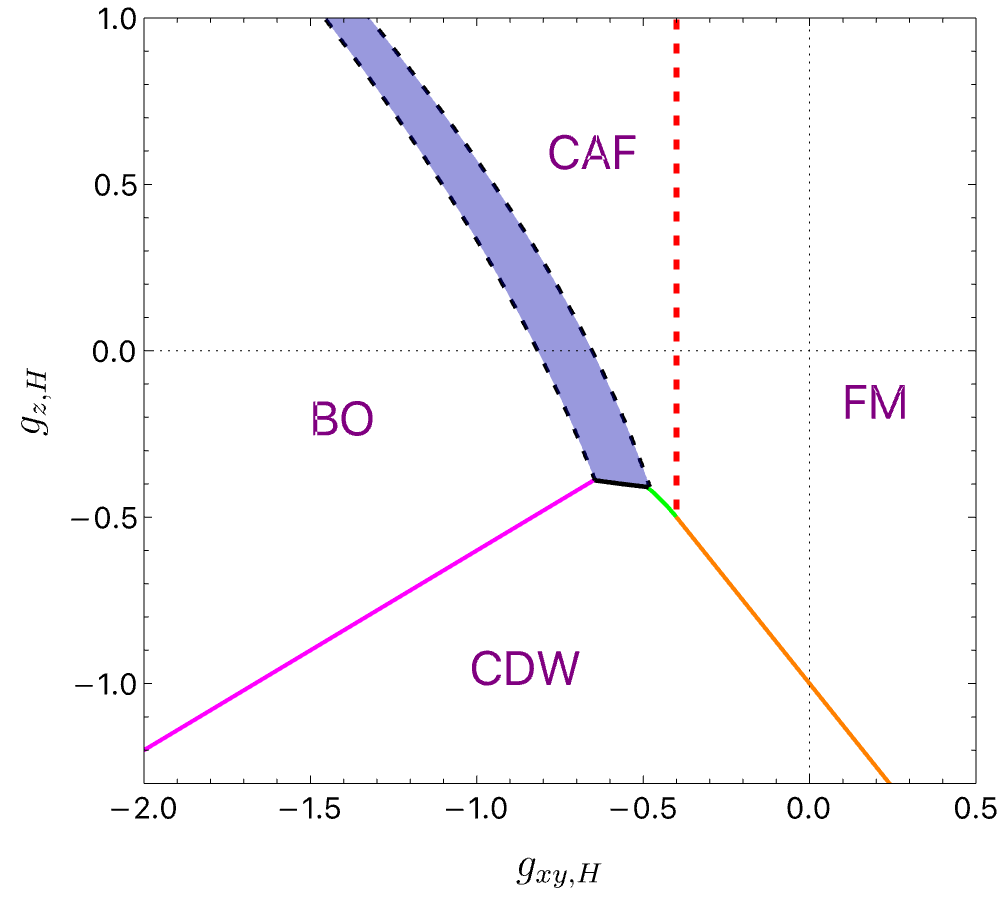}
    \caption{Phase diagram for  $E_Z=1.0,g_{z,F}=0.75g_{z,H},g_{xy,F}=1.25g_{xy,H},E_V=0.0$. The ratios of couplings ensure $g_{xy,F}/g_{xy,H}>1$, but $g_{z,F}/g_{z,H}<1$. The B/CAF coexistence region (shaded blue) still occurs but there is no coexistence near the CDW/FM phase boundary, which remains first-order. }
    \label{fig:PD2}
\end{figure}

\begin{figure}[H]
    \centering
    \includegraphics[width=0.875\columnwidth]{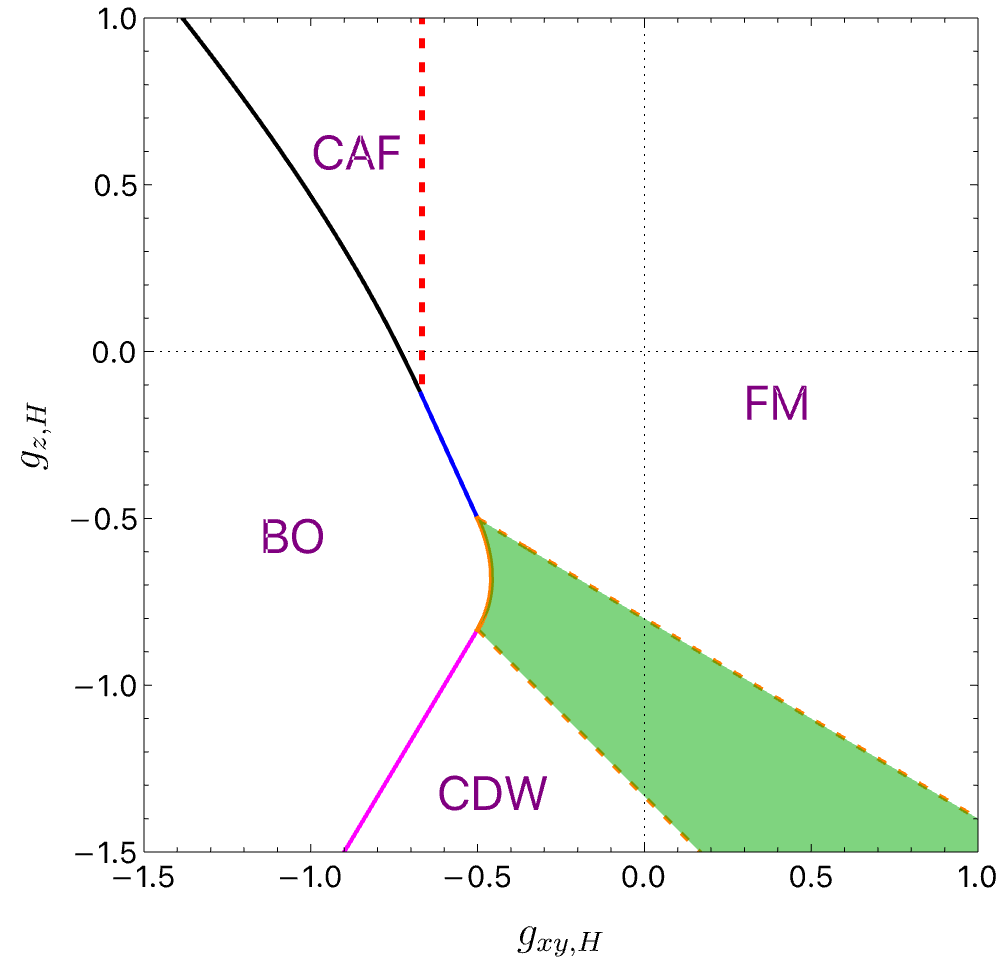}
    \caption{Phase diagram for  $E_Z=1.0,g_{z,F}=1.25g_{z,H},g_{xy,F}=0.75g_{xy,H},E_V=0.0$. Now the conditions are no longer met for the B/CAF phase to occur. However, the SVE+ phase (shaded green) does occur near the CDW/FM phase boundary, which is now split into two second-order lines. }
    \label{fig:PD3}
\end{figure}

\begin{figure}[H]
    \centering
    \includegraphics[width=0.875\columnwidth]{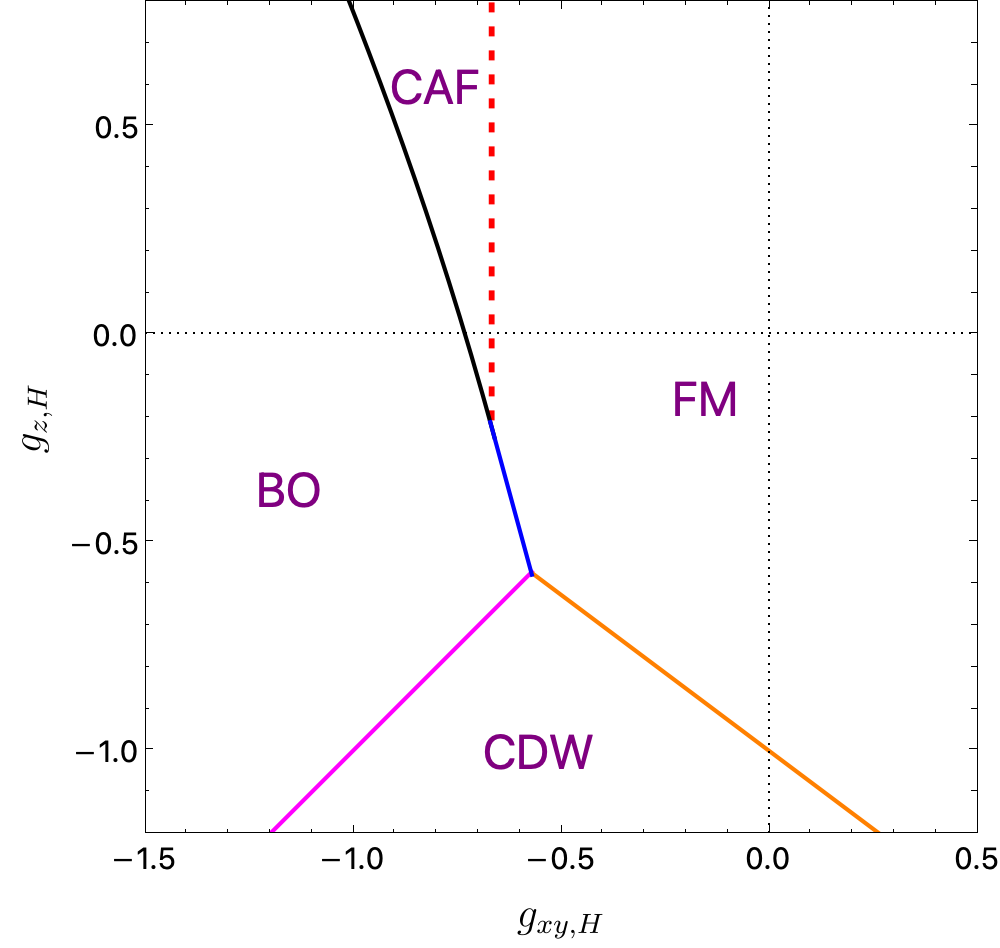}
    \caption{Phase diagram for  $E_Z=1.0,g_{z,F}=0.75g_{z,H},g_{xy,F}=0.75g_{xy,H},E_V=0.0$. There is no coexistence.}
    \label{fig:PD4}
\end{figure}

\cref{fig:PD2} shows the phase diagram for the case $g_{z,F}/g_{z,H}<1,\ g_{xy,F}/g_{xy,H}>1$. As mentioned above, this satisfies the condition for coexistence between CAF and BO order (the B/CAF phase), but fails to meet the condition for the existence of the SVE phase between the CDW and FM phases. The B/CAF phase with its attendant SVEY order is still present, though reduced in extent. Similarly, \cref{fig:PD3} has $g_{z,F}/g_{z,H}>1,\ g_{xy,F}/g_{xy,H}<1$, allowing the SVE+ phase to exist but forbidding coexistence between BO and CAF order. 
Finally, \cref{fig:PD4} has $g_{z,F}/g_{z,H}<1,\ g_{xy,F}/g_{xy,H}<1$, disallowing any coexistence. The reason for the difference in detail between the Kharitonov phase diagram \cref{fig:PDK} and \cref{fig:PD4} is that the boundaries between phases sometimes depend on the Hartree coupling and sometimes on the Fock coupling, as seen by the ground state energies in \cref{eq:E_simple}.

It is of experimental interest to ask how the system evolves with increasing Zeeman coupling when all interaction parameters are fixed, and $E_V=0$. This corresponds to applying a tilted field to the system \cite{young2014:nu0}, keeping the perpendicular component of $B$ constant. The evolution of the order parameters depends on the values chosen for $g_{z,H/F},g_{xy,H/F}$. Clearly, if the system is in the FM phase already at $E_Z=0$, there will be no further evolution with increasing $E_Z$. Similarly, if the system is in the CAF phase at $E_Z=0$, it can only evolve into the FM with increasing $E_Z$. Since these possibilities have been thoroughly explored in the past literature, we will ignore them in favor of more interesting possibilities. 

\begin{figure}[H]
    \centering
    \includegraphics[width=0.88\columnwidth]{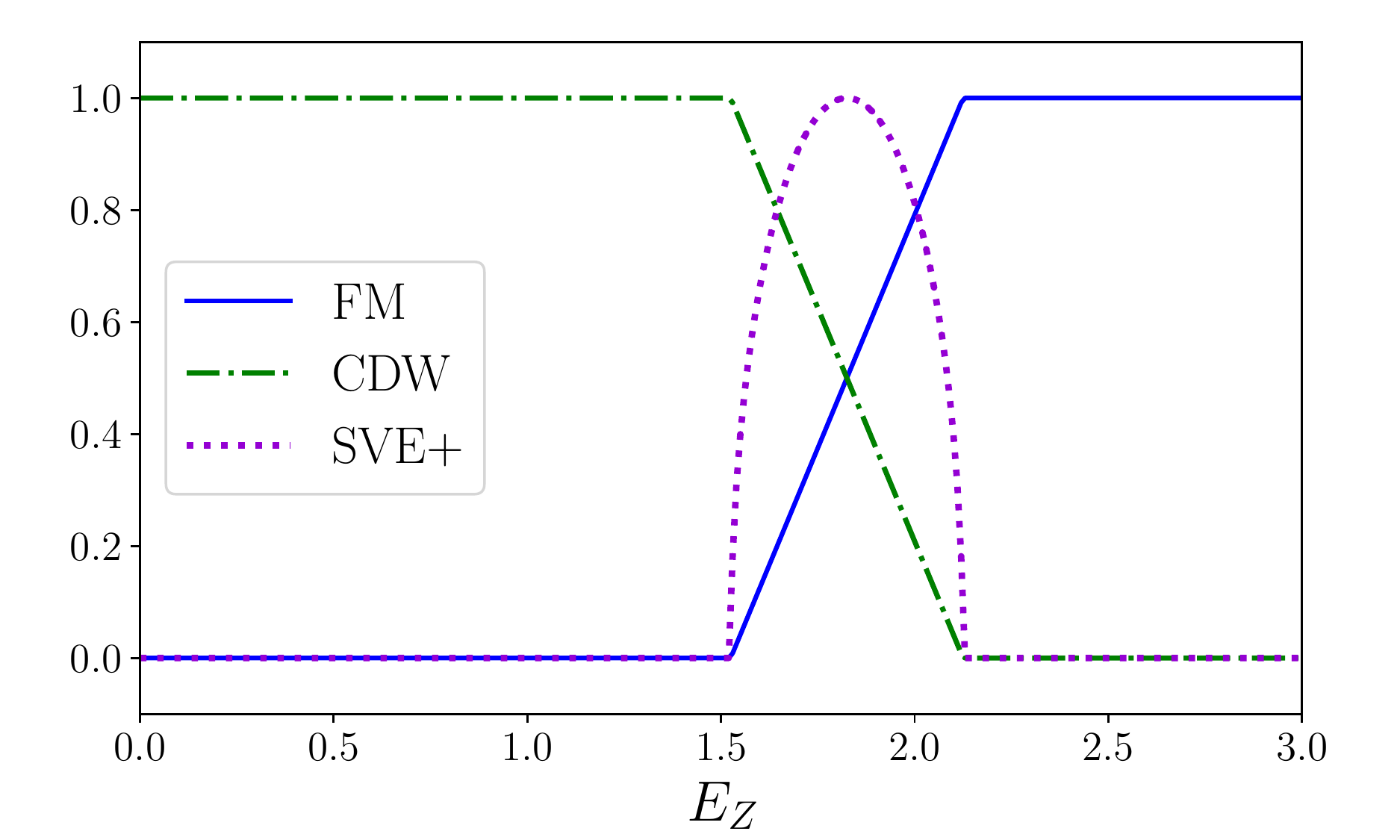}
    \caption{Evolution of order parameters vs $E_Z$ for  $g_{z,H}=-1.2,g_{z,F}=1.25g_{z,H},g_{xy,H}=-0.5,g_{xy,F}=1.25g_{xy,H},E_V=0.0$. The ratios of the H and F couplings corresponds to \cref{fig:PD1}. The system is in CDW phase at $E_Z=0$. As $E_Z$ increases there is a second-order transition to a phase where FM and CDW order coexist. This coexistence is mediated by the presence of the SVE+ order parameter, which spontaneously breaks the $U(1)_s$ and $U(1)_v$ symmetries.  For large enough $E_Z$ the system goes into FM phase.}
    \label{fig:OP1_CDW}
\end{figure}

\begin{figure}[h]
    \centering
    \includegraphics[width=0.875\columnwidth]{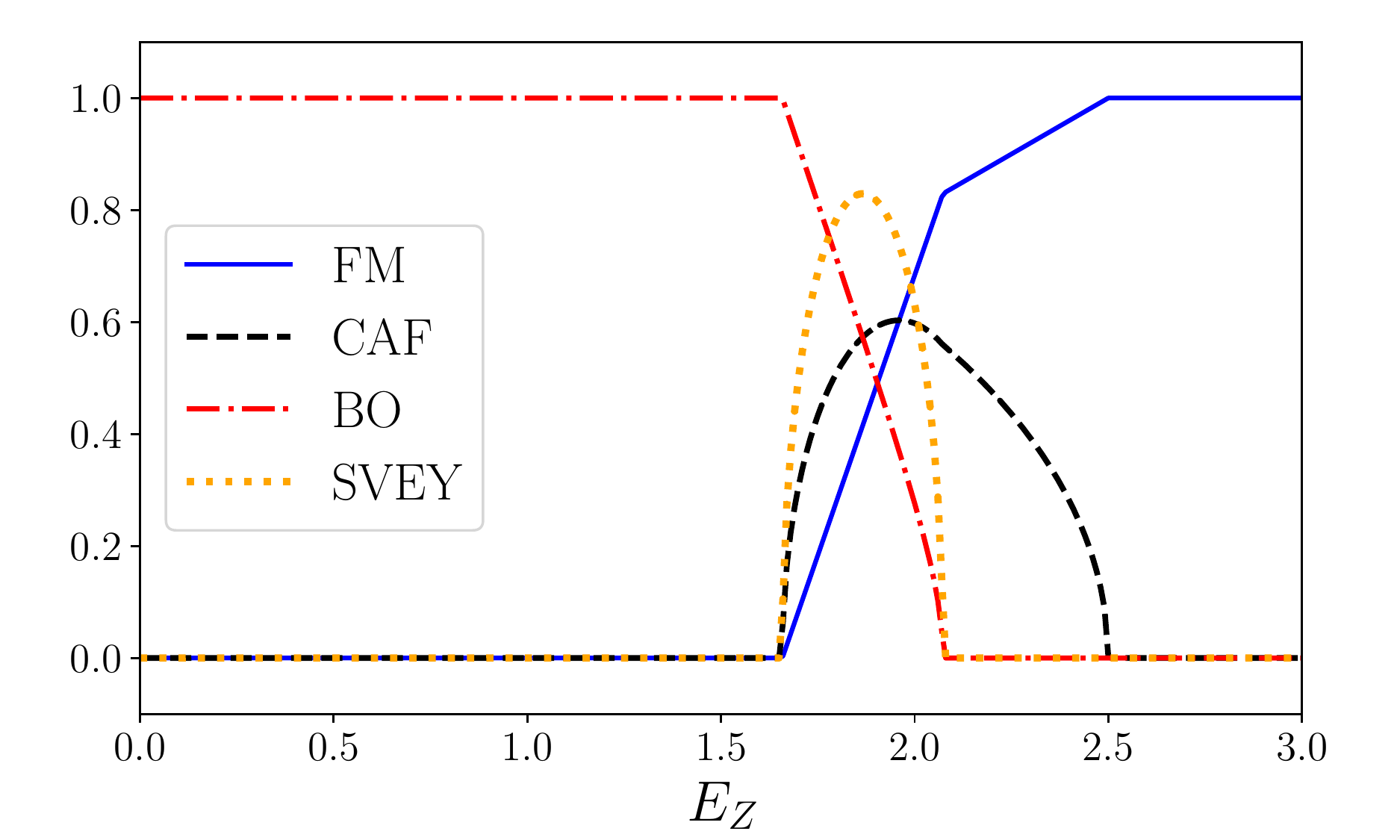}
    \caption{Evolution of order parameters vs $E_Z$ for  $g_{z,H}=-0.5,g_{z,F}=1.25g_{z,H},g_{xy,H}=-1.0,g_{xy,F}=1.25g_{xy,H},E_V=0.0$.The ratios of the H and F couplings corresponds to the phase diagram of \cref{fig:PD1}. The system is in BO phase  at $E_Z=0$. As $E_Z$ increases the system undergoes a second-order transition to the intermediate B/CAF phase, where BO, CAF, and SVEY order coexist. As $E_Z$ increases further, there is another second-order transition to the pure CAF phase, and eventually a third second-order transition to the FM phase at large $E_Z$.}
    \label{fig:OP1_BO}
\end{figure}

Let us first consider the case $g_{z,F}/g_{z,H}>1,\ g_{xy,F}/g_{xy,H}>1$, which corresponds to the phase diagram of \cref{fig:PD1}. \cref{fig:OP1_CDW} shows the evolution of order parameters with $E_Z$ when the system is in the CDW phase at vanishing Zeeman coupling. In addition to the FM and CDW order parameters, we also show the SVE+ order parameter.  As  $E_Z$ increases, the system undergoes a phase transition from the CDW phase into the SVE+ phase (coexisting FM, CDW, and SVE+ order), and then into the FM phase. Alternatively, as shown in \cref{fig:OP1_BO} the system could start in the BO phase at $E_Z=0$. In this case the system first goes into the B/CAF phase (which also has SVEY order), then into the pure CAF phase, and finally into the FM phase at large $E_Z$.

\begin{figure}[h]
    \centering
    \includegraphics[width=0.875\columnwidth]{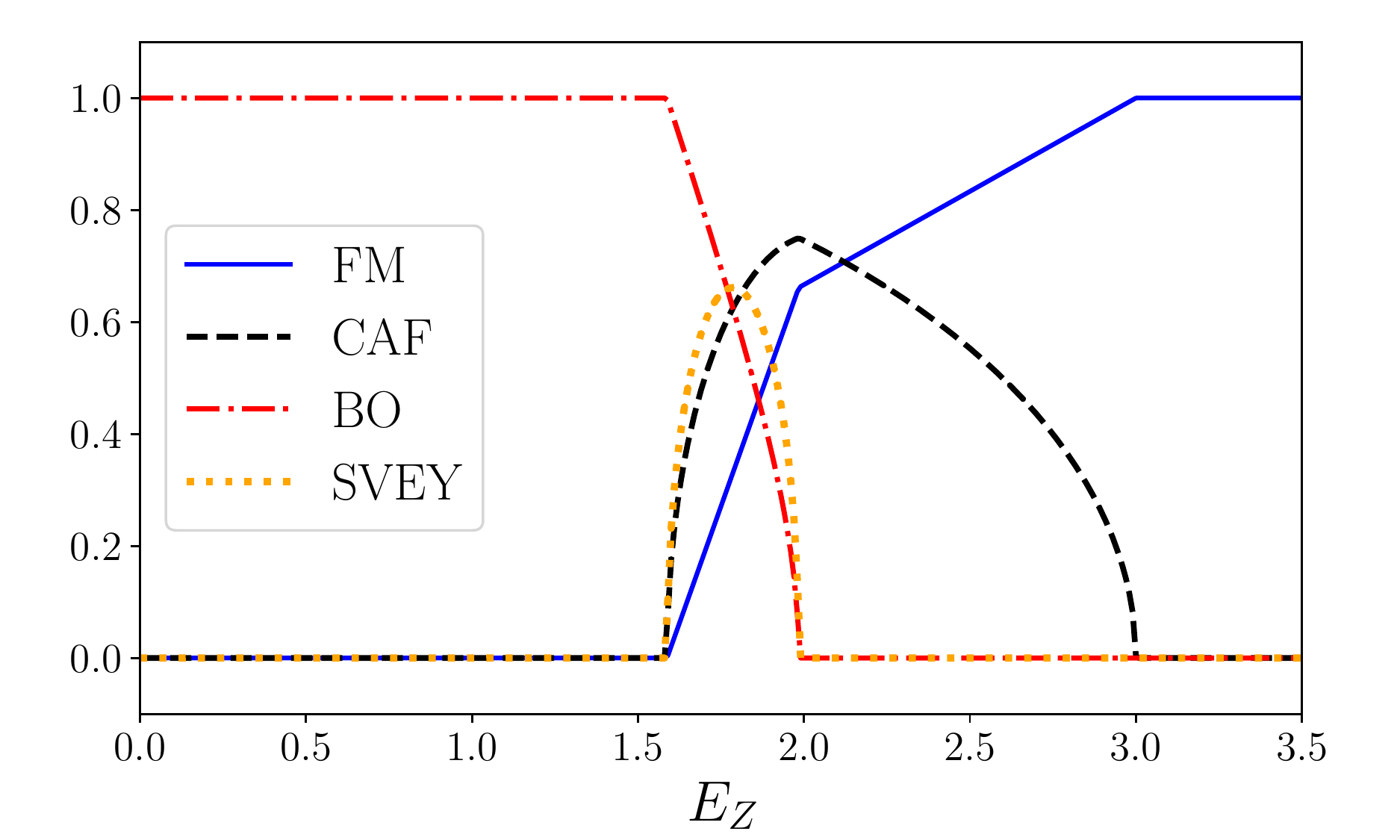}
    \caption{Evolution of order parameters vs $E_Z$ for  $g_{z,H}=-0.2,g_{z,F}=0.75g_{z,H},g_{xy,H}=-1.2,g_{xy,F}=1.25g_{xy,H},E_V=0.0$. The ratios of the H and F couplings correspond to the phase diagram of \cref{fig:PD2}. The system is in the BO phase at $E_Z=0$.  As $E_Z$ increases there is a second-order transition to the B/CAF phase (with BO, CAF, FM, and SVEY order). For larger $E_Z$, there is another second-order transition to the pure CAF phase. Finally, the system goes through another second-order transition into the FM phase at very large $E_Z$.}
    \label{fig:OP2_BO}
\end{figure}

Coming next to the case of $g_{z,F}/g_{z,H}<1,\ g_{xy,F}/g_{xy,H}>1$, which corresponds to the phase diagram of \cref{fig:PD2}, \cref{fig:OP2_BO} shows the evolution of order parameters as a function of $E_Z$ for $g_{z,H}=-0.2, g_{xy,H}=-1.2$. The system starts in the BO phase at vanishing Zeeman coupling, undergoes a second-order transition to the B/CAF coexistence phase, undergoes yet another second-order transition to the pure CAF phase, and finally goes into the FM phase at large $E_Z$. 

\begin{figure}[h]
    \centering
    \includegraphics[width=0.875\columnwidth]{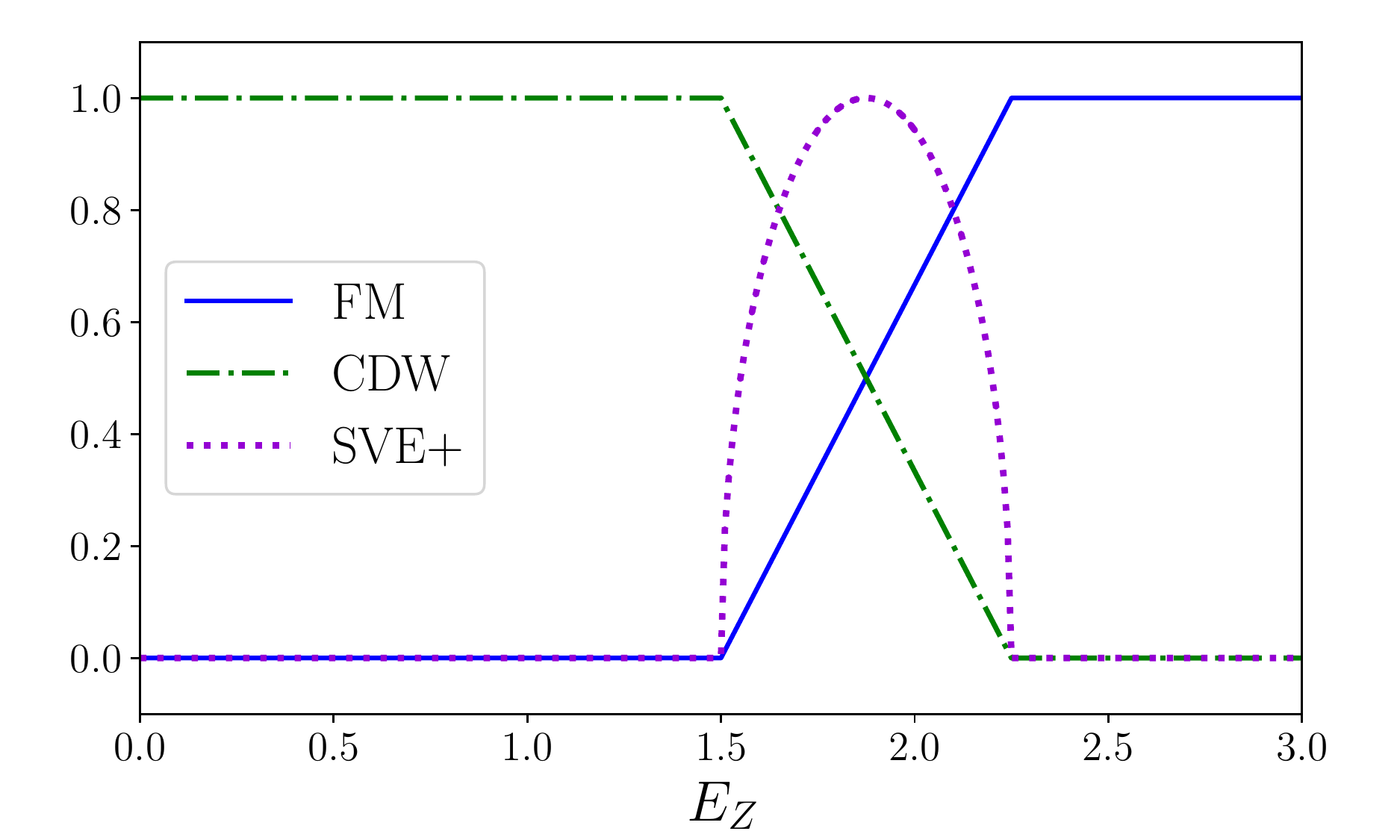}
    \caption{Evolution of order parameters vs $E_Z$ for  $g_{z,H}=-1.5,g_{z,F}=1.25g_{z,H},g_{xy,H}=-0.5,g_{xy,F}=0.75g_{xy,H},E_V=0.0$. The ratios of the H and F couplings correspond to the phase diagram of \cref{fig:PD3}. The system is in the CDW phase at $E_Z=0$. As $E_Z$ increases, the system enters the SVE+ phase (coexisting CDW, FM, and SVE+ order) via a second-order transition.  For larger $E_Z$ there is another second-order transition to the FM phase.}
    \label{fig:OP3_CDW}
\end{figure}

Finally, we consider the case  $g_{z,F}/g_{z,H}<1,\ g_{xy,F}/g_{xy,H}>1$, which corresponds to the phase diagram of \cref{fig:PD3}. \cref{fig:OP3_CDW} shows the evolution of the order parameter as a function of $E_Z$ for $g_{z,H}=-1.5,\ g_{xy,H}=-0.5$. The system starts in the CDW phase at vanishing $E_Z$, makes a second-order phase transition to the SVE+ phase at intermediate $E_Z$, and finally goes into the FM phase via another second-order phase transition at large $E_Z$.

\subsection{Nonvanishing Valley Zeeman Coupling and $g_{\mu,F}/g_{\mu,H}>0$}
\label{subsec:evnon0_ratioplus}
In this subsection we consider how the phase diagrams change when $E_V>0$. We will still stay ``close" to the USR model, assuming $g_{\mu,F}/g_{\mu,H}>0$. The most obvious change is already present in the USR limit: The BO phase is replaced by the B/CO phase, in which both bond order and CDW order coexist. This is analogous to the replacement of the antiferromagnetic phase at $E_Z=0$ by the CAF phase at arbitrarily small $E_Z$. Concurrently, the first-order phase transition between the CDW and BO phases in the USR phase diagram \cref{fig:PDK} is converted into a second-order transition. These are all previously known results \cite{Goerbig_nu0_Skyrmion_Zoo_2021,Murthy_BLG_2017}.

\begin{figure}[H]
    \centering
    \includegraphics[width=0.875\columnwidth]{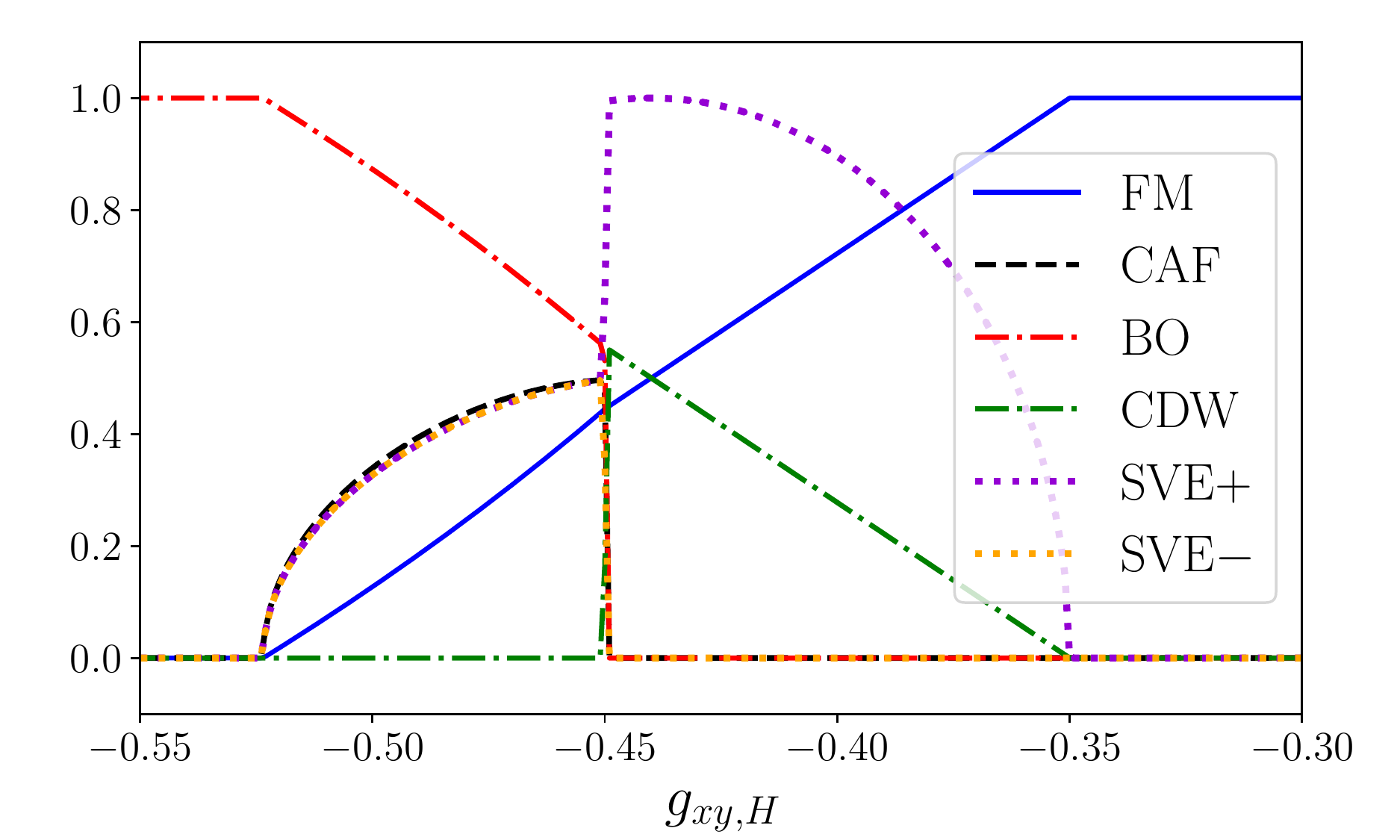}
    \caption{Evolution of order parameters  at $E_V=0$ as a function of $g_{xy,H}$ for  $g_{z,H}=-0.45,g_{z,F}=1.25g_{z,H},g_{xy,F}=1.25g_{xy,H}$. The ratios of the H and F couplings correspond to the phase diagram of \cref{fig:PD1}. It can be seen that the  system undergoes a first-order transition from the B/CAF to the F/CDW phase.}
    \label{fig:Zero_Ev_section}
\end{figure}

\begin{figure}[H]
    \centering
    \includegraphics[width=0.875\columnwidth]{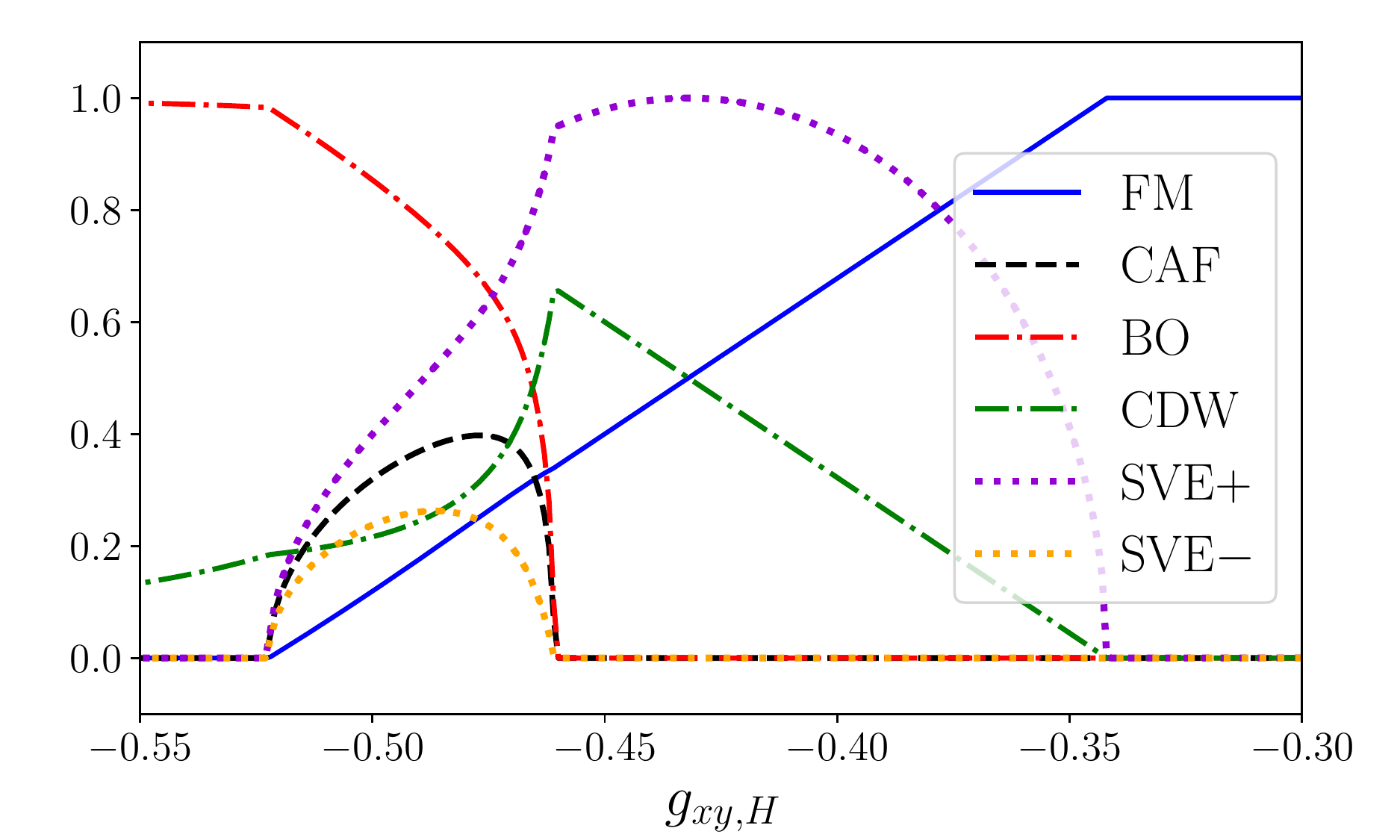}
    \caption{Evolution of order parameters as a function of $g_{xy,H}$ for  $g_{z,H}=-0.45,g_{z,F}=1.25g_{z,H},g_{xy,F}=1.25g_{xy,H}$ for vanishingly small $E_V=0.01$. The ratios of the H and F couplings correspond to the phase diagram of \cref{fig:PD1}. The  first-order transition between the B/CAF and SVE+ phases has now become second-order.}
    \label{fig:tiny_Ev_section}
\end{figure}

The introduction of small $E_V$ does not modify the SVE+ phase (whenever it occurs). While $E_V>0$ does not change the symmetries broken in the B/CAF phase, it does change the phase quantitatively. Let us first look at \cref{fig:Zero_Ev_section}, which shows the order parameters at $E_V=0$ along a horizontal cut across \cref{fig:PD1} at $g_{z,H}=-0.45$. Recall that the B/CAF phase, at $E_V=0$, has BO, CAF, FM, and SVEY order. The SVEY order parameter changes discontinuously at the first-order phase transition with the SVE+ phase.  All these features change when one adds a tiny $E_V$. In \cref{fig:tiny_Ev_section} we show the evolution of order parameters along the same  horizontal cut ($g_{zH}=-0.45$) when $E_V=0.01$. It can be seen that the SVE+ phase does not change character qualitatively. However, the B/CAF phase now changes significantly. Firstly, it acquires a nonzero CDW order parameter, because the B/CO phase also has CDW order. More importantly, it now has both SVEX and SVEY order, and undergoes  a second-order phase transition  to the SVE+ phase. Thus, even a very tiny $E_V$ can affect certain phases and phase transitions significantly. 

\begin{figure}[H]
    \centering
    \includegraphics[width=0.875\columnwidth]{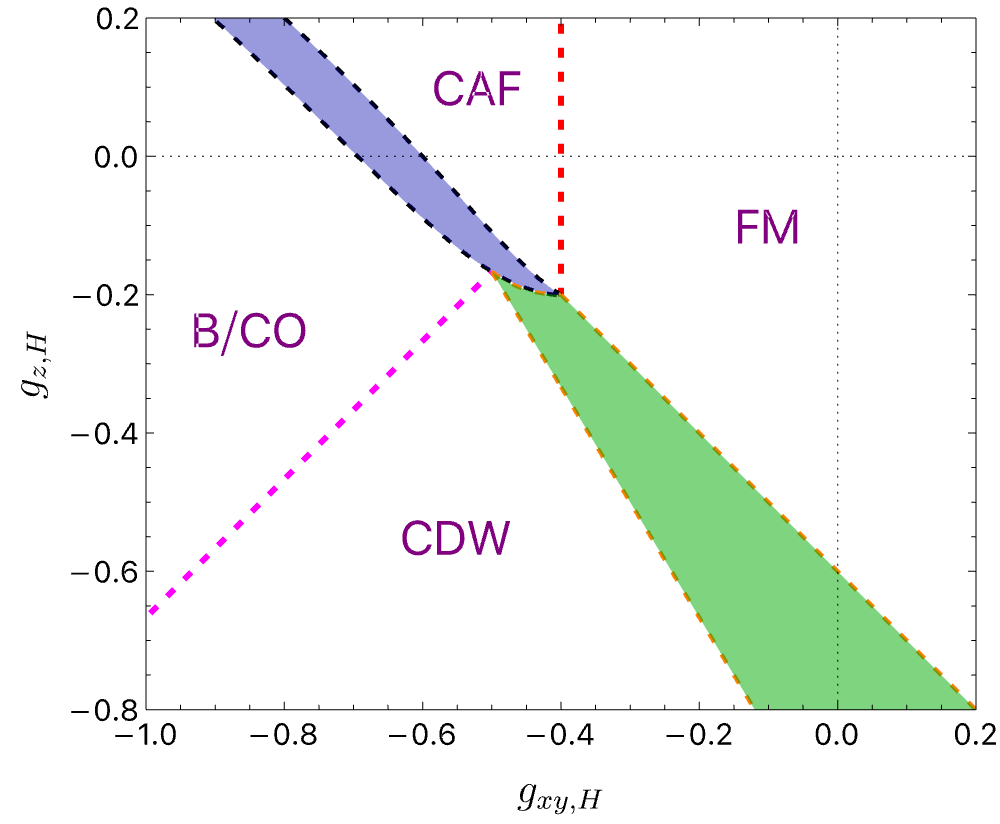}
    \caption{For this figure the coupling ratios of $g_{xy}$ and $g_z$ are same as in \cref{fig:PD1} and we have considered a small valley Zeeman $E_V=0.25$.By comparing this figure with  \cref{fig:PD1} one can see that for this choice of coupling ratios, the presence of finite $E_V$ doesn't qualitatively change the nature of the phase diagram.}
    \label{fig:PD5}
\end{figure}

Next we turn a moderate value of $E_V=0.25$. \cref{fig:PD5} shows the phase diagram for the ratio of Hartree and Fock parts of the couplings being $g_{z,F}/g_{z,H}=g_{xy,F}/g_{xy,H}=1.25$, the same as in \cref{fig:PD1}. The topology of the phase diagram is identical to that of \cref{fig:PD1}, with the solitary change that the first-order transition between the B/CAF and SVE+ phases has been replaced by a second-order transition. The B/CAF region has shrinks, while the SVE+ region expands.

\begin{figure}[H]
    \centering
    \includegraphics[width=0.875\columnwidth]{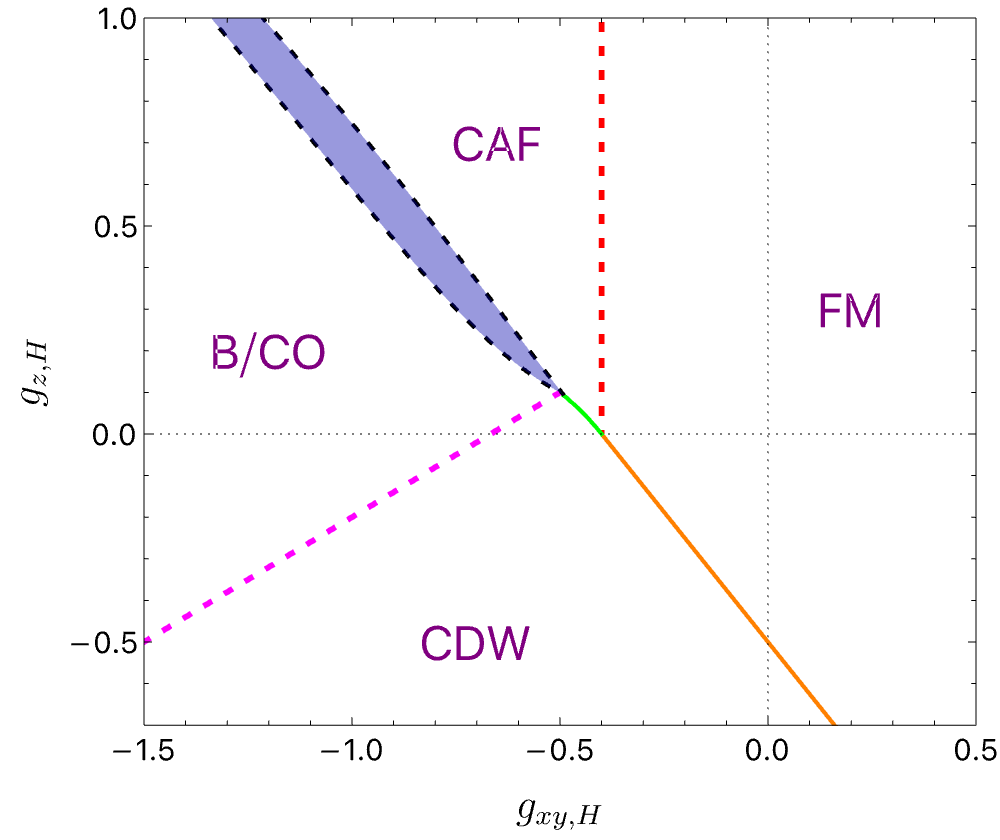}
    \caption{For this figure the coupling ratios of $g_{xy}$ and $g_z$ are same as in \cref{fig:PD2} and we have considered a valley Zeeman field $E_V=0.5$.By comparing this figure with \cref{fig:PD2} one can see that for this choice of coupling ratios, the presence of finite $E_V$ reduce the coexistence phase area of B/CAF. And here CDW and CAF phases, CDW and FM phases are separated by first order lines.}
    \label{fig:PD6}
\end{figure}

\begin{figure}[H]
    \centering
    \includegraphics[width=0.875\columnwidth]{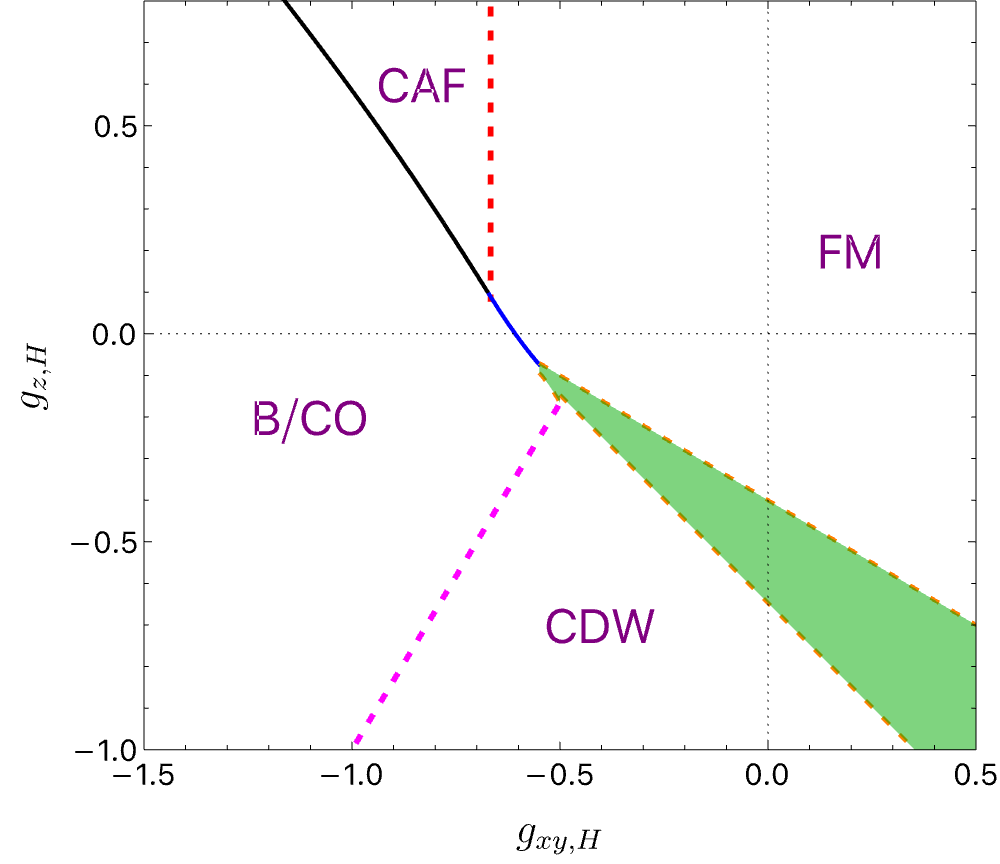}
    \caption{Phase diagram at $E_V=0.5$ for the coupling ratios $g_{z,F}/g_{z,H}=1.25,\ g_{xy,F}/g_{xy,H}=0.75$.  Comparing this  with the phase diagram for the same ratios at $E_V=0$ (\cref{fig:PD3}) one can see that for this choice of coupling ratios, $E_V$ has suppressed the B/CAF phase completely and shrunk the SVE+ phase. }
    \label{fig:PD7}
\end{figure}
\begin{figure}[H]
    \centering
    \includegraphics[width=0.875\columnwidth]{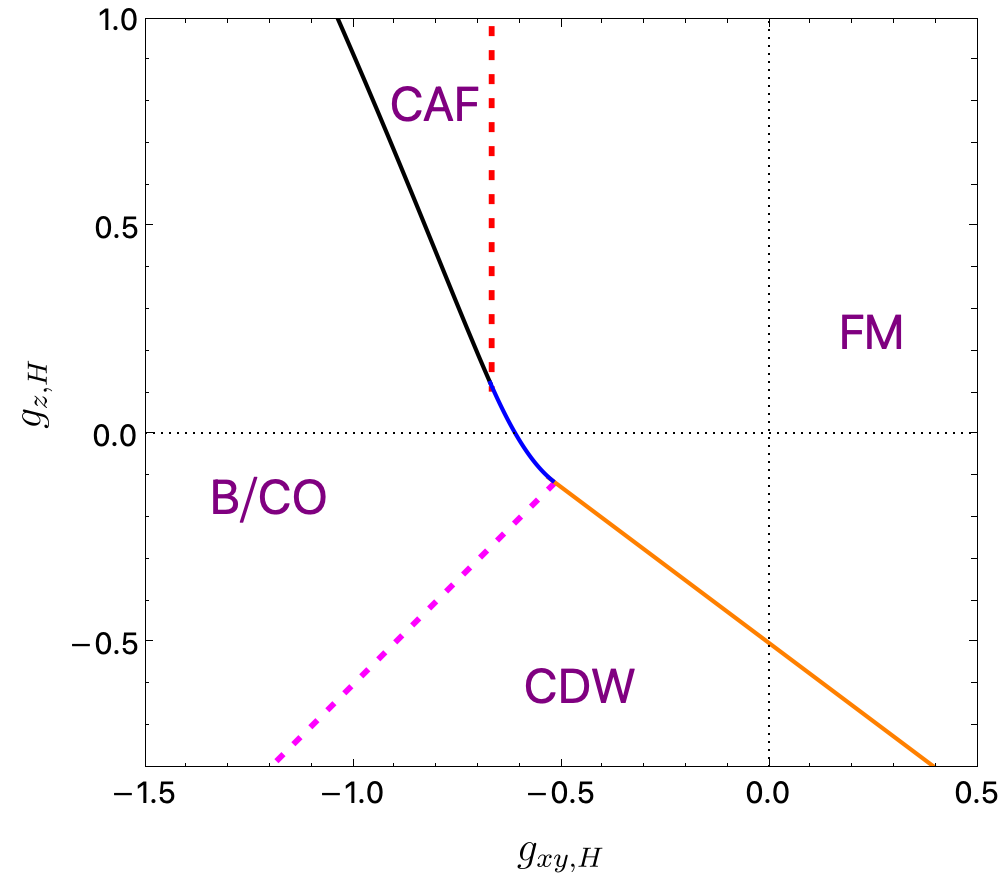}
    \caption{Phase diagram at $E_V=0.5$ for the coupling ratios $g_{\mu,F}/g_{\mu,H}=0.75$. There is no qualitative change as compared to the corresponding phase diagram at $E_V=0$ (\cref{fig:PD4}) }
    \label{fig:PD8}
\end{figure}

\cref{fig:PD6} shows the phase diagram at $E_V=0.5$ when the ratios are $g_{z,F}=0.75g_{z,H};\ \ g_{xy,F}=1.25g_{xy,H}$. As in the case of the previous figure, some of the phase boundaries move, but the topology remains the same as at $E_V=0$. The same is true for  $g_{z,F}/g_{z,H}=1.25;\ \ g_{xy,F}/g_{xy,H}=0.75$, shown in \cref{fig:PD7} and $g_{\mu,F}/g_{\mu,H}=0.75$ shown in \cref{fig:PD8}. 

\begin{figure}[H]
    \centering
    \includegraphics[width=0.875\columnwidth]{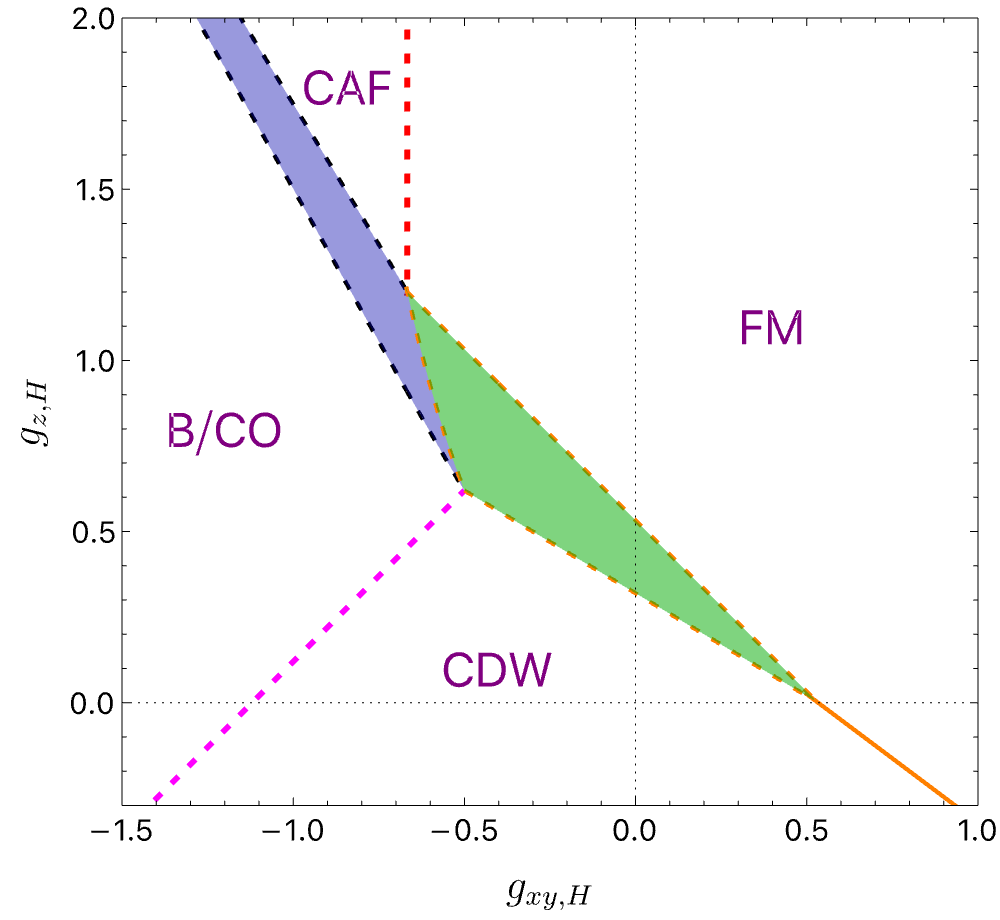}
    \caption{Phase diagram for $E_Z=1$, and $E_V=1.4$. The ratios of the Hartree and Fock couplings are $g_{z,F}=0.75g_{z,H};\ \ g_{xy,F}=0.75g_{xy,H}$ as in \cref{fig:PD4} and \cref{fig:PD8}. In contrast to those cases  where there is no coexistence anywhere in the phase diagram, we do obtain the coexistence phases B/CAF (blue shaded region) and SVE+ (green shaded region). }
    \label{fig:PD9}
\end{figure}

More interesting phenomena occur if one increases $E_V$ to an even larger value. Recall that for $g_{z,F}/g_{z,H}=g_{xy,F}/g_{xy,H}=0.75$, there is no coexistence anywhere in the phase diagram (\cref{fig:PD4}) at $E_V=0$, or at $E_V=0.5$ (\cref{fig:PD8}). The phase diagram for this ratio of the H and F parts of the couplings, at $E_V=1.4$, is shown in \cref{fig:PD9}. It can be seen that a sufficiently large $E_V$ can induce coexistence, even when it does not occur at $E_V=0$. A similar phenomenon is seen in previous work in BLG \cite{Murthy_BLG_2017}. 

\begin{figure}[H]
    \centering
    \includegraphics[width=0.875\columnwidth]{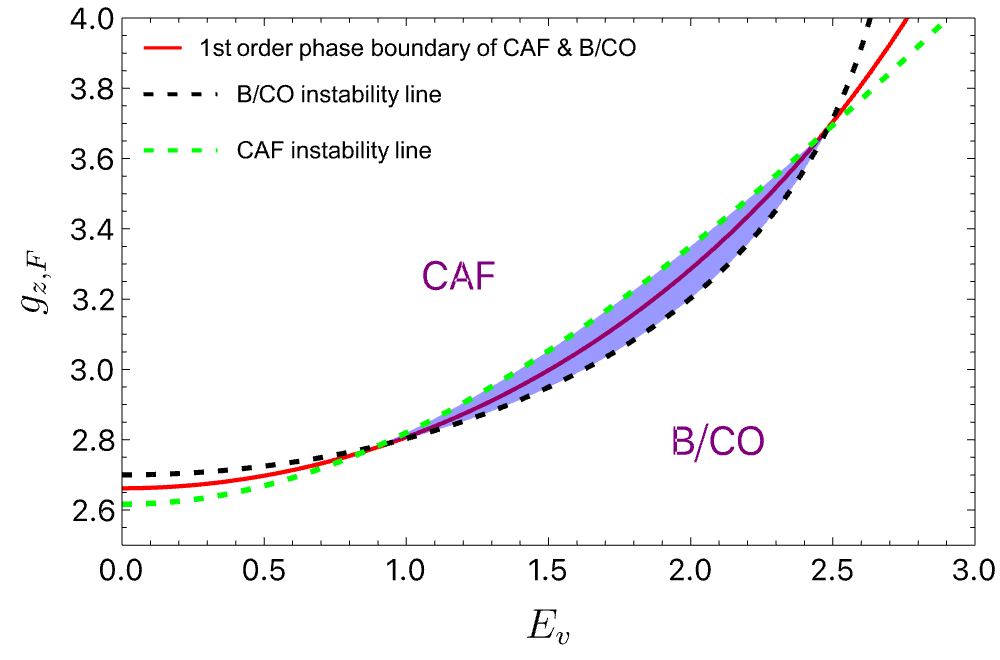}
    \caption{The coupling constants are fixed such that at $E_V=0$ there is no B/cAF phase. We choose $E_Z=1,\ g_{z,H}=3.4,\ g_{xy,H}=-2.5,\ g_{xy,F}=-2.1$. Varying the other two parameters, $g_{z,F},\ E_V$, we see that there is an intermediate regime in both in which the B/CAF phase does appear.  }
    \label{fig:large_EV_BCAF}
\end{figure}
\cref{fig:large_EV_BCAF} shows a slightly different way of looking at the occurrence of the B/CAF phase as $E_V$ varies. We have fixed the Zeeman coupling at $E_Z=1$, and the  $xy$ interactions such that $g_{xy,F}/g_{xy,H}<1$, implying that coexistence will not occur for $E_V=0$. The coupling $g_{z,H}$ is also fixed. We show the range of parameters in $E_V,\ g_{z,F}$ where the B/CAF phase occurs.  As can be seen, there is an intermediate range of $E_V$ and $g_{z,F}$ where the B/CAF phase appears.

\begin{figure}[H]
    \centering
    \includegraphics[width=0.875\columnwidth]{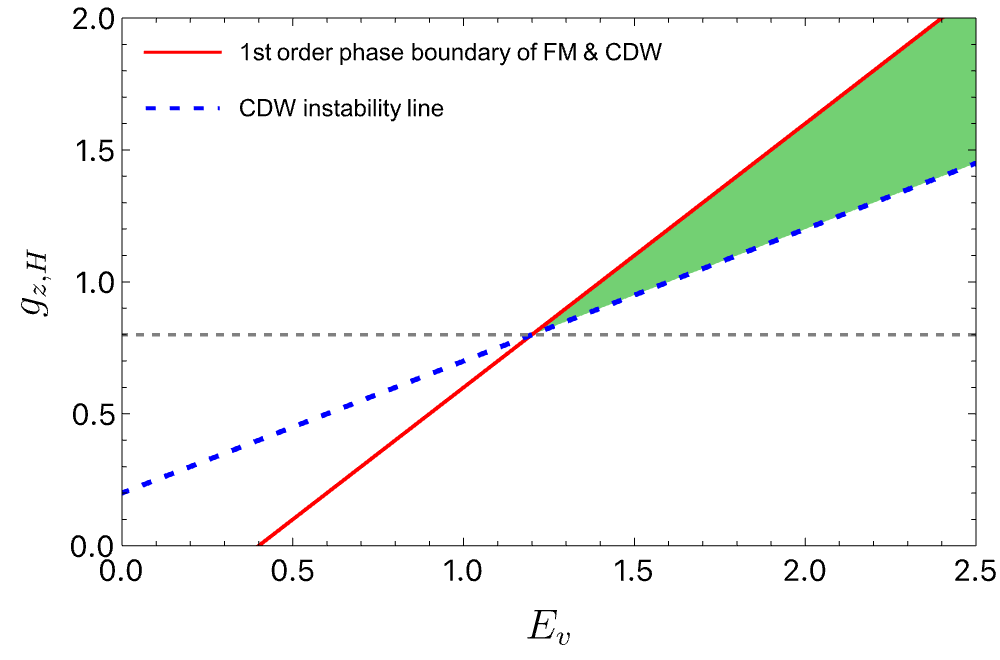}
    \caption{The parameters are $E_Z=1.0,g_{z,F}=0.8,g_{xy,F}=-0.6$. One can see from this figure also that only when $(|g_{z,H}| > |g_{z,F}|)$ (for both $g_{z,H},g_{z,F}>0$), their can be a coexistence phase of (FM+CDW) for some range of $E_V$. The vertical dashed line denotes $g_{z,H}=g_{z,F}$.}
    \label{fig:Large_EV_SVE+}
\end{figure}

\cref{fig:Large_EV_SVE+} shows the fate of the SVE+ phase at large $E_V$. Once again, we choose $g_{z,F}/g_{z,H}<1$, which implies that the SVE+ phase does not occur for $E_V=0$. We see that for large enough $E_V$, the SVE+ phase is stabilized. 

\begin{figure}[H]
    \centering
    \includegraphics[width=0.875\columnwidth]{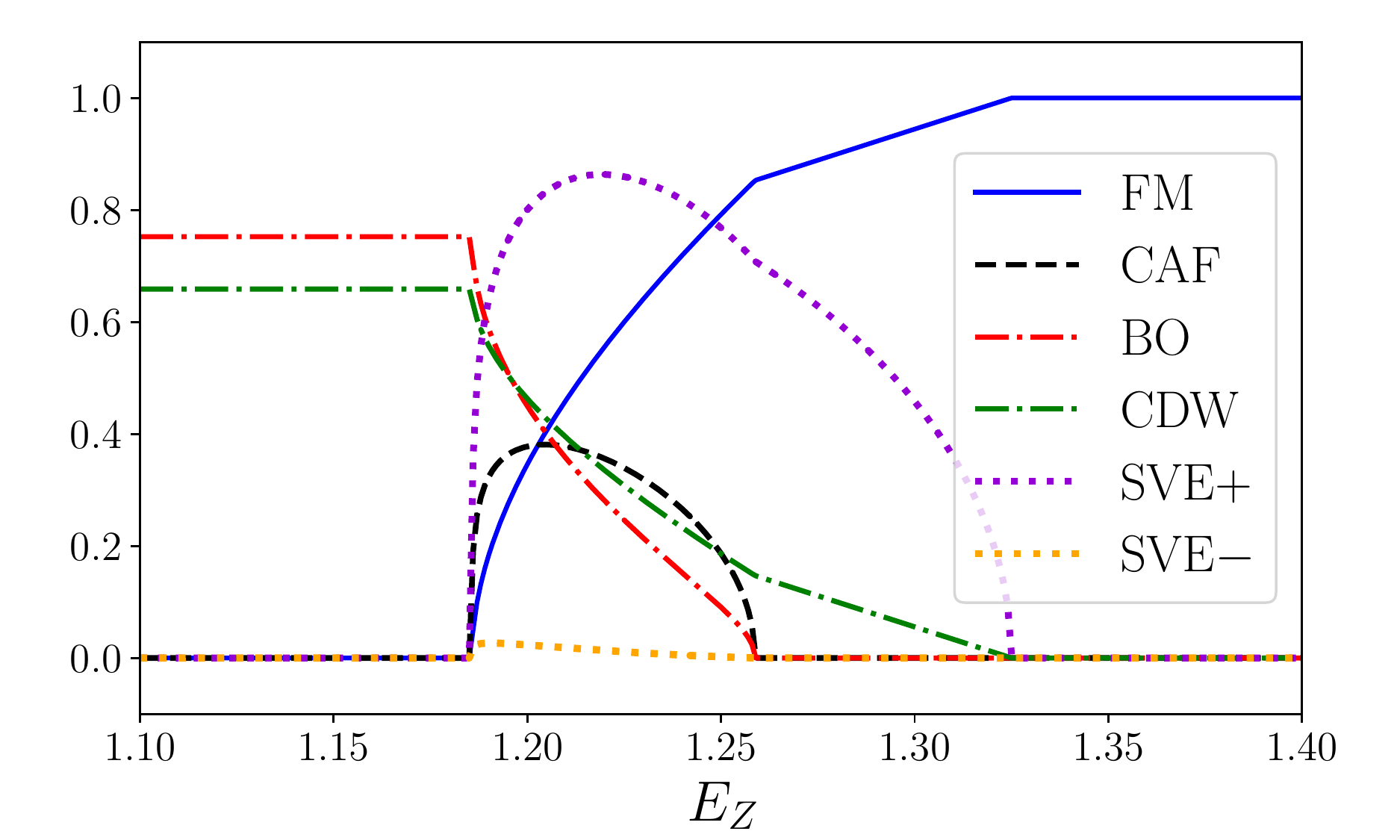}
    \caption{Evolution of order parameters vs $E_Z$ for  $g_{z,H}=0.9,g_{z,F}=0.75g_{z,H},g_{xy,H}=-0.8,g_{xy,F}=0.75g_{xy,H},E_V=1.4$. The ratios of the H and F couplings correspond to the phase diagram of \cref{fig:PD9}. The system is in the B/CO phase at $E_Z=0$.  As $E_Z$ increases there is a second-order transition to the B/CAF phase (with BO, CDW, CAF, FM, SVEX and SVEY order). For larger $E_Z$, there is another second-order transition to the SVE+ phase(with FM and CDW order). The system finally goes into the FM phase at very large $E_Z$.}
    \label{fig:large_EV_OP_evolution}
\end{figure}

It is also interesting to consider the evolution of the order parameters as $E_Z$ increases in \cref{fig:PD9}.  If one starts in the B/CO phase, the expected sequence of phases as $E_Z$ increases for small or vanishing $E_V$, provided the B/CAF phase occurs, is B/CO$\ \rightarrow\ $B/CAF$\ \rightarrow\ $CAF$\ \rightarrow\ $FM. However, for $E_Z=1.4$ the sequence is different: BO$\ \rightarrow\ $B/CAF$\ \rightarrow\ $SVE+$\ \rightarrow\ $FM. This is shown in \cref{fig:large_EV_OP_evolution}. 

\subsection{Hartree and Fock parts of $g_{z/xy}$ having opposite signs}
\label{subsec:evzero_ratiominus}

It is conceivable that for strong LL-mixing, the renormalizations of the interactions could be large enough to make the signs of the Hartree and Fock parts of $g_{z/xy}$ opposite. For completeness we present some phase diagrams for this type of situation in this subsection. 

\begin{figure}[H] 
    \centering
    \includegraphics[width=0.875\columnwidth]{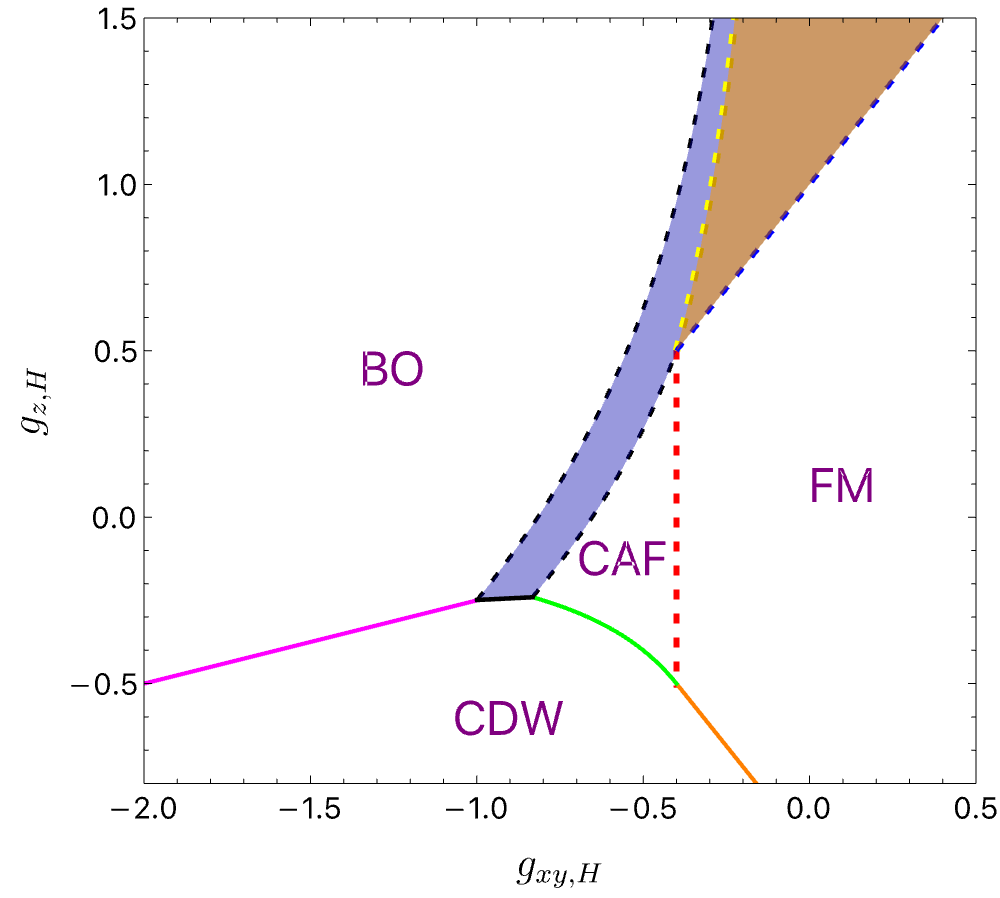}
    \caption{Phase diagram for  $E_Z=1.0,E_V=0.0,g_{z,F}=-1.0*g_{z,H},g_{xy,F}=1.25*g_{xy,H}$. The CAF now occurs in a finite region of the parameter space. There are two coexistence phases. The blue shaded region is the B/CAF phase with BO, CAF, and SVEY order. The region shaded brown shows a new type of coexistence between FM and SVEX/SVEY order. }
    \label{fig:PD10}
\end{figure}

\begin{figure}[H] 
    \centering
    \includegraphics[width=0.875\columnwidth]
    {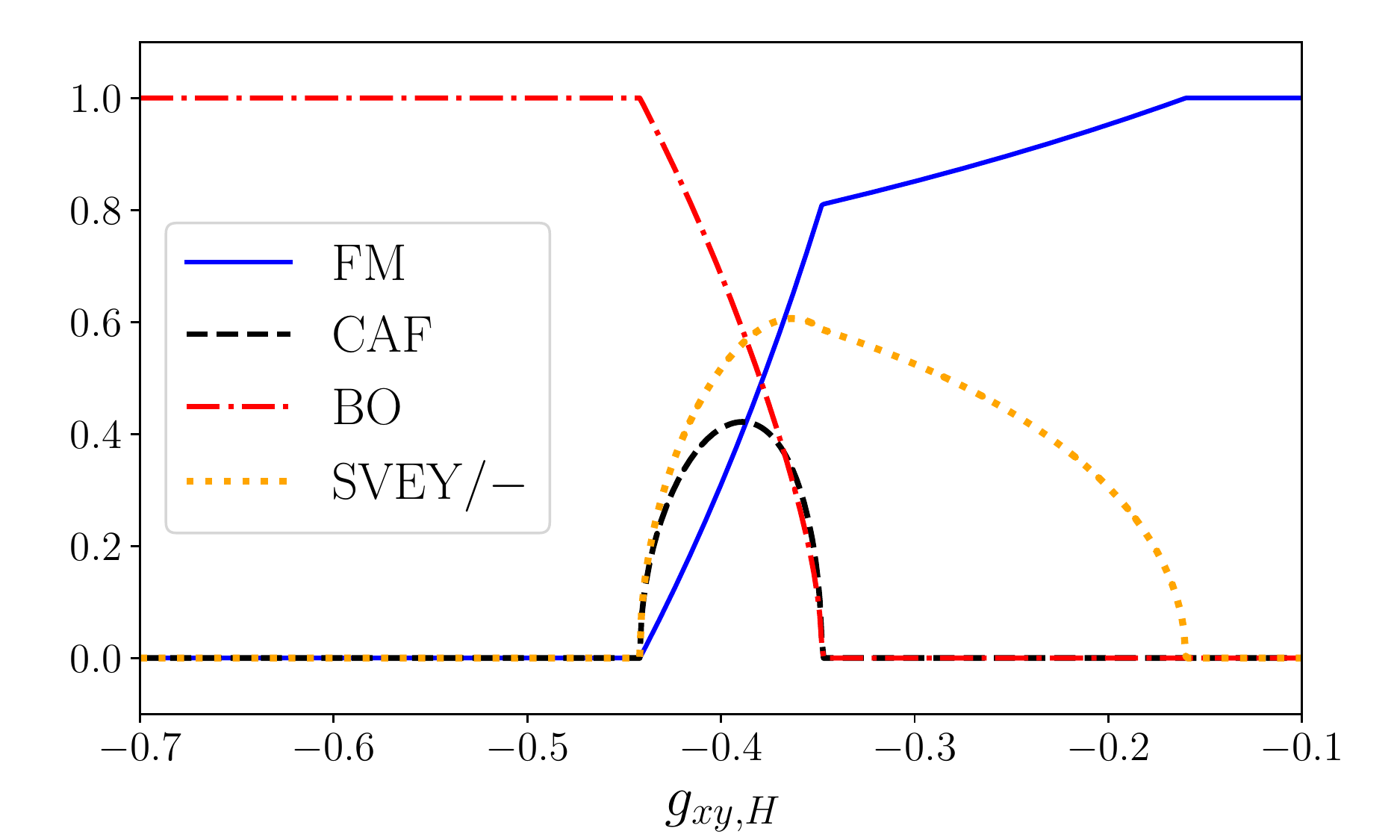}
    \caption{Nonzero order parameters along a horizontal cut of the phase diagram of \cref{fig:PD10} at $g_{z,H}=0.8$.  All other coupling constants are the same as in \cref{fig:PD10}. One starts in the BO phase at large negative $g_{xy,H}$. There is a second-order transition into the B/CAF phase with BO, CAF, and SVEY order. Next, there is another second-order transition into the FSVE phase showing the coexistence of FM and SVEX/SVEY order, the two being degenerate. We have chosen to plot SVE-$=\langle|\tau_x\sigma_x-\tau_y\sigma_y|\rangle/2$, which is continuous across the transition. Finally, for larger $g_{xy,H}$ the system goes into the FM phase. }
    \label{fig:OP_n_gZ}
\end{figure}

First we consider the case $g_{z,F}/g_{z,H}=-1$, but $g_{xy,F}/g_{xy,H}>1$. The most obvious change is in the topology of the phase diagram. The CAF phase is completely surrounded by other phases. In addition, there are two coexistence phases. The blue shaded region is the B/CAF phase familiar from the previous subsections. It has BO, CAF, and SVEY order. The brown shaded region represents a new type of coexistence phase which does not occur when $g_{\mu,F}/g_{\mu,H}>0$. This phase has FM order coexisting with SVEX/SVEY order (all the ground states generated from SVEX by $U(1)_s\otimes U(1)_v$ are degenerate), with no other order parameters being present. This state spontaneously breaks the $U(1)_s$ and $U(1)_v$ symmetries, but preserves $U(1)_{sv}$. \cref{fig:OP_n_gZ} shows the order parameters along a horizontal cut in the phase diagram of \cref{fig:PD10} at $g_{z,H}=1$.  We will call this the FSVE phase. 


When $E_V>0$, the FSVE phase acquires both SVEX and SVEY order in addition to FM and CDW order (which is natural since $E_V>0$). Thus, the FSVE  phase remains different from the B/CAF phase (because there is no CAF order in the FSVE phase) and the SVE+ phase, which requires a precise equality of the SVEX and SVEY order parameters. More details are in the appendices.  

\begin{figure}[h] 
    \centering
    \includegraphics[width=0.875\columnwidth]{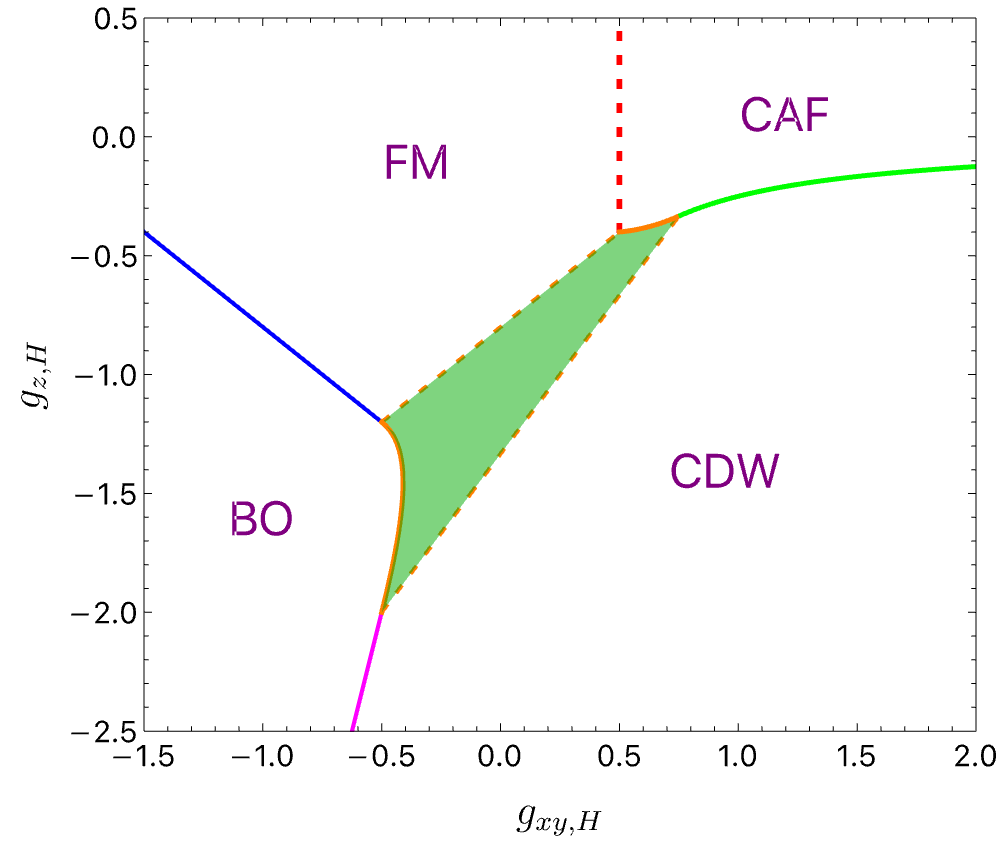}
    \caption{Phase diagram for the ratios $g_{z,F}/g_{z,H}=1.25>0$ and $g_{xy,F}/g_{xy,H}=-1<0$. The other couplings are  $E_Z=1,\ E_V=0$. While the topology has changed, the SVE+ phase that interpolates between the CDW and FM phases occurs here as well.}
    \label{fig:PD11}
\end{figure}
As a second example, we show the case when $g_{z,F}/g_{z,H}=1.25>0$ and $g_{xy,F}/g_{xy,H}=-1<0$, which is shown in Fig.\ref{fig:PD11}. As expected for negative ratios between the H and F parts of the coupling, the phases are drastically rearranged in the phase diagram. However, the nature of the phases remains the same as in \cref{subsec:evzero}. There is no B/CAF phase, but the SVE+ phase does occur, interpolating between the CDW and FM phases. 

A more complete set of figures for other cases when one or both of the H and F coupling ratios are negative appears in the appendices.

\section{Caveats, Conclusions, and Open Questions}
\label{sec:conclusions}

Monolayer graphene at charge neutrality in the quantum Hall regime is an example of quantum Hall ferromagnetism \cite{Shivaji_Skyrmion,QHFM_Fertig_1989,QHFM_Yang_etal_1994,QHFM_Moon_etal_1995}. While earlier theoretical \cite{kharitonov2012:nu0} and experimental work \cite{young2014:nu0,Magnon_transport_Yacoby_2018} seemed to suggest a simple canted antiferromagnetic phase at perpendicular magnetic field, recent scanning tunneling observations \cite{li2019:stm,STM_Yazdani2021visualizing,STM_Coissard_2022} clearly show Kekul\'e and even charge density wave order. 

A major assumption underlying most earlier theoretical work \cite{alicea2006:gqhe,Herbut1,Herbut2,kharitonov2012:nu0} is that the residual interactions (other than Coulomb), being descended from lattice-scale couplings, should be ultra-short-range on the scale of the magnetic length $\ell$. Recently, based on renormalization group ideas, it was argued  \cite{Das_Kaul_Murthy_2022} that the effective low-energy interactions will naturally acquire the length scale $\ell$ even if they were ultra-short-range microscopically. It should be noted that this argument does not depend on whether one considers the $N=0$ manifold (the ZLLs) or some other manifold of Landau levels of graphene. For such generic interactions, coexistence between CAF and Kekul\'e order in the physically relevant region of the coupling constant space was shown to occur \cite{Das_Kaul_Murthy_2022} in a robust regime of couplings. 

Our goal in this paper was to present a more complete phase diagram of MLG at charge neutrality (and more generally, in any half-filled manifold of Landau levels). As in previous theoretical work, we assume a clean system with four-Fermi interactions only. Momentum conservation constrains the residual four-Fermi interactions to have a $U(1)$ valley symmetry (reduced to a $Z_3$ upon including higher-Fermi interactions, which we ignore). This still leaves two coupling functions $v_{z}(\bq),v_{xy}(\bq)$ to be determined. In the Hartree-Fock  approximation, restricting to ground states respecting translation invariance up to an intervalley coherence, the energetics can be captured by a Hartree coupling and a Fock coupling for each of the two types. After the inclusion of the Zeeman and valley Zeeman couplings $E_Z,\ E_V$, one has six coupling constants to deal with.   

Our main physically relevant finding is that coexistence between CAF and B/CO order (the bond order always coexists with CDW order when $E_V>0$)  occurs in a large region of the coupling constant space when $g_{xy,H},g_{xy,F}<0$. Depending on the particular values of the interaction couplings, the valley Zeeman coupling may either enhance or suppress coexistence. In fact, in a certain range of couplings, a re-entrant transition from a pure B/CO through a coexistence phase back to a pure B/CO is possible upon increasing $E_V$. The fact that coexistence is generic is consistent with the ubiquity of B/CO order seen in STM observations \cite{li2019:stm,STM_Yazdani2021visualizing,STM_Coissard_2022}.

We found two other coexistence phases in regions of the phase diagram which may not be directly relevant to graphene. The first, called the SVE+ phase, interpolates between the CDW and FM phases, and also has a spin-valley entangled SVE+ order parameter. The corresponding phase in BLG was found in previous work \cite{Murthy_BLG_2017}. The second one, called the FSVE phase, occurs when $g_{z,F}$ is allowed to be of the opposite sign as $g_{z,H}$, which could conceivably occur for strong Landau-level mixing. For generic $E_V>0$, the FSVE phase has FM, CDW, SVEX  and SVEY order, but no CAF or bond order. 

Our results are complete given our assumptions, but they come with caveats. We have ignored spontaneous translation symmetry breaking beyond that required by intervalley coherence, static disorder, and quantum/thermal fluctuations. Let us consider each in turn. Intervalley coherence is allowed in our approach, and is a signal of minimal translation symmetry breaking because it implies a new reciprocal lattice vector $\bK-\bK'$. In combination with the reduction of the $U_v(1)$ symmetry to $Z_3$ upon including higher-Fermi interactions, this leads to the Kekul\'e order, which does break lattice translation symmetry with a three-fold enlargement of the unit cell. Thus, this type of translation symmetry breaking in implicitly included in our approach. Since the STM experiments see only this minimal type of translation symmetry breaking, we believe our assumption of translation invariance up to an intervalley coherence is not a serious limitation. Next, static disorder is present in all samples. When B/CO order is present, it is expected to be pinned by the local value of disorder. In fact, since the B/CO order couples to static disorder while the CAF order does not, one expects the region of the coupling constant space where B/CO order is present to increase as disorder increases. Otherwise, static disorder is expected to have a quantitative effect on the transport gaps, but leave the nature of the state unaffected. 

Now we turn to quantum fluctuations. The Hartree-Fock approximation has an excellent track record in describing quantum Hall ferromagnets at zero temperature \cite{QHFM_Fertig_1989,Shivaji_Skyrmion,QHFM_Yang_etal_1994,QHFM_Moon_etal_1995}. Near a second-order phase boundary, HF will always predict a mean-field transition. Quantum fluctuations will shift the phase boundary, and correct the critical behavior to the appropriate universality class; for example, the transition from the B/CO to the coexistence phase with the CAF order should have the universality class of the three-dimensional $XY$ model.  If a phase occurs in a very narrow sliver of coupling constant space, one might envisage quantum fluctuations making it disappear. However, since all the phases we find are robust, occurring over substantial ranges of coupling constants, we expect quantum fluctuations to alter the phase boundaries in detail, but not affect the phase diagram qualitatively. An important potential exception is the first-order transition seen in many parameter regimes between the FM and CDW phases. When $E_Z,\ E_V>0$, neither of these phases breaks any symmetry of the Hamiltonian. One cannot rule out a first-order transition without a change in symmetry between the two phases: An example is the liquid-gas transition. However, a phase transition between the FM and CDW phases is not necessary, since both have the same symmetry. Quantum fluctuations may destroy the first-order line in favor of a smooth crossover between the CDW and FM. Consider parameter regimes when the SVE+ phase intervenes between the CDW and FM mean-field phases. Since the SVE+ phase breaks $U(1)_s$ and $U(1)_v$ spontaneously, a second-order phase transition is allowed between it and the FM or CDW regions. Quantum fluctuations may change the topology of the phase diagram to make the SVE+ phase an island in the middle of the crossover between the CDW-dominated and FM-dominated  regions.  

Finally, we turn to $T>0$ and thermal fluctuations. Recall that spontaneous intervalley coherence is subject to a $Z_3$ symmetry, and hence does not result in a Goldstone mode. Long-range B/CO order is expected to be survive to a critical temperature $T_c>0$. The $U_{spin}(1)$ symmetry is an almost exact symmetry, being broken only by the tiny spin-orbit coupling ($\approx 10\mu eV$) in graphene \cite{SOC_Graphene_Paco2006,SOC_Graphene_MacD2006,SOC_Graphene_Yao2007,SOC_Graphene_Sandler2007}. Thus, setting spin-orbit coupling to zero, any spontaneous breaking of the $U(1)_s$ symmetry leads to a gapless Goldstone mode at $T=0$.  For $T>0$ the system is in the universality class of the two-dimensional $XY$-model. Long-range order is absent at any nonzero $T$, and there should be a Berezinskii-Kosterlitz-Thouless (BKT) transition \cite{KT,Berezinskii_1972} at $T_{KT}$, below which there is power-law order. $T_{KT}$ should vanish as the CAF order parameter vanishes, because the stiffness vanishes as well. Thus, there is an intermediate temperature regime in the B/CAF phase which is above $T_{KT}$ but below the $T_c$ of the $Z_3$ bond order.

Let us turn to experimental signatures of coexistence. STM experiments can directly measure the B/CO order; thus the key question is how to detect CAF order. Any order parameter that spontaneously breaks $U(1)_s$ will lead to a gapless Goldstone mode. Thus, the CAF phase and all three coexistence phases we have found would support a Goldstone mode at $T=0$. Magnon transmission experiments \cite{Magnon_transport_Yacoby_2018,Magnon_Transport_Assouline_2021,Magnon_transport_Zhou_2022} reveal the presence of  magnetic excitations, but because the magnons are created in ferromagnetic regions with a gap of $E_Z$, such experiments are unable to reveal whether the magnetic excitations at $\nu=0$ are gapless. In Bernal-stacked bilayer graphene, where a CAF state is also expected to be present, a very recent experiment has confirmed the presence of the gapless Goldstone mode \cite{JunZhu_2021}. If such an experiment can be carried out for MLG it would be direct confirmation of the spontaneous breaking of spin-rotation symmetry $U(1)_s$. More broadly, in the context of transport, a measurement of the bulk thermal conductivity below $T_{KT}$ should reveal the Goldstone mode. Additionally, the BKT transition itself should have a signature in $R_{xx}$ \cite{Rxx_KT_Eisenstein}.

A more indirect way to probe the CAF order parameter is to examine the detailed structure of spin/valley skyrmions, which can be induced by external charges. A  thorough analysis of skyrmions in the ultra-short-range model of charge-neutral graphene was carried out very recently \cite{Goerbig_nu0_Skyrmion_Zoo_2021}. Using this framework, an analysis of the B/CO texture near a charge defect shows \cite{STM_Yazdani2021visualizing} that it is consistent with theory, assuming that the true ground state is pure B/CO. A phase with coexistence between B/CO and CAF order will have skyrmions that differ in detail from those of the pure B/CO phase. 

There are two broad open questions. Firstly, given a microscopic model at some intermediate energy scale much larger than $\hbar\omega_c$, how does one reliably deduce the effective coupling functions in the manifold of the $n=0$ Landau levels? Kharitonov \cite{kharitonov2012:nu0}, following earlier RG treatments \cite{RG1_Graphene_Aleiner_2007,RG2_Graphene_Aleiner_2008}, carried out just such a procedure, under the assumption that the couplings (other than Coulomb) remained ultra-short-range under RG, implying a finite number of couplings to renormalize. Based on a general fermionic RG procedure \cite{Shankar_RG_1994} which includes all low-energy interactions, there has been quite a bit of previous work attempting to integrate out higher Landau levels perturbatively \cite{RG_Murthy_Shankar_2002,RG_Bishara_Nayak_2009,RG_Sodemann_MacD2013,RG_Peterson_Nayak_2013,RG_Peterson_Nayak_2014}. While these works restricted themselves to the Coulomb interaction, it should be straightforward to extend them to include all symmetry-allowed interactions. 

The second broad question is complementary to the first. Given an experimental sample, is there a complete set of measurements that could determine the couplings $g_{z,H},g_{z,F},g_{xy,H},g_{xy,F}$? Given that $E_Z$ is determined by the total field, and $E_V$ can be deduced from zero-magnetic-field gap measurements at charge neutrality, this would fully determine the effective theory at the mean-field level. One way to approach this is via a detailed investigation of skyrmions \cite{Goerbig_nu0_Skyrmion_Zoo_2021}. As long as the size of the skyrmions is much larger than the magnetic length, a nonlinear sigma model approach is capable of capturing their structure and energetics. The parameters that enter the nonlinear sigma model are exactly those that enter the mean-field theory, with the exception of the stiffness, which is determined by the dominant Coulomb interaction. 

Last, but not least, let us briefly consider fractionally filled states in the $n=0$ manifold of Landau levels in graphene. For the case of $SU(4)$ Coulomb interactions plus ultra short range residual interactions, it is possible to construct variational states with integer and/or fractional fillings in the different flavors and compute their energies \cite{Sodemann_MacDonald_2014,Hegde_2022}. Determining whether this construction can be extended to generic residual interactions of arbitrary range is an important open question. 

We look forward to addressing these and other questions in the near future.

\begin{acknowledgements}
SJD would like to acknowledge the Infosys funding for final year 
students. He also wants to acknowledge ICTS for its hospitality and kind support towards
academic collaboration.
AD was supported by the German-Israeli Foundation Grant No. I-1505-303.10/2019, DFG MI 658/10-2,
DFG RO 2247/11-1, DFG EG 96/13-1, and CRC 183 (project C01).  AD also thanks
the Israel planning and budgeting committee (PBC) and the Weizmann Institute of Science,
the Dean of Faculty fellowship, and the Koshland Foundation for financial
support. S.R. and G.M. would like to thank the VAJRA scheme of SERB, India for its support. 
R.K.K. was supported in part by NSF DMR-2026947. G.M. would like to acknowledge partial support from the US-Israel Binational Science Foundation (grant no. 2016130). G.M. and R.K.K. are grateful for the wonderful environment at the Aspen Center for Physics (NSF grant PHY-1607611). 
\end{acknowledgements}

\appendix
\section{Different Phases and Hessian}
\label{sec:appintro}
For charge-neutral graphene, assuming that interactions are ultra-short-range (USR), there are
four phases, namely, the ferromagnet (FM), the canted antiferromagnet (CAF), the bond-ordered 
phase (BO), and the  charge density wave (CDW) phase. We call these phases ``simple", because
they can all be described by a single nontrivial angle which is known analytically in terms of
the couplings. As shown in the main text, when one removes the USR restriction on the
interactions, other coexistence phases become possible. Among them is a spin-valley
entangled phase SVE+, which can also be described by a single angle, and is also ``simple". 

A primary tool in our investigation of the phase diagram is a study of the stability of a given
ground state. Recall that each candidate state is described by filling in two linear combinations
of the four spin-valley degenerate states at each guiding center. We reproduce the equations
from the main text here for convenience, \cite{Doucot_Goerbig_Skyrmion_2008,Lian_Rosch_Goerbig_2016,Lian_Goerbig_2017,Atteia_Goerbig_2021}
\begin{align}
    |f_1\rangle=&\cos{\frac{\alpha_1}{2}}|\bn\rangle\otimes|\bs\rangle+e^{i\beta_1}\sin{\frac{\alpha_1}{2}}|-\bn\rangle\otimes|-\bs\rangle\\
    |f_2\rangle=&\cos{\frac{\alpha_2}{2}}|\bn\rangle\otimes|-\bs\rangle+e^{i\beta_2}\sin{\frac{\alpha_2}{2}}|-\bn\rangle\otimes|\bs\rangle
\end{align}
where $\bn=\sin{\theta_p}\cos{\phi_p}\he_x+\sin{\theta_p}\sin{\phi_p}\he_y+\cos{\theta_p}\he_z$,
and $\bs=\sin{\theta_s}\cos{\phi_s}\he_x+\sin{\theta_s}\sin{\phi_s}\he_y+\cos{\theta_s}\he_z$
are unit vectors indicating the directions of the state on the valley and spin Bloch spheres,
respectively. The spinors $|\bn\rangle$ and $|\bs\rangle$ are defined in the standard way
\begin{equation}
    |\bn\rangle=\left(\begin{array}{c}
    \cos{\frac{\theta_p}{2}}\\
    e^{i\phi_p}\sin{\frac{\theta_p}{2}}\end{array}\right);\\ |\bs\rangle=\left(\begin{array}{c}
    \cos{\frac{\theta_s}{2}}\\
    e^{i\phi_s}\sin{\frac{\theta_s}{2}}\end{array}\right)
\end{equation}
As shown in the main text, the $U(1)_s$ symmetry allows us to set $\phi_s=0$, and the $U(1)_v$ symmetry allows us to set $\phi_p=0$. The $SU(2)_s$ symmetry of the interactions forces the HF energy to depend only on $\beta_1+\beta_2$. The energy depends on $\beta_1+\beta_2$ only via $\cos(\beta_1+\beta_2)$, which appears linearly in the energy. Therefore, we can restrict consideration to the two discrete possibilities $\beta_1+\beta_2=0,\ \pi$ mod $2\pi$.  The HF energy of a given state thus depends only on the four continuously varying angles $\alpha_1,\alpha_2,\theta_p,\theta_s$.

We now indicate how to determine the values of the four angles for the ``simple" states shown in \cref{subsec:ansatz_Hessian}. We begin by examining the expressions for the order parameters as functions of the four angles:
\begin{align}
    FM&=\frac{\langle\sigma_z\rangle}{2}=\frac{(\cos[\alpha_1] - \cos[\alpha_2]) \cos[\theta_s]}{2},\nonumber \\
    CAF&=\frac{\langle\tau_z\sigma_x\rangle}{2}=\frac{1}{2}\Bigg[\cos[\theta_s] \Big(\cos[\beta_1] \sin[\alpha_1]\nonumber\\
    &+ \cos[\beta_2] \sin[\alpha_2]\Big) \sin[\theta_p]\Bigg],\nonumber \\ 
   BO&=\frac{\langle\tau_x\rangle}{2}=\frac{(\cos[\alpha_1] + \cos[\alpha_2]) \sin[\theta_p]}{2},\nonumber \\
   CDW&=\frac{\langle\tau_z\rangle}{2}=\frac{(\cos[\alpha_1] + \cos[\alpha_2]) \cos[\theta_p]}{2}
\end{align}
First we focus on the CAF phase. In this phase the BO and CDW order parameters should vanish identically, implying  that $\alpha_2=\pi-\alpha_1$. Examining the FM and CAF order-parameters, we see that we can choose $\theta_s=0$ and $\theta_p=\pi/2$ because $\cos{\theta_s}$ and $\cos{\theta_p}$ appear as a overall normalization factors which can be set to  one. After imposing these constraints on the angles, the HF energy will depend only upon the angles $\alpha_1$, $\beta_1$, $\beta_2$, and can be expressed as 
\begin{align}
    E_{HF}=&\frac{1}{4} \Big(-8 E_Z \cos (\alpha_1)\nonumber\\
    &+2 \sin ^2(\alpha_1) (g_{xy,F}-g_{z,F})\cos (\beta_1+\beta_2)\nonumber\\
    &-\cos (2 \alpha_1) (3 g_{xy,F}+g_{z,F})\nonumber\\
    &-5 g_{xy,F}-3 g_{z,F}\Big).
\label{eq:ven_CAF}
\end{align}
The CAF phase occurs for $g_{xy,F}<0$, $g_{z,F}>0$. This restricts $\cos(\beta_1+\beta_2)=1$, allowing us to choose $\beta_1=\beta_2=\pi$. The CAF and FM order-parameters are $FM=\cos[\alpha_1]$ , $CAF=\sin[\alpha_1]$. Here $\alpha_1$ is the single non-trivial angle which varies through the CAF phase. It has the functional dependence $\alpha_1=\cos^{-1}\left[\frac{-E_Z}{2g_{xy,F}}\right]$, which we found by minimizing \cref{eq:ven_CAF}. Clearly when $E_Z > 2|g_{xy,F}|$, the angle $\alpha_1$ will be fixed to zero and this corresponds to the FM phase. The parameterization of the FM phase is subsumed in the above. 

Now we turn our attention to the B/CO phase which generally occurs for $E_V>0$. Since the CDW and BO phases are restricted versions of the B/CO phase, their parameterizations are subsumed in that of the B/CO phase. In the B/CO phase the FM and CAF order-parameters should vanish identically, allowing us to choose $\alpha_1=\alpha_2=0$. The B/CO phase is a singlet, which means that the direction of ${\mathbf s}$ can be chosen arbitrarily, allowing us to fix $\theta_s=0$. With this  choice of angles the BO and CDW order-parameters are $BO=\sin[\theta_p]$, $CDW=\cos[\theta_p]$. Thus, in this phase the angle $\theta_p$ is the non-trivial angle. To find its functional dependence, we examine the HF energy with the constraints on angles $\alpha_1=\alpha_2=\theta_s=0$, which is
\begin{align}
    E_{HF}=&-2 E_V \cos (\theta_p)+\frac{1}{2} g_V \cos (2\theta_p)\nonumber\\
    &-\frac{g_{xy,F}}{2}+g_{xy,H}-\frac{g_{z,F}}{2}+g_{z,H},
\label{eq:ven_BO}
\end{align}
where $g_V=2g_{z,H}-g_{z,F}-2g_{xy,H}+g_{xy,F}$. Minimizing \cref{eq:ven_BO} leads to the functional dependence $\theta_p=\cos ^{-1}\left[\frac{E_V}{g_V}\right]$ for the B/CO phase. For $E_V>g_V$, the angle $\theta_p$ will be fixed to zero, which corresponds to the pure CDW phase.  For $E_V=0$, the angle $\theta_p=\pi/2$, which corresponds to the BO phase. 

Having described how to fix the angles for the simple states, we now turn to the instabilities of these states. 

Any HF state must be at least a local extremum. Thus, all the first derivatives of the energy $E_{HF}$ with respect to $\alpha_1,\alpha_2,\theta_p,\theta_s$ must vanish. To look for instabilities we need  to compute the second derivatives of $E_{HF}$ with respect to the four angles (the Hessian matrix). 
\begin{equation}
{\cal E}(\alpha_1,\alpha_2,\theta_p,\theta_s)=
\begin{pmatrix}
    \frac{\dE}{\partial^2{\alpha_1}} & \frac{\dE}{\partial{\alpha_1}\partial{\alpha_2}} & \frac{\dE}{\partial{\alpha_1}\partial{\theta_p}} & \frac{\dE}{\partial{\alpha_1}\partial{\theta_s}}
    \\ 
    \frac{\dE}{\partial{\alpha_2}\partial{\alpha_1}} & \frac{\dE}{\partial^2{\alpha_2}} & \frac{\dE}{\partial{\alpha_2}\partial{\theta_p}} & \frac{\dE}{\partial{\alpha_2}\partial{\theta_s}}
    \\ 
    \frac{\dE}{\partial{\theta_p}\partial{\alpha_1}} &  \frac{\dE}{\partial{\theta_p}\partial{\alpha_2}} &  \frac{\dE}{\partial^2{\theta_p}} &   \frac{\dE}{\partial{\theta_p}\partial{\theta_s}}
    \\ 
    \frac{\dE}{\partial{\theta_s}\partial{\alpha_1}} &  \frac{\dE}{\partial{\theta_s}\partial{\alpha_2}} &  \frac{\dE}{\partial{\theta_s}\partial{\theta_p}} & \frac{\dE}{\partial^2{\theta_s}}
\end{pmatrix}.
\end{equation}

The eigenvalues of the Hessian determine the stability of the given state. An eigenvalue crossing zero signals an instability of the given state. For ``simple" states, one can analytically obtain the Hessian matrix and its eigenvalues. We will use the Hessian eigenvalues of ``simple" states to map out the gross features of the phase diagram. Finer details of the phase diagrams are obtained by self-consistent iterative HF. 

In the following sections we will present explicit expressions for the Hessian and its
eigenvalues and attendant instabilities in the CAF 
(\cref{sec:appHessCAF}), the B/CO (\cref{sec:appHessBO}), the FM (\cref{sec:appHessFM}), the CDW
(\cref{sec:HessCDW}), and the SVE (\cref{sec:appHessSVE}) phases.
In \cref{sec:appHessHF_sign} we present phase diagrams when the Hartree and Fock parts of
either/both of the couplings have opposite signs, which may be relevant for very strong
Landau-level mixing. 
\section{CAF phase}
\label{sec:appHessCAF}

The CAF phase occurs only for $g_{xy,F}<0$ and $E_Z<2|g_{xy,F}|$. It is described by the
following values of the angles
\begin{align}
    \alpha_1=&\cos^{-1}\left[\frac{-E_Z}{2g_{xy,F}}\right]=\pi-\alpha_2\\
    \theta_p=&\frac{\pi}{2},\; \theta_s=0\\
    \beta_1=&\beta_2=\pi
    \label{eq:S_CAFangles}
\end{align} 

The energy of the CAF phase is 
\begin{equation}
    E_\text{CAF}=\frac{E_Z^2}{2 g_{xy,F}}-g_{z,F}.
\end{equation}
The Hessian matrix for the CAF state has the following block diagonal form:
\begin{equation}
    {\cal E}_\text{CAF}=
    \left(
\begin{array}{cc}
    A_{3 \times 3} & 0_{3 \times 1} \\
    0_{1 \times 3} & -\frac{E_Z^2}{g_{xy,F}}
\end{array}
\right)_{4 \times 4},
\end{equation}
where 

\begin{widetext}
\begin{align}
&A=\nonumber\\
&\left(
\begin{array}{ccc}
 \frac{1}{4} \left(\frac{E_Z^2 (g_{xy,F}-g_{xy,H})}{g_{xy,F}^2}-6 g_{xy,F}+4 g_{xy,H}+2 g_{z,F}\right) &
   \frac{1}{2} (g_{xy,F}+2 g_{xy,H}+g_{z,F})-\frac{E_Z^2 (g_{xy,F}+g_{xy,H})}{4 g_{xy,F}^2} & -E_V
   \sqrt{1-\frac{E_Z^2}{4 g_{xy,F}^2}} \\
 \frac{1}{2} (g_{xy,F}+2 g_{xy,H}+g_{z,F})-\frac{E_Z^2 (g_{xy,F}+g_{xy,H})}{4 g_{xy,F}^2} & \frac{1}{4}
   \left(\frac{E_Z^2 (g_{xy,F}-g_{xy,H})}{g_{xy,F}^2}-6 g_{xy,F}+4 g_{xy,H}+2 g_{z,F}\right) & -E_V
   \sqrt{1-\frac{E_Z^2}{4 g_{xy,F}^2}} \\
 -E_V \sqrt{1-\frac{E_Z^2}{4 g_{xy,F}^2}} & -E_V \sqrt{1-\frac{E_Z^2}{4 g_{xy,F}^2}} &
   \frac{\left(E_Z^2-4 g_{xy,F}^2\right) (g_{xy,F}-g_{z,F})}{2 g_{xy,F}^2} \\
\end{array}
\right).
\end{align}

Thus we find the instability equations of the CAF phase are
\begin{subequations}
\begin{align}
    g_{xy,F}&=-\frac{E_Z}{2} & \\
    E_V&= \frac{\sqrt{g_{xy,F}-g_{z,F}} \sqrt{\frac{E_Z^2 g_{xy,H}}{2}+g_{xy,F}^3-g_{xy,F}^2 (2
   g_{xy,H}+g_{z,F})}}{g_{xy,F}}.
\end{align}
\end{subequations}
\end{widetext}
The first instability corresponds to the second-order line between the FM and CAF phases, while the second corresponds to the instability in the B/CAF phase.


\section{BO phase}
\label{sec:appHessBO}
The angles $\beta_1$ and $\beta_2$ never appear in the expression for the HF energy of the BO state.  Thus, $\beta_1$ and $\beta_2$ are undetermined at this point. The other angles are 
\begin{align}
\alpha_1=&\alpha_2=\theta_s=0\\
\theta_p=&\cos ^{-1}\left[\frac{E_V}{g_V}\right],
\end{align} 
with
\begin{equation}
g_V=2g_{z,H}-g_{z,F}-2g_{xy,H}+g_{xy,F}.
\end{equation}
The energy in the BO phase is 
\begin{equation}
    E_\text{BO}=-\frac{E_V^2}{g_V}-g_{xy,F}+2 g_{xy,H}.
\end{equation}
The Hessian matrix and its eigenvalues do depend on $\beta_1+\beta_2$, with the eigenvalues depending on $\cos[2(\beta_1+\beta_2)]$. The most severe constraint on the region of stability occurs when $\cos[2(\beta_1+\beta_2)]=1$. Below, we choose $\beta_1=\beta_2=0$, and obtain the form of the Hessian
\begin{equation}
{\cal E}_\text{BO}=
\left(
\begin{array}{cc}
    \frac{A_{2 \times 2}}{2} & 0_{2 \times 2} \\
    0_{2 \times 2} & B_{2 \times 2}
\end{array}
\right)_{4 \times 4}
\end{equation}
where

\begin{widetext}
\begin{equation}
A=
\left(
\begin{array}{cc}
 \frac{E_V^2 (g_{z,F}-g_{xy,F})}{(g_{xy,F}-2 g_{xy,H}-g_{z,F}+2 g_{z,H})^2}+2 E_Z+g_{xy,F}-4
   g_{xy,H}-g_{z,F} & (g_{xy,F}-g_{z,F}) \left(1-\frac{E_V^2}{(g_{xy,F}-2 g_{xy,H}-g_{z,F}+2
   g_{z,H})^2}\right) \\
 (g_{xy,F}-g_{z,F}) \left(1-\frac{E_V^2}{(g_{xy,F}-2 g_{xy,H}-g_{z,F}+2 g_{z,H})^2}\right) & \frac{E_V^2
   (g_{z,F}-g_{xy,F})}{(g_{xy,F}-2 g_{xy,H}-g_{z,F}+2 g_{z,H})^2}-2 E_Z+g_{xy,F}-4 g_{xy,H}-g_{z,F} \\
\end{array}
\right),
\end{equation}
and
\begin{equation}
B=
\left(
\begin{array} {cc}
    - \frac {2 E_V^2} {g_ {xy, F} - 2 g_ {xy, H} - g_{z, F} + 
    2 g_{z, H}} + 2 g_ {xy, F} - 4 g_{xy, H} - 2 g_{z, F} + 4 g_{z, H} & 0 \\
    0 & 0 \\
\end{array}
\right).
\end{equation}

The instability lines in the BO phase are
\begin{subequations}
\begin{align}
    g_{z,H}=& \frac{1}{2} (E_V-g_{xy,F}+2 g_{xy,H}+g_{z,F}) & \\
    E_V=&\sqrt{\frac{\left(E_Z^2-2 g_{xy,H} (-g_{xy,F}+2 g_{xy,H}+g_{z,F})\right) (g_{xy,F}-2
   g_{xy,H}-g_{z,F}+2 g_{z,H})^2}{2g_{xy,H} (g_{xy,F}-g_{z,F})}}.
\end{align}
\end{subequations}
 \end{widetext}
These correspond to the instability of the CDW phase and the instability of the B/CAF
phase, respectively.
%


\section{FM phase}
\label{sec:appHessFM}
The FM phase is described by the angles
\begin{subequations}
\begin{align}
\alpha_1&=0,&\ \alpha_2=\pi\\
\theta_p=&\frac{\pi}{2},\ \theta_s=0\\
\beta_1=&\beta_2=\pi
\end{align}
\end{subequations}
The ground state energy of the FM phase is 
\begin{equation}
    E_\text{FM}=-2 (E_Z+g_{xy,F})-g_{z,F}.
\end{equation} The form of
the hessian matrix of the FM phase is
\begin{equation}
{\cal E}_\text{FM}=
\left(
\begin{array}{cc}
    \frac{A_{2 \times 2}}{2} & 0_{2 \times 2} \\
    0_{2 \times 2} & B_{2 \times 2}
\end{array}
\right)_{4 \times 4},
\end{equation}
where
\begin{equation}
A=\left(
\begin{array}{cc}
 2 E_Z+3 g_{xy,F}+g_{z,F} & g_{z,F}-g_{xy,F} \\
 g_{z,F}-g_{xy,F} & 2 E_Z+3 g_{xy,F}+g_{z,F} \\
\end{array}
\right),
\end{equation} 
and 
\begin{equation}
B=\left(
\begin{array}{cc}
 0 & 0 \\
 0 & 2 E_Z \\
\end{array}
\right).
\end{equation}
The instability lines in the FM phase are 
\begin{subequations}
\begin{align}
    g_{xy,F}=&-\frac{E_Z}{2} & \\
    g_{z,F}=&-E_Z-g_{xy,F}.
\end{align}
\end{subequations}
The first is the second-order line between the FM and CAF phases, while the second is the
instability towards the CDW phase. Note that this is not the actual position of the dominant  
instability, which should depend on $E_V$ as well. The reason is that  $E_V$ does not appear
in the Hessian matrix for the FM phase. Note that one eigenvalue is always zero, which means
one direction in angle space is flat. While we believe that some higher-order derivative of
the ground state energy must reveal the instability towards the SVE+ phase, we have not
pursued this issue because we can find the instability easily from the SVE+ side.  

\section{CDW phase}
\label{sec:HessCDW}
The CDW phase is described by the angles 
\begin{equation}
    \alpha_1=\alpha_2=\theta_p=\theta_s=0.
\end{equation}
The angles $\beta_1,\ \beta_2$ never appear in the expression of the ground state energy, which is 
\begin{equation}
E_\text{CDW}=-2 E_V-g_{z,F}+2 g_{z,H}.
\end{equation} 
The angles $\beta_1$ and $\beta_2$ also do not appear in the Hessian matrix, which has
the following form

\begin{widetext}
\begin{align}
&{\cal E}_\text{CDW}=\nonumber\\
&\left(
\begin{array}{cccc}
 E_V+E_Z-g_{xy,F}+g_{z,F}-2 g_{z,H} & 0 & 0 & 0 \\
 0 & E_V-E_Z-g_{xy,F}+g_{z,F}-2 g_{z,H} & 0 & 0 \\
 0 & 0 & 2 (E_V-g_{xy,F}+2 g_{xy,H}+g_{z,F}-2 g_{z,H}) & 0 \\
 0 & 0 & 0 & 0 \\
\end{array}
\right).
\end{align}
\end{widetext}
Note that one eigenvalue is always zero. This does not indicate instability
but rather the fact that one of the four continuously varying angles does not appear
in the Hessian.  The lines of instability of the CDW phase are
\begin{subequations}
\begin{align}
    g_{z,H}=&\frac{1}{2} (E_V-g_{xy,F}+2 g_{xy,H}+g_{z,F})\\
    g_{z,H}=&\frac{1}{2} (E_V-E_Z-g_{xy,F}+g_{z,F}).
\end{align}
\end{subequations}
The first is the instability toward the B/CO phase, while the second is the instability towards the SVE+ phase. \\

\section{SVE+ phase}
\label{sec:appHessSVE}

The SVE+ phase occurs near the boundary of the FM and CDW phases. In addition to the FM and CDW
order parameters, this phase also has a nonzero expectation value of 
\begin{equation}    SVE+=\langle\left[\tau_x\sigma_x+\tau_y\sigma_y\right]/2\rangle
\end{equation}
Ordering the rows and columns as $\bK\uparrow,~\bK\downarrow,~\bK'\uparrow,~\bK'\downarrow$,
the projector matrix has the form 
\begin{equation}
    \Delta_{SVE+}\left(\xi\right)=\left(
\begin{array}{cccc}
 1 & 0 & 0 & 0 \\
 0 & \cos \left(\frac{\xi }{2}\right)^2 & -\frac{\sin (\xi )}{2} & 0 \\
 0 & -\frac{\sin (\xi )}{2} & \sin \left(\frac{\xi }{2}\right)^2 & 0 \\
 0 & 0 & 0 & 0 \\
\end{array}
\right),
\end{equation}
with
\begin{equation}
 \xi=\cos^{-1}\left(\frac{E_Z-E_V+g_{xy,F}+g_{z,H}}{g_{z,F}-g_{z,H}}\right).
\end{equation}
Clearly, the phase exists only for the cosine argument having a magnitude smaller than unity. 
The angles describing this phase are
\begin{subequations}
\begin{align}
\alpha_1=&0;\ \alpha_2=\xi\\
\theta_p=&\theta_s=0\\
\beta_1=&\beta_2=\pi
\end{align}
\end{subequations}



\begin{widetext}
The HF energy is $E_{SVE+}=$
\begin{equation}
\frac{E_V^2-2 E_V (E_Z+g_{xy,F}+g_{z,F})+E_Z^2+2
   E_Z (g_{xy,F}-g_{z,F})+4 g_{z,H}
   (E_Z+g_{xy,F})+g_{xy,F}^2+2 g_{z,F}
   (g_{z,H}-g_{xy,F})-g_{z,F}^2}{2 (g_{z,F}-g_{z,H})}
\end{equation}

The Hessian matrix is
\begin{equation}
{\cal E}_{SVE+}=
\left(
\begin{array}{cc}
    A_{2 \times 2} & 0_{2 \times 2} \\
    0_{2 \times 2} & B_{2 \times 2}
\end{array}
\right)_{4 \times 4}
\end{equation}
where,
\begin{equation}
A=\left(
  \begin {array} {cc}
          \frac {E_V (g_ {xy, F} + g_ {z, F}) - (E_Z + g_ {xy, F} + 
             g_ {z, F}) (g_ {xy, F} - g_ {z, F} + 2
                 g_ {z, H})} {g_ {z, F} - g_ {z, H}} & 0  \\
       0 & \frac {(E_V - E_Z - g_ {xy, F} - g_ {z, F}) (E_V - E_Z - 
       g_ {xy, F} + g_ {z, F} - 2
           g_ {z, H})} {g_ {z, F} - g_ {z, H}} \\
    \end {array}
   \right)
\end{equation}
and 
\begin{equation}
B=
\left(
\begin{array}{cc}
    C & 0 \\
    0 & D
\end{array}
\right)
\end{equation}
with 
$C=- \frac {(E_V - E_Z - g_ {xy, F} - g_ {z, F}) \left (E_V (g_ {xy, F} - g_ {xy, H}) + E_Z(-g_ {xy, F} + g_ {xy, H} + g_ {z, F} - g_ {z, H}) - g_ {xy, F}^2 + g_ {xy, F} g_ {xy, H} + g_ {xy, F} g_ {z, F} - 2 g_ {xy, F}g_ {z, H} + g_ {xy, H} g_ {z, F} \right)} {(g_ {z, F} - g_ {z, H})^2}$ and $D= \frac {E_Z (E_V - E_Z - g_ {xy, F} + g_ {z, F} - 2 g_ {z, H})} {g_ {z, F} - g_ {z, H}}$.

\vspace{0.2cm}
The instabilities of SVE+ phase are given by
\begin{subequations}
\begin{align}
    g_{z,F}=&E_V-E_Z-g_{xy,F}\\
    g_{z,H}=&\frac{1}{2} ( E_V- E_Z- g_{xy,F}+ g_{z,F})\\
    g_{z,H}=&\frac{(g_{xy,F}-g_{xy,H}) (E_V-E_Z-g_{xy,F})+g_{z,F}(E_Z+g_{xy,F}+g_{xy,H})}{E_Z+2 g_{xy,F}}\\
    E_V=&\frac{(E_Z+g_{xy,F}+g_{z,F}) (g_{xy,F}-g_{z,F}+2 g_{z,H})}{g_{xy,F}+g_{z,F}}.
\end{align}
\end{subequations}
The first corresponds to the instability towards the FM phase, while the second corresponds to
the instability towards the CDW phase. The third is the instability towards the B/CAF phase,
while the fourth seems to not be relevant. 
\end{widetext}

\twocolumngrid
\section{Hartree and Fock couplings having opposite signs}
\label{sec:appHessHF_sign}

When Landau-level mixing is strong, it could happen that the Hartree and Fock parts of a given coupling are of opposite sign. We will present sample results for the three cases (i) $g_{z,F}/g_{z,H}<0;\ g_{xy,F}/g_{xy,H}>0$. (ii) $g_{z,F}/g_{z,H}>0;\ g_{xy,F}/g_{xy,H}<0$. (iii) $g_{z,F}/g_{z,H}<0;\ g_{xy,F}/g_{xy,H}<0$.

We have found that if the ratio F/H of a particular coupling is negative, the magnitudes of H and F parts of that particular coupling seem to be irrelevant to the phase diagram. This is in contrast to the case when the ratio F/H is positive, where it matters a great deal whether the ratio is bigger or smaller than unity. 

Consider first case (i): $g_{z,F}/g_{z,H}<0;\ g_{xy,F}/g_{xy,H}>0$. We set $g_{z,F}=-g_{z,H}$, and vary the ratio $g_{xy,F}/g_{xy,H}$. 

\begin{figure}[H]
    \centering
    \includegraphics[width=0.925\columnwidth]{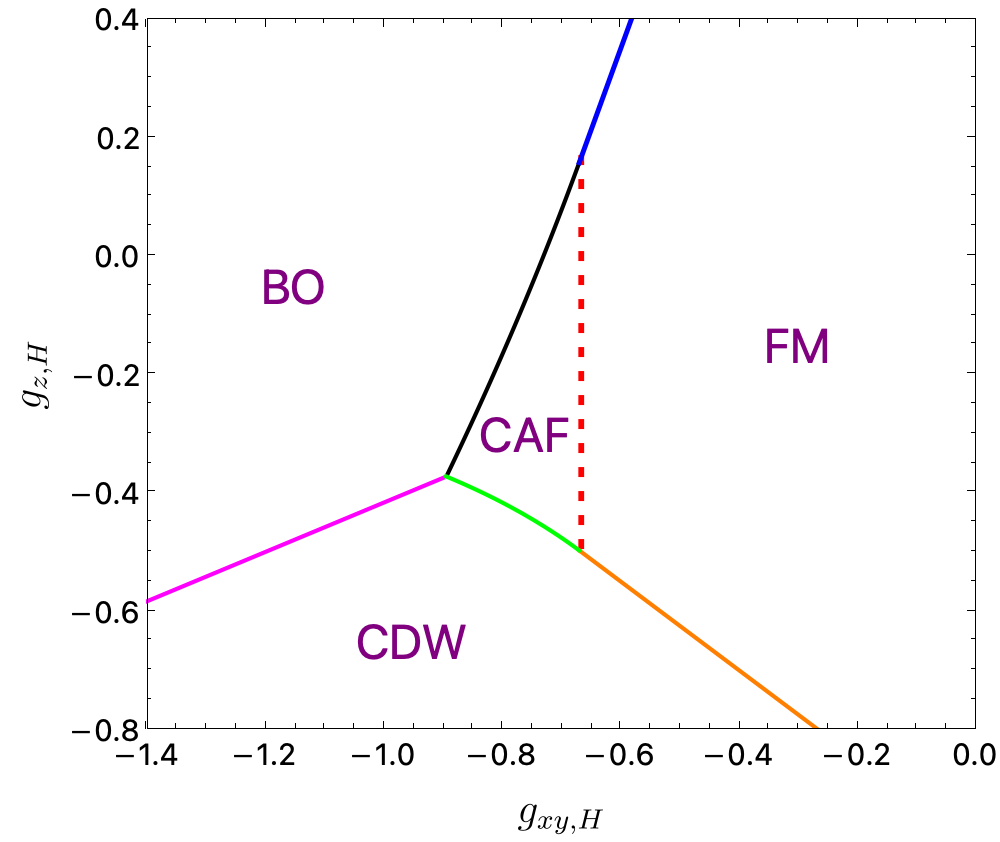}
    \caption{Phase diagram for $g_{z,F}/g_{z,H}=-1$, and $g_{xy,F}/g_{xy,H}=0.75$, $E_Z=1$ and $E_V=0.0$. The topology of the phase diagram changes; in particular, the CAF region no longer extends to infinity, but is enclosed by the other phases. There is no coexistence anywhere in the phase diagram. }  
    \label{fig:PDS1}
\end{figure}

\begin{figure}[H]
    \centering
    \includegraphics[width=0.925\columnwidth]{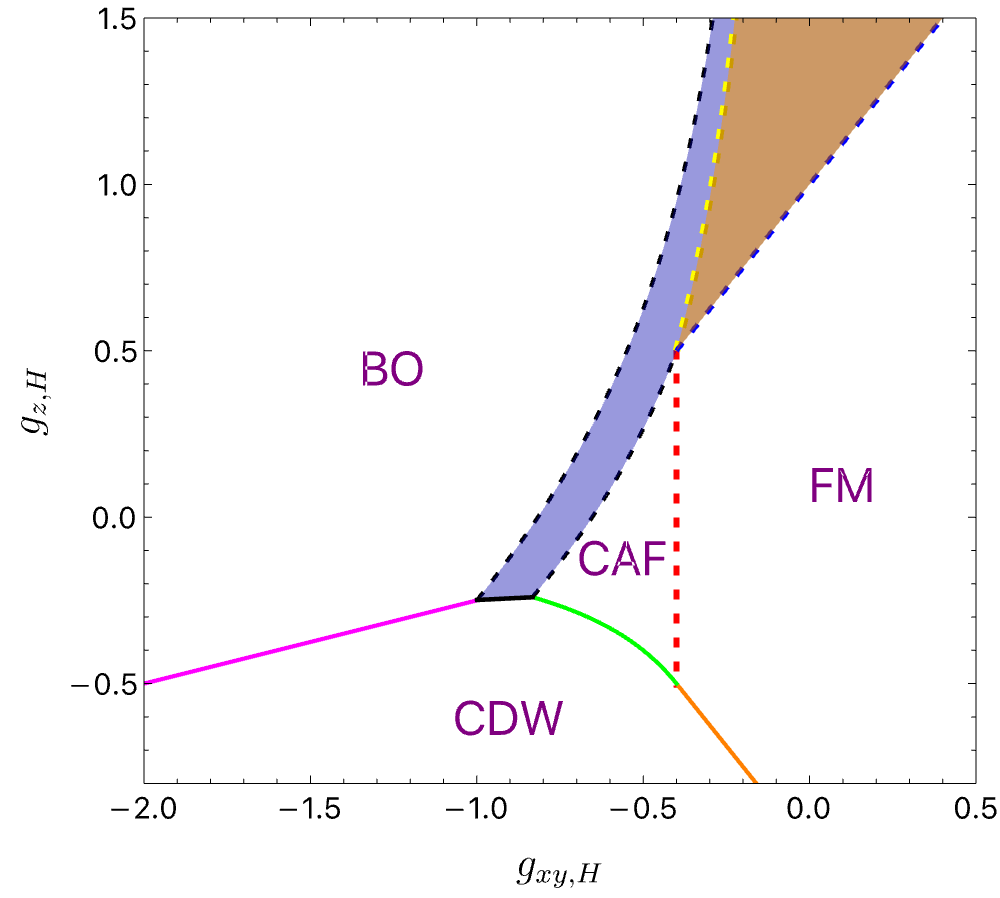}
    \caption{Phase diagram for $g_{z,F}/g_{z,H}=-1$, and $g_{xy,F}/g_{xy,H}=1.25$, $E_Z=1$ and $E_V=0.0$. The blue shaded region is the B/CAF phase, which has BO, FM, CAF, and SVEY order. The brown shaded region is a new coexistence phase with FM and either of SVEX or SVEY order. }
    \label{fig:PDS2}
\end{figure}

The phase diagram for $g_{xy,F}/g_{xy,H}<1$ is shown in \cref{fig:PDS1}. The topology of the phase diagram changes dramatically, but there is no coexistence anywhere. The phase diagram for $g_{xy,F}/g_{xy,H}>1$ is shown in \cref{fig:PDS2}), in which there are two coexistence regions. The first one is the B/CAF phase, with BO, FM, CAF, and SVEY order, which we already encountered when $g_{z,F}/g_{z,H}>0$. The second phase, which we call the FSVE phase, only occurs when $g_{z,F}/g_{z,H}<0$. 

\begin{figure}[H]
    \centering
    \includegraphics[width=0.925\columnwidth]{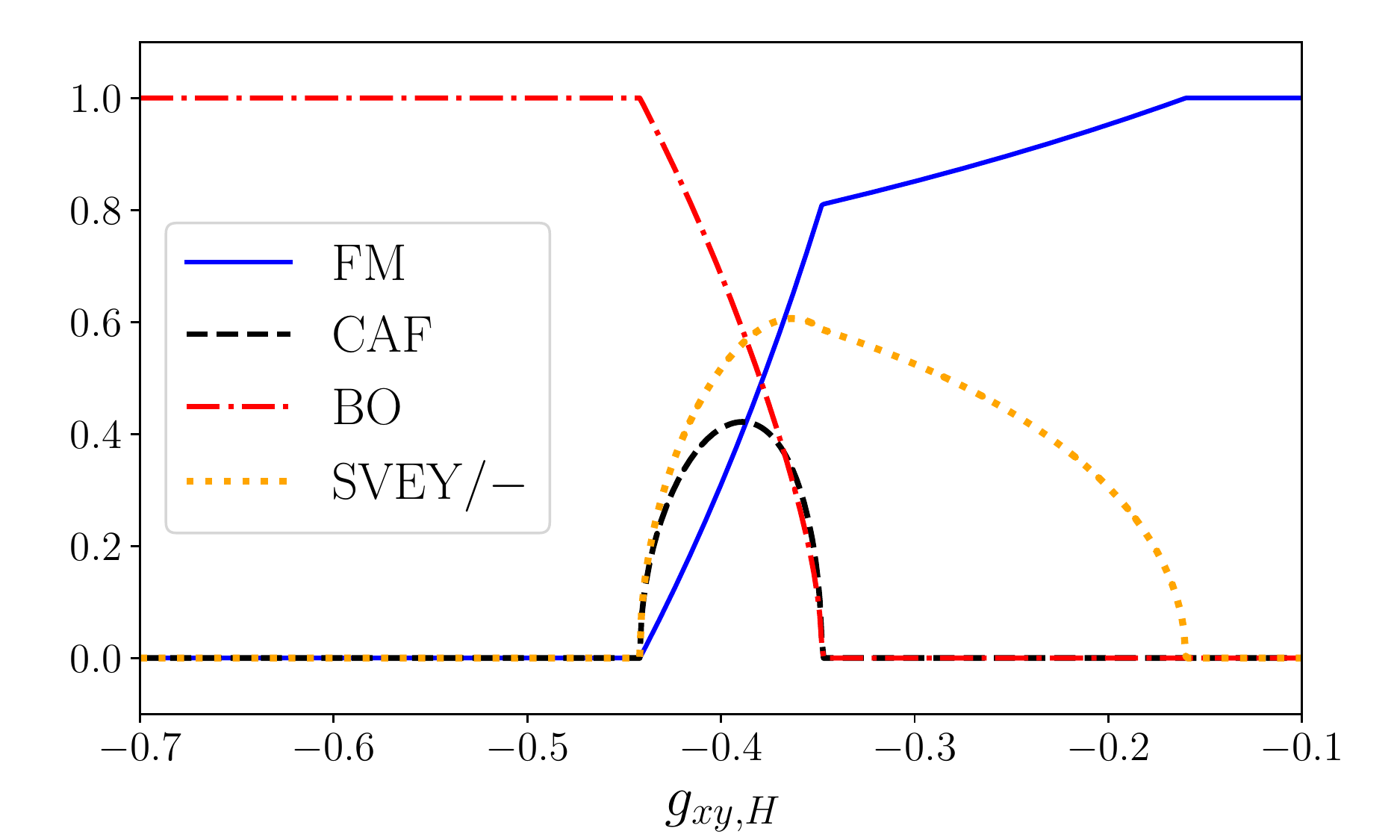}
    \caption{Nonzero order parameters vs $g_{xy,H}$ for $g_{z,H}=0.8$. All other coupling constants are the same as in \cref{fig:PDS2}. The B/CAF phase shows BO, FM, and SVEY (=SVE-) order, while the FSVE phase shows FM and SVEY order.}
    \label{fig:OPS1}
\end{figure}

At $E_V=0$ this phase shows the coexistence of FM and either SVEX or SVEY order (they are degenerate). \cref{fig:OPS1} shows the order parameters along a horizontal section of the phase diagram at $g_{z,H}=0.8$. 

\begin{figure*}
    \centering
    \includegraphics[width=\textwidth]{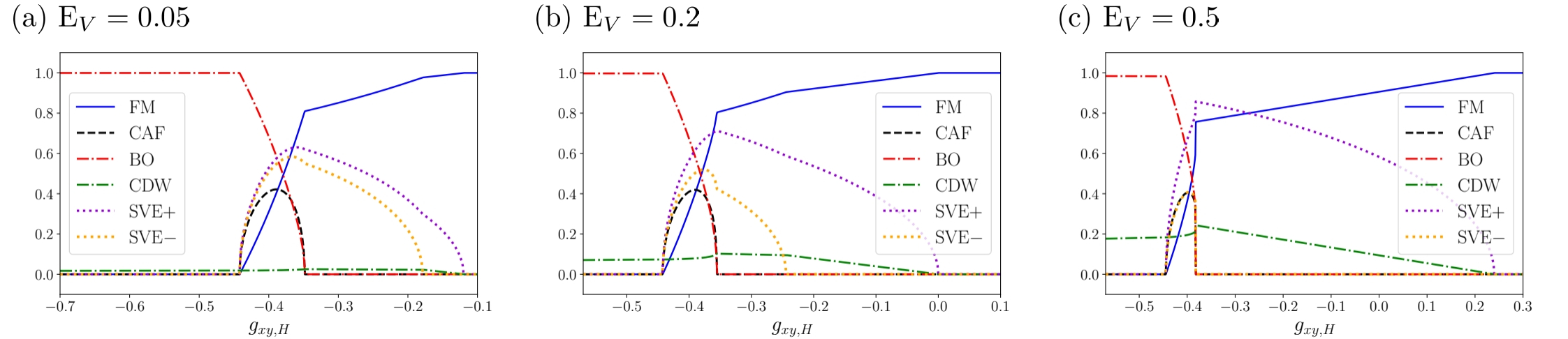}
    \caption{Order parameters vs $g_{xy,H}$ for various values of $E_V$, with 
    $E_Z=1.0,~g_{z,F}=-1.0,~g_{z,H}=0.8,~g_{xy,F}=1.25g_{xy,H}$. (a) For a tiny $E_V=0.05$ the FSVE phase splits into two phases. One of them has CDW, FM, and unequal SVEX and SVEY order parameters, while the other is the familiar SVE+ phase with CDW, FM, and SVE+ order.(b) As  $E_V$ increases to $0.2$, the B/CAF and FSVE phases shrink, while the SVE+ phase expands at their expense.  (c) At $E_V=0.5$ the FSVE phase has vanished, leaving behind the B/CAF and SVE+ phases. }
    \label{fig:nonzeroEvorder}
\end{figure*}

\cref{fig:nonzeroEvorder} shows the changes that occur on this section when $E_V$ is turned on.  The B/CAF phase changes in a familiar manner, acquiring a CDW order parameter as well as unequal SVEX and SVEY order parameters in addition to the BO and CAF order already present. The FSVE phase splits into two phases, both having some CDW order. The first has FM and unequal SVEX and SVEY order parameters, while the second is the familiar SVE+ phase. 

\begin{figure}[H]
    \centering
    \includegraphics[width=0.925\columnwidth]{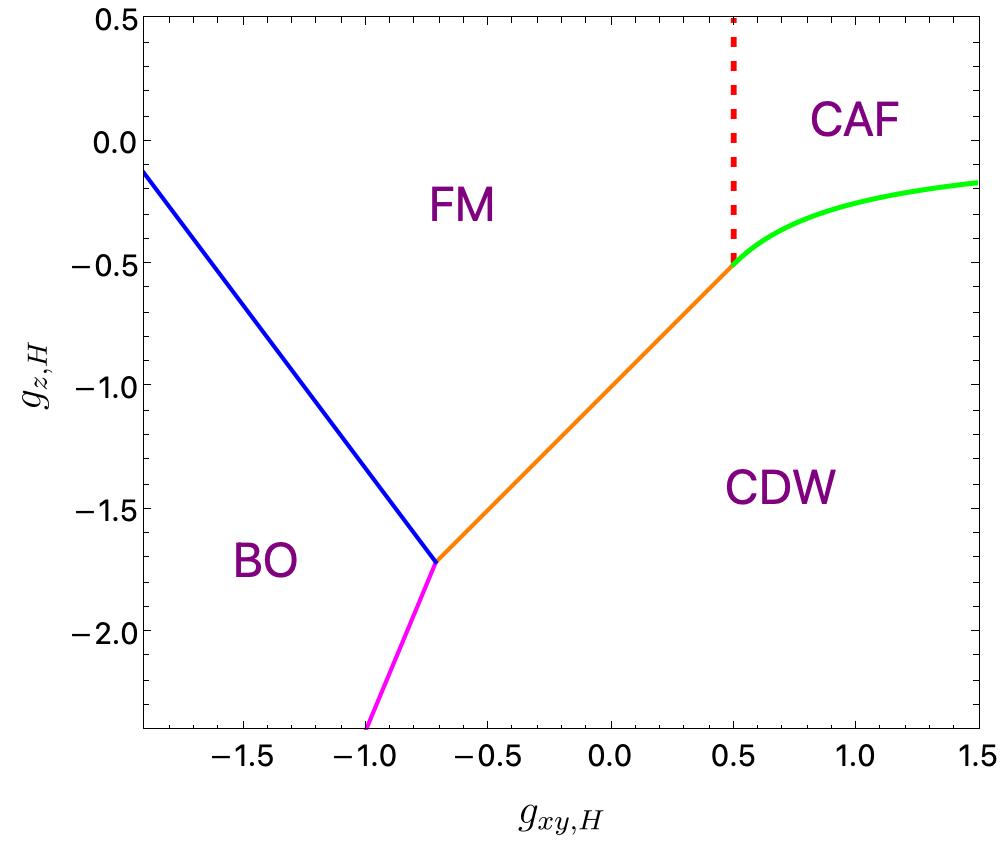}
    \caption{For this phase diagram we have considered $E_Z=1.0,E_V=0.0,g_{z,F}=0.75*g_{z,H},g_{xy,F}=-1.0*g_{xy,H}$. One can clearly see that as here $g_{xy,F}=-1.0*g_{xy,H}$ CAF is the stable phase for $g_{xy,H}>\frac{E_Z}{2}$.}
    \label{fig:PDS3}
\end{figure}

\begin{figure}[H]
    \centering
    \includegraphics[width=0.925\columnwidth]{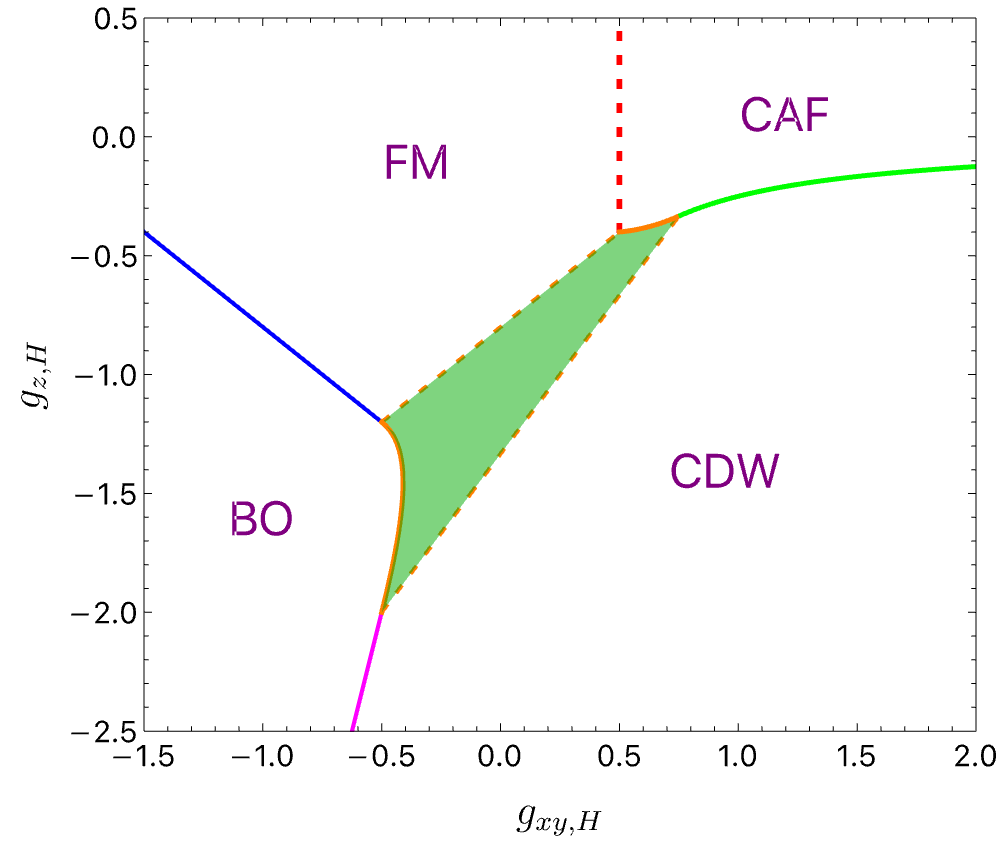}
    \caption{Phase diagram for  $E_Z=1.0,E_V=0.0,g_{z,F}=1.25*g_{z,H},g_{xy,F}=-1.0*g_{xy,H}$. The topology of the phase diagram is different from the earlier cases, but now the SVE+ coexistence phase appears between the FM and CDW phases.  }
    \label{fig:PDS4}
\end{figure}

Let us now go on to case (ii), $g_{z,F}/g_{z,H}>0;\ g_{xy,F}/g_{xy,H}<0$. We choose  $g_{xy,F}=-1.0*g_{xy,H}$. \cref{fig:PDS3} shows the phase diagram for $g_{z,F}/g_{z,H}<1$. The topology again changes dramatically, but there are no coexistence phases. \cref{fig:PDS4} shows the phase diagram for 
$g_{z,F}/g_{z,H}>1$.  The familiar SVE+ phase interpolates between the FM and CDW phases.

\begin{figure}[H]
    \centering
    \includegraphics[width=0.925\columnwidth]{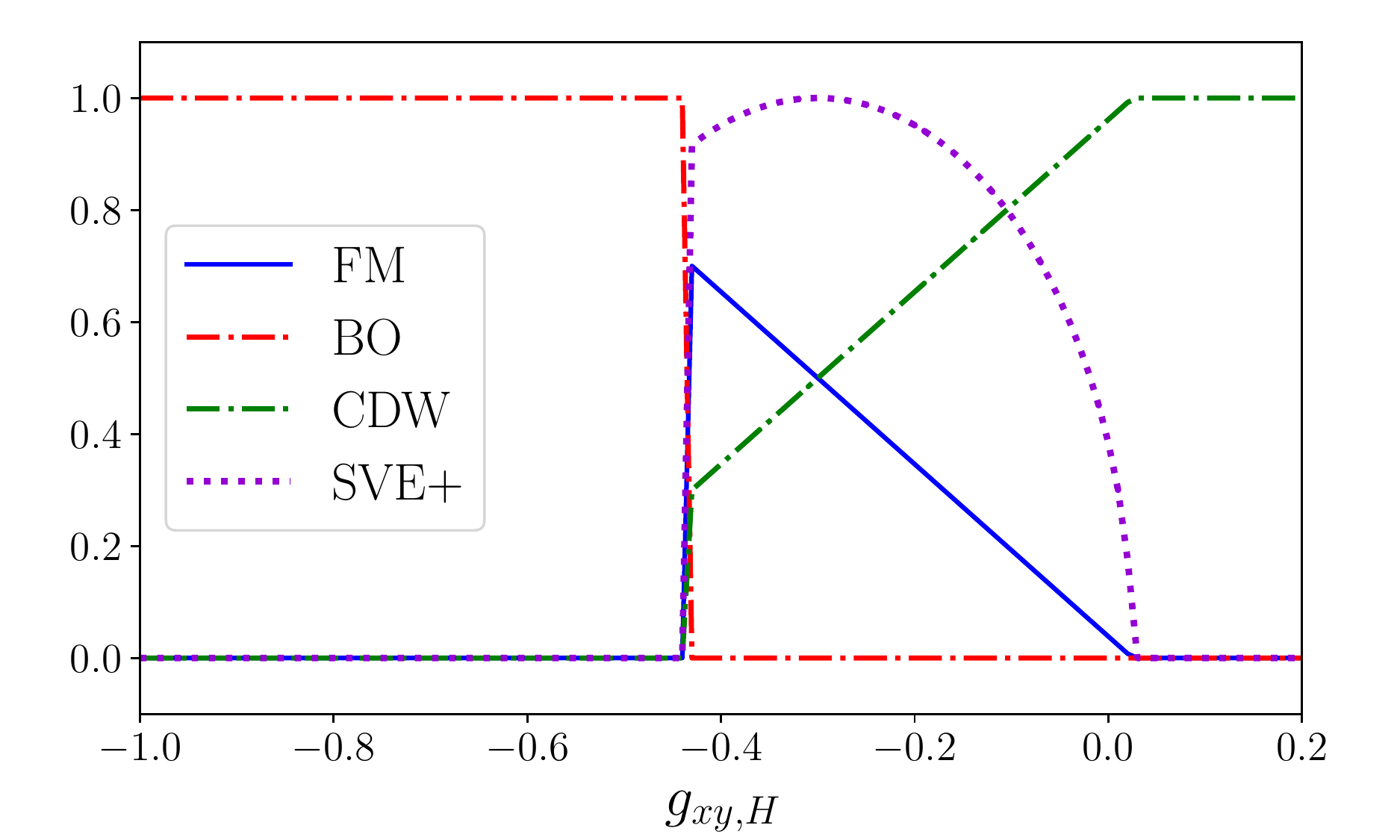}
    \caption{Nonzero order parameters vs $g_{xy,H}$ for $g_{z,H}=-1.3$. All other coupling 
    constants are chosen as in \cref{fig:PDS4}. The system is in the BO phase at the extreme
    left, makes a first-order transition into the SVE+ phase, and finally, a second-order
    transition into the CDW phase at the extreme right. }
    \label{fig:OPS2}
\end{figure}

\cref{fig:OPS2} shows the order parameters vs $g_{xy,H}$ along a section of \cref{fig:PDS4} at $g_{z,H}=-1.3$. One sees the first-order transition from the BO to the SVE+ phase, and the subsequent second-order transition into the CDW phase.

\begin{figure}[H]
    \centering
    \includegraphics[width=0.925\columnwidth]{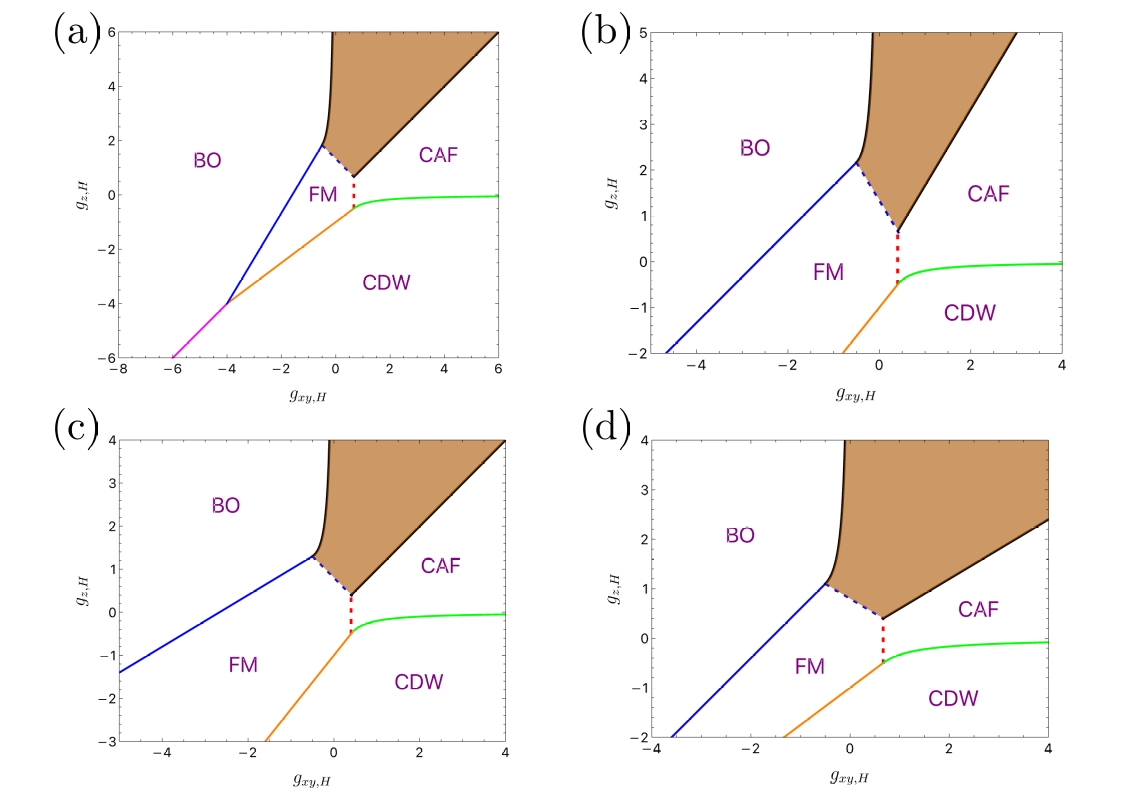}
    \caption{Phase diagrams for case (iii) $g_{z,F}/g_{z,H}<0;\ g_{xy,F}/g_{xy,H}<0$. 
    (a) $E_Z=1.0,E_V=0.0,g_{z,F}=-0.75*g_{z,H},g_{xy,F}=-0.75*g_{xy,H}$
    (b) $E_Z=1.0,~E_V=0.0,~g_{z,F}=-0.75*g_{z,H},~g_{xy,F}=-1.25*g_{xy,H}$
    (c) $E_Z=1.0,~E_V=0.0,~g_{z,F}=-1.25*g_{z,H},~g_{xy,F}=-1.25*g_{xy,H}$
    (d) $E_Z=1.0,~E_V=0.0,~g_{z,F}=-1.25*g_{z,H},~g_{xy,F}=-0.75*g_{xy,H}$. The brown shaded region denotes the FSVE phase in all the figures.}
    \label{fig:SignPD}
\end{figure}

Finally we turn to case (iii), $g_{z,F}/g_{z,H}<0;\ g_{xy,F}/g_{xy,H}<0$. The various panels
of \cref{fig:SignPD} show the phase diagrams for different F/H ratios. The common feature is
the presence of the FSVE coexistence phase.

\onecolumngrid

\twocolumngrid
\bibliographystyle{apsrev}
\bibliography{hall}



\end{document}